%% file: arXiv2.tex
\documentclass[12pt,letterpaper]{article}

\usepackage[letterpaper,left=1.25in,right=1.25in,top=1.25in,bottom=1.25in]{geometry}
\usepackage{setspace}
\usepackage{graphicx}
\usepackage{enumitem}
\usepackage{microtype}
\usepackage{etoolbox}
\usepackage{subfigure}
\usepackage{booktabs}
\usepackage{comment}
\usepackage{wrapfig}
\usepackage{arydshln}
\usepackage{multirow}
\usepackage{url}
\usepackage{natbib}
\usepackage{bibunits}
\usepackage{authblk}
\usepackage{amsmath}
\usepackage{amssymb}
\usepackage{mathtools}
\usepackage{amsfonts}
\usepackage{bm}
\usepackage{bbm}
\RequirePackage[hypertexnames=false,colorlinks,citecolor=blue,linkcolor=blue,urlcolor=blue]{hyperref}
\AtBeginEnvironment{thebibliography}{\sloppy}
\usepackage{algorithm}
\usepackage{algorithmic}

\def\eqref#1{equation~\ref{#1}}
\def\1{\bm{1}}

\def\rmd{{\mathrm{d}}}

\def\bmzero{{\bm{0}}}

\def\bmtheta{{\bm{\theta}}}

\def\bms{{\bm{s}}}

\def\bmv{{\bm{v}}}
\def\bmw{{\bm{w}}}

\def\bmalpha{{\bm{\alpha}}}
\def\bmbeta{{\bm{\beta}}}

\def\bmdelta{{\bm{\delta}}}

\def\bmtheta{{\bm{\theta}}}

\def\bmrho{{\bm{\rho}}}

\def\bmphi{{\bm{\phi}}}

\def\bmSigma{{\bm{\Sigma}}}

\DeclareMathAlphabet{\mathsfit}{\encodingdefault}{\sfdefault}{m}{sl}
\SetMathAlphabet{\mathsfit}{bold}{\encodingdefault}{\sfdefault}{bx}{n}

\def\calA{{\mathcal{A}}}
\def\calB{{\mathcal{B}}}

\def\calD{{\mathcal{D}}}

\def\calF{{\mathcal{F}}}
\def\calG{{\mathcal{G}}}
\def\calH{{\mathcal{H}}}
\def\calI{{\mathcal{I}}}

\def\calN{{\mathcal{N}}}

\def\calR{{\mathcal{R}}}

\def\calU{{\mathcal{U}}}

\def\calX{{\mathcal{X}}}
\def\calY{{\mathcal{Y}}}
\def\calZ{{\mathcal{Z}}}

\def\bbE{{\mathbb{E}}}

\def\bbI{{\mathbb{I}}}

\def\bbP{{\mathbb{P}}}

\def\bbR{{\mathbb{R}}}

\newcommand{\Var}{\mathrm{Var}}

\DeclareMathOperator*{\argmin}{arg\,min}

\DeclareMathOperator{\sign}{sign}

\newcommand{\p}[1]{\left(#1\right)}
\newcommand{\sqb}[1]{\left[#1\right]}
\newcommand{\cb}[1]{\left\{#1\right\}}

\newcommand{\bigp}[1]{\big(#1\big)}
\newcommand{\bigsqb}[1]{\big[#1\big]}
\newcommand{\bigcb}[1]{\big\{#1\big\}}

\newcommand{\Bigp}[1]{\Big(#1\Big)}
\newcommand{\Bigsqb}[1]{\Big[#1\Big]}

\newcommand{\Biggp}[1]{\Bigg(#1\Bigg)}

\newcommand{\Biggcb}[1]{\Bigg\{#1\Bigg\}}

\newcommand{\abs}[1]{\left|#1\right|}

\newcommand{\norm}[1]{\left\|#1\right\|}

\newcommand{\annot}[2]{\underbrace{#1}_{\text{#2}}}

\newcommand{\BigExp}[1]{\mathbb{E}\Big[#1\Big]}

\mathtoolsset{showonlyrefs=true}

\usepackage{amsthm}

\theoremstyle{plain}

\newtheorem{theorem}{Theorem}[section]

\newtheorem{corollary}[theorem]{Corollary}
\newtheorem{proposition}[theorem]{Proposition}

\newtheorem{definition}{Definition}[section]
\newtheorem{assumption}{Assumption}[section]

\newtheorem*{remark}{Remark}

\usepackage{multirow}
\usepackage{wrapfig}
\usepackage{subfigure}

\renewcommand{\eqref}[1]{(\ref{#1})}

\usepackage{xcolor}
\newcount\Comments  % 0 suppresses notes to selves in text
\newcommand{\kibitz}[2]{\ifnum\Comments=1\textcolor{#1}{#2}\fi}

\mathtoolsset{showonlyrefs=false}
\mathtoolsset{showonlyrefs=true}
\allowdisplaybreaks
\setlist[itemize]{topsep=2pt,itemsep=1pt,parsep=0pt,partopsep=0pt}
\setlist[enumerate]{topsep=2pt,itemsep=1pt,parsep=0pt,partopsep=0pt}
\title{Generalized Riesz Regression: A Unified Framework for Debiased Machine Learning with Riesz Representer Fitting under Bregman Divergence}
\author{Masahiro Kato\thanks{Email: \texttt{mkato-csecon@g.ecc.u-tokyo.ac.jp}. I thank Xiaohong Chen for insightful comments. This study builds on our previous study, ``Direct Bias-Correction Term Estimation for Propensity Scores and Average Treatment Effect Estimation,'' available on arXiv (arXiv:2509.22122) \citep{Kato2025directbias} and OpenReview \citep{Kato2025directbias2}. This study refines and generalizes those results and incorporates material from our subsequent studies, ``A Unified Theory for Causal Inference: Direct Debiased Machine Learning via Bregman-Riesz Regression'' (arXiv:2510.26783) \citep{Kato2025unifiedtheory} and ``Direct Debiased Machine Learning via Bregman Divergence Minimization'' (arXiv:2510.23534). We do not intend to submit the earlier versions for publication.}$\,$}
\affil{The University of Tokyo\\Mizuho-DL Financial Technology, Co., Ltd.\\Osaka Metropolitan University}
\date{\today}
\defaultbibliographystyle{tmlr}
\defaultbibliography{Bibtex/others,Bibtex/causalinference,Bibtex/rdd,Bibtex/semiparametric,Bibtex/nonparametric,Bibtex/highdimensional,Bibtex/experimentaldesign,Bibtex/reinforcement,Bibtex/weaksupervised,Bibtex/machinelearning,Bibtex/neuralnet,Bibtex/conditionalaveragetreatmenteffect,Bibtex/policylearning,Bibtex/statistics,Bibtex/bandits,Bibtex/finance,Bibtex/conformalinference,Bibtex/optimaltransport,Bibtex/treatment_choice,Bibtex/sequentialtesting,Bibtex/nlp,Bibtex/diffusionprocess}

\begin{document}
\begin{bibunit}
\maketitle

\input{Main}

\clearpage
\putbib
\end{bibunit}

\clearpage
\thispagestyle{empty}
\begin{center}
{\Large\bfseries Appendix to\\[0.4em]
``Generalized Riesz Regression: A Unified Framework for Debiased Machine Learning with Riesz Representer Fitting under Bregman Divergence''\par}
\vspace{1.2em}
{\large Masahiro Kato\par}
\vspace{0.4em}
{The University of Tokyo, Mizuho-DL Financial Technology, Co., Ltd., and Osaka Metropolitan University\par}
\end{center}
\clearpage

\begin{bibunit}
\input{Supplementary}

\clearpage
\putbib
\end{bibunit}
\end{document}

%% file: Main.tex
\begin{abstract}
Estimating the Riesz representer is central to debiased machine learning, yet the generator and representer model determine which regression directions their first-order conditions protect. We introduce generalized Riesz regression, which minimizes a Bregman divergence made observable by the Riesz identity. Squared and Kullback--Leibler-type choices recover Riesz regression, tailored loss minimization, and density-ratio objectives. For any twice-differentiable generator and differentiable representer model, the first-order conditions impose empirical Riesz equations in model-dependent tangent directions. Compatibility aligns those directions with regressors chosen in advance. These equations give sharp control of the systematic Neyman error and an orthogonal-score identity under exact balance. We derive convergence rates for sparse models linear in dual coordinates, reproducing kernel Hilbert space models, and neural networks, with generator curvature entering the sparse rate. We establish asymptotic normality under Donsker conditions or nuisance estimation via cross-fitting. Applications include treatment effects, average marginal effects, and covariate shift.
\end{abstract}

\section{Introduction}
The Riesz representer plays a central role in debiased machine learning for causal and structural parameter estimation \citep{Chen2015sievesemiparametric,Chernozhukov2022automaticdebiased}. Examples include the Average Treatment Effect (ATE) \citep{Imbens2015causalinference}, the Average Marginal Effect (AME), the Average Policy Effect (APE), and parameters under covariate shift \citep{Shimodaira2000improvingpredictive}. The Riesz representer arises from the Riesz representation theorem for a typically linear parameter functional and is closely connected to the semiparametric efficiency bound \citep{Newey1994theasymptotic}. An estimator of the Riesz representer supplies the adjustment term in the efficient influence function.

In many applications, the Riesz representer has a closed-form expression in terms of other nuisance parameters, but estimating those components and then transforming them does not directly target the representer. For the ATE, the usual construction estimates the propensity score and takes its inverse. Under covariate shift, the representer is a density ratio,\footnote{Throughout, \emph{density ratio} is the conventional name for the Radon--Nikodym derivative between two probability laws. Continuous random variables and Lebesgue densities are not required. On a discrete support, the same object is the ratio of probability mass functions.} while separate estimation of the source and target laws compounds two estimation problems.

Direct estimators instead use the target functional to estimate the representer itself. Examples include Riesz regression \citep{Chen2014sieveinference,Chernozhukov2021automaticdebiased}, entropy balancing \citep{Hainmueller2012entropybalancing}, stable balancing weights \citep{Zubizarreta2015stableweights}, tailored loss minimization \citep{Zhao2019covariatebalancing}, calibrated estimation \citep{Tan2019regularizedcalbrated}, and direct density-ratio estimation \citep{Sugiyama2012densityratio}. These methods were developed for different problems, but their first-order conditions can be compared through empirical Riesz equations. The central question is which functions enter those equations and therefore which components of regression error receive direct protection in the orthogonal score.

This study develops a general answer. Generalized Riesz regression fits a representer model under a Bregman divergence and uses the Riesz identity to remove the unknown representer from the population objective. For an arbitrary smooth pair, the first-order conditions act on model-dependent tangent directions. Compatibility aligns those directions with regressors chosen in advance, after which the empirical Riesz equations control the corresponding component of the Neyman error. The formulation applies to general linear functionals and to signed representers on sign-specific components determined by the target functional.

\subsection{Setup}
\label{sec:setup}
We first describe a baseline setup that covers many causal and structural parameters with a single i.i.d.\ sample. Extensions to multi-sample settings, such as covariate shift adaptation, are presented in Section~\ref{sec:application}.

We denote the observation by $W\coloneqq(X,Y)$, where $Y\in\calY\subseteq\bbR$ is an outcome and $X\in\calX\subseteq\bbR^k$ is a regressor vector. Let $P_0$ be the distribution of $W$, and let $\calD\coloneqq\bigcb{W_i}_{i=1}^n$ be an i.i.d.\ sample from $P_0$. We denote the regression function by $\gamma_0(x)\coloneqq\bbE_{P_0}[Y\mid X=x]$ and omit the subscript $P_0$ when the dependence is clear.

\paragraph{Parameter functional.}
Our goal is to estimate a parameter of interest of the form $\theta_0\coloneqq\bbE[m(W,\gamma_0)]$, where $m(W,\gamma)$ is defined for generic regression functions $\gamma\colon\calX\to\calY$. We assume that the map $\gamma\mapsto m(W,\gamma)$ is linear almost surely. This condition holds for the functionals considered below.

\paragraph{Riesz representer.}
Let $\calH\coloneqq\cb{\gamma\colon\calX\to\bbR\mid\bbE[\gamma(X)^2]<\infty}$ be $L_2(P_X)$ with inner product $\langle f,g\rangle\coloneqq\bbE[f(X)g(X)]$. If $\gamma\mapsto\bbE[m(W,\gamma)]$ is linear and continuous on $\calH$, the Riesz representation theorem gives a unique $\alpha_0\in\calH$ such that
\begin{align}
\label{eq:riesz_identity}
\bbE[m(W,\gamma)]=\bbE[\alpha_0(X)\gamma(X)]
\end{align}
for every $\gamma\in\calH$. We call $\alpha_0$ the Riesz representer \citep{Chen2015sievesemiparametric,Chernozhukov2022automaticdebiased}, and Appendix~\ref{appdx:nonlinear_estimands_gamma} treats nonlinear maps through local Riesz representers.

\paragraph{Neyman orthogonal score.}
Let $\eta_0\coloneqq(\gamma_0,\alpha_0)$ and define
\begin{align}
\label{eq:orthogonal_score}
\psi(W;\eta,\theta)\coloneqq m(W,\gamma)+\alpha(X)(Y-\gamma(X))-\theta.
\end{align}
It holds that $\bbE[\psi(W;\eta_0,\theta_0)]=0$. The Gateaux derivative with respect to $\alpha$ along a direction $b$ equals $\bbE[b(X)(Y-\gamma_0(X))]=0$, while the derivative with respect to $\gamma$ vanishes by Equation~\eqref{eq:riesz_identity}. Therefore, the score is Neyman orthogonal at $\eta_0$, and its population error has the exact form
\begin{align}
\label{eq:orth_score_bias_identity}
\bbE[\psi(W;\eta,\theta_0)]
=\bbE[(\alpha_0(X)-\alpha(X))(\gamma(X)-\gamma_0(X))].
\end{align}

Given nuisance parameter estimators $\widehat\eta=(\widehat\gamma,\widehat\alpha)$, the Augmented Riesz Weighted (ARW) estimator is
\begin{align}
\label{eq:arw_estimator}
\widehat\theta^{\mathrm{ARW}}
\coloneqq\frac1n\sum_{i=1}^n\Bigp{m(W_i,\widehat\gamma)+\widehat\alpha(X_i)(Y_i-\widehat\gamma(X_i))}.
\end{align}
We also use the regression adjustment estimator $\widehat\theta^{\mathrm{RA}}=n^{-1}\sum_i m(W_i,\widehat\gamma)$ and the Riesz weighted estimator $\widehat\theta^{\mathrm{RW}}=n^{-1}\sum_i\widehat\alpha(X_i)Y_i$. In ATE estimation, these estimators correspond to regression adjustment, inverse probability weighting, and augmented inverse probability weighting. Section~\ref{sec:comptmle} gives the relation to targeted maximum likelihood estimation (TMLE).

\subsection{Contributions of This Study}
\label{sec:contribution}
Generalized Riesz regression connects representer fitting, balancing, and orthogonal-score construction by identifying the regression directions protected by the first-order conditions. The Riesz identity converts a Bregman divergence from the unknown representer into an observable objective. Squared distance recovers Riesz regression, an unnormalized Kullback--Leibler divergence recovers tailored loss minimization, and the same formulation contains direct density-ratio objectives \citep{Chernozhukov2021automaticdebiased,Zhao2019covariatebalancing}. Table~\ref{tbl:dre_rre} records the principal correspondences.

\paragraph{Protected directions and score control.}
Proposition~\ref{prop:arbitrary_pair_foc} identifies the tangent directions generated by any smooth pair. Proposition~\ref{prop:compatible_characterization} characterizes the pairs that align those directions with regressors selected before estimation, and Theorem~\ref{thm:autocovariance} gives the resulting exact or regularization-controlled empirical Riesz equations. Equation~\eqref{eq:automatic_neyman_min} gives sharp dual-norm control of the systematic score error on the protected linear space. Exact balance yields the finite-sample orthogonal-score identity in Theorem~\ref{thm:automaticneymanorthogonalization}, while Equation~\eqref{eq:inexact_balance_bound} isolates approximation error outside that space.

\paragraph{Rates, inference, and specification.}
Theorem~\ref{thm:sparse_dual_rate} derives a high-dimensional rate for sparse models linear in dual coordinates and displays the generator curvature constant. We also establish rates for RKHS and neural-network models, asymptotic normality of the ARW estimator, and asymptotic linearity of the Riesz weighted estimator under approximate balance. Propositions~\ref{prop:summary_space_projection} and~\ref{prop:generator_condition_numbers} distinguish the function space protected by balance from the curvature governing representer sensitivity. These results lead to diagnostics based on empirical imbalance, function-space coverage, and weight concentration.

Sections~\ref{sec:generalizedrieszregression}--\ref{sec:convanalysis} develop the objective, balancing theory, score identities, specification principles, and inference. Sections~\ref{sec:application} and~\ref{sec:experiments} present applications and experiments, and the appendices contain proofs, extended comparisons, and additional applications.

\section{Generalized Riesz Regression}
\label{sec:generalizedrieszregression}
We estimate the Riesz representer by fitting a representer model to the true representer under a Bregman divergence. The Bregman divergence contains squared distance and Kullback--Leibler-type divergences as special cases. We use the name generalized Riesz regression because the choice of divergence is tied to the link function of the representer model through the balancing conditions in Section~\ref{sec:automaticcovariatebalancing}.

Let $\alpha\colon\calX\to\calA$ be a candidate representer and let $g\colon\calA\to\bbR$ be differentiable and strictly convex. The set $\calA$ is the range allowed for the representer. Some losses are defined on all of $\bbR$, whereas others require the sign of the representer to be known from the target functional.

\subsection{Bregman Divergence}
For $x\in\calX$, define
\[
\mathrm{BD}^{\dagger}_g(\alpha_0(x)\mid\alpha(x))
\coloneqq g(\alpha_0(x))-g(\alpha(x))-\partial g(\alpha(x))(\alpha_0(x)-\alpha(x)).
\]
The population target minimizes the expectation of this divergence over a model class $\calH_\alpha$. Although $\alpha_0$ is unknown, Equation~\eqref{eq:riesz_identity} gives the equivalent objective
\begin{align}
\label{eq:population_bregman}
\mathrm{BD}_g(\alpha)
\coloneqq\bbE\Bigsqb{-g(\alpha(X))+\partial g(\alpha(X))\alpha(X)-m(W,(\partial g)\circ\alpha)}.
\end{align}
The equality holds up to the constant $\bbE[g(\alpha_0(X))]$. We estimate $\alpha_0$ by
\begin{align}
\label{eq:empbregman}
\widehat\alpha\in\argmin_{\alpha\in\calH_\alpha}
\left\{\widehat{\mathrm{BD}}_g(\alpha)+\lambda J(\alpha)\right\},
\end{align}
where $\widehat{\mathrm{BD}}_g$ is the sample analogue of Equation~\eqref{eq:population_bregman}. If $\alpha_0\in\calH_\alpha$ and the model respects the domain of $g$, the population minimizer is $\alpha_0$. Under misspecification, the loss determines the Bregman projection of $\alpha_0$ onto $\calH_\alpha$.

\subsection{Special Cases of the Bregman Divergence}
The squared, unnormalized Kullback--Leibler, binary Kullback--Leibler, Basu power, and positive--unlabeled losses give the principal examples used in this study. We refer to the corresponding estimators as SQ-Riesz, UKL-Riesz, BKL-Riesz, BP-Riesz, and PU-Riesz regression. Table~\ref{tbl:dre_rre} records their relation to existing methods.

\subsection{SQ-Riesz Regression}
Let $C\in\bbR$ be a constant and consider the convex function
\[g^{\text{SQ}}(\alpha) = (\alpha - C)^2.\]
This generator is motivated by the squared loss. The constant $C$ shifts the affine parameterization. If the model contains an intercept, changing $C$ only reparameterizes the same affine class. Automatic regressor balancing is determined by the linearity of $\partial g\circ\alpha$ and not by $C$ alone. The derivative of $g^{\text{SQ}}(\alpha)$ with respect to $\alpha$ is
\begin{align*}
  \partial g^{\text{SQ}}(\alpha) &= 2(\alpha - C).
\end{align*}

Substituting $g^{\text{SQ}}$ into Equation~\eqref{eq:population_bregman} gives
\[\text{BD}_g\bigp{\alpha} \coloneqq \BigExp{\alpha(X)^2 - 2m\bigp{W, \bigp{\alpha(\cdot) - C}}}.\]
Then, the estimation problem can be written as
\begin{align}
\label{eq:squaredloss}
  \widehat{\alpha} \coloneqq \argmin_{\alpha \in \calH_\alpha}\widehat{\text{BD}}_{g^{\text{SQ}}}\bigp{\alpha} + \lambda J(\alpha),
\end{align}
where
\[\widehat{\text{BD}}_{g^{\text{SQ}}}\bigp{\alpha} \coloneqq \frac{1}{n}\sum^n_{i=1}\p{ \alpha(X_i)^2 - 2m\bigp{W_i, \bigp{\alpha(\cdot)}}}.\]
Terms that do not depend on $\alpha$ are omitted from the empirical objective. By linearity of $m$, the contribution of the constant $C$ is also independent of $\alpha$\footnote{The original population Bregman divergence objective using $g(\alpha) = (\alpha - C)^2$ in Equation~\eqref{eq:population_bregman} is given as
\begin{align*}
  \text{BD}_{g^{\text{SQ}}}\bigp{\alpha} &= \bbE\sqb{- (\alpha(X) - C)^2 + 2(\alpha(X) - C)\alpha(X) - 2m\bigp{W, (\alpha(\cdot) - C)}}\\
  &= \bbE\sqb{ \alpha(X)^2 - C^2 - 2m\bigp{W, \bigp{\alpha(\cdot) - C}}}.
\end{align*}
}.
This estimation method corresponds to Riesz regression in debiased machine learning \citep{Chernozhukov2021automaticdebiased} and least-squares importance fitting (LSIF) in density ratio estimation \citep{Kanamori2009aleastsquares}. An appropriate choice of $\calH$ also recovers nearest-neighbor matching, as shown by \citet{Kato2025nearestneighbor} using the argument of \citet{Lin2023estimationbased}. 

\subsection{UKL-Riesz Regression}
\label{sec:empbalancing}
We next consider a KL-divergence-motivated generator. Let $C<\inf_x|\alpha(x)|$ and define
\[g^{\text{UKL}}(\alpha) = (|\alpha| - C)\log\p{|\alpha| - C} - |\alpha|.\]
The constant $C$ determines the sign-specific domain and the compatible exponential link and must be specified jointly with the representer model. In the sigmoid ATE model in Equation~\eqref{eq:riesz_model1}, $C=1$ reproduces the usual inverse propensity representation. The derivative of $g^{\text{UKL}}(\alpha)$ with respect to $\alpha$ is
\begin{align*}
  \partial g^{\text{UKL}}(\alpha) &= \sign(\alpha)\log\p{|\alpha| - C}.
\end{align*}

Substituting $g^{\text{UKL}}$ into Equation~\eqref{eq:population_bregman} gives\footnote{
This Bregman divergence objective is derived as follows:
\begin{align*}
  \text{BD}_{g^{\text{UKL}}}\bigp{\alpha} &= \bbE\Bigsqb{- \p{|\alpha(X)| - C}\log\p{|\alpha(X)| - C} + |\alpha(X)| + \sign\p{\alpha(X)}\alpha(X) \log\p{|\alpha(X)| - C}\\
  &\ \ \ \ \ \ \ \ \ \ \ \ \ \ \ \ \ \ \ \ \ \ \ \ \ \ \ \ \ \ \ \ \ \ \ \ \ \ \ \ \ \ \ \ \ \ \ \ \ \ \ \ \ \ \ \ \ \ \ \ - m\Bigp{W, \sign\bigp{\alpha(\cdot)} \log\p{|\alpha(\cdot)| - C}}}.
\end{align*}
}:
\[\text{BD}_{g^{\text{UKL}}}\bigp{\alpha} \coloneqq \BigExp{C\log\p{|\alpha(X)| - C} + |\alpha(X)| - m\Bigp{W, \sign\bigp{\alpha(\cdot)} \log\p{|\alpha(\cdot)| - C}}}.\]

We estimate $\alpha_0$ by minimizing the empirical objective:
\[
\widehat{\alpha} \coloneqq \argmin_{\alpha \in \calH_\alpha}\widehat{\text{BD}}_{g^{\text{UKL}}}\bigp{\alpha} + \lambda J(\alpha),
\]
where
\begin{align*}
\widehat{\text{BD}}_{g^{\text{UKL}}}(\alpha)
={}&\frac{1}{n}\sum_{i=1}^n\Bigp{C\log(|\alpha(X_i)|-C)+|\alpha(X_i)|}\\
&-\frac{1}{n}\sum_{i=1}^n m\Bigp{W_i,\sign(\alpha)\log(|\alpha|-C)}.
\end{align*}

The UKL generator more closely corresponds to the standard KL divergence than the binary KL generator introduced in the next subsection. In density-ratio estimation, the corresponding constrained formulation is the Kullback--Leibler importance estimation procedure (KLIEP). Its equivalence to UKL divergence minimization is also known as Silverman's trick \citep{Silverman1978densityratios,KatoMinami2023unifiedperspective}, and the constraint formulation admits a dual interpretation. In ATE estimation, tailored loss minimization corresponds to UKL minimization, whose dual yields entropy balancing weights \citep{Hainmueller2012entropybalancing}.

\subsection{BKL-Riesz Regression}
We next introduce a binary Kullback--Leibler loss, which is related to logistic regression. Let $0<C<\inf_x|\alpha(x)|$ and define
\[
g^{\mathrm{BKL}}(\alpha)
\coloneqq(|\alpha|-C)\log(|\alpha|-C)-(|\alpha|+C)\log(|\alpha|+C).
\]
Its derivative is
\[
\partial g^{\mathrm{BKL}}(\alpha)
=\sign(\alpha)\log\left(\frac{|\alpha|-C}{|\alpha|+C}\right).
\]
Substitution into Equation~\eqref{eq:population_bregman} gives
\begin{align}
\label{eq:bkl_objective_main}
\mathrm{BD}_{g^{\mathrm{BKL}}}(\alpha)
=\bbE\left[C\log(|\alpha(X)|^2-C^2)
-m\left(W,\sign(\alpha)\log\left(\frac{|\alpha|-C}{|\alpha|+C}\right)\right)\right].
\end{align}
The empirical estimator minimizes the sample analogue of Equation~\eqref{eq:bkl_objective_main} plus $\lambda J(\alpha)$. In density-ratio estimation, the loss is connected to logistic classification and case--control likelihoods \citep{Qin1998inferencesfor,Cheng2004semiparametricdensity}. The loss, link, and target functional must be considered together: the usual Bernoulli likelihood with a sigmoid propensity-score model does not automatically impose the ATE Riesz equations.

\subsection{BP-Riesz Regression}
Basu's power (BP) divergence bridges the squared loss and KL divergence \citep{Basu1998robustandefficient}. Let $C<\inf_x|\alpha(x)|$ and define
\[g^{\text{BP}}(\alpha) \coloneqq \frac{\bigp{|\alpha| - C}^{1 + \omega} - \bigp{|\alpha| - C}}{\omega} - |\alpha|.\]
Its derivative is
\[\partial g^{\text{BP}}(\alpha) = \p{1 + \frac{1}{\omega}}\sign(\alpha)\Bigp{\bigp{|\alpha| - C}^{\omega} - 1}.\]
Substitution into the Bregman objective yields BP-Riesz regression. The constant $C$ determines the sign-specific domain and the compatible power link and is therefore part of the representer specification rather than a separate condition for balance.

The population and empirical objectives follow by substituting $g^{\mathrm{BP}}$ and its derivative into Equations~\eqref{eq:population_bregman} and~\eqref{eq:empbregman}. The Online Appendix gives the expanded expression. We estimate $\alpha_0$ by minimizing the empirical objective plus $\lambda J(\alpha)$.

Basu's power divergence bridges the squared loss and the (U)KL divergence. When $\omega=1$, the BP generator agrees with a squared generator up to affine terms on each sign-specific component of its domain. As $\omega\to0$, it converges to the UKL generator. This follows because
\[\lim_{\omega \to 0}\frac{\bigp{|\alpha| - C}^\omega - 1}{\omega} = \log\bigp{|\alpha| - C}.\]
UKL-Riesz regression corresponds to an exponential or sigmoid model for the Riesz representer. Misspecification of that model can reduce estimation accuracy. \citet{Sugiyama2012densityratio} compare the corresponding behavior of squared and Kullback--Leibler losses in density-ratio estimation. BP-Riesz regression provides an intermediate objective and preserves the dual form used for automatic regressor balancing.

\subsection{PU-Riesz Regression}
PU-Riesz regression uses the loss from positive--unlabeled learning. Let $\widetilde C>0$, restrict $\alpha$ to $(0,1)$, and define
\[
g^{\mathrm{PU}}(\alpha)
\coloneqq \widetilde C\log(1-\alpha)
+\widetilde C\alpha\bigp{\log(\alpha)-\log(1-\alpha)}.
\]
The constant $\widetilde C$ corresponds to the class prior in case--control positive--unlabeled learning and has a different role from $C$ in the preceding losses. The derivative is
\[
\partial g^{\mathrm{PU}}(\alpha)
=\widetilde C\bigp{\log(\alpha)-\log(1-\alpha)}.
\]
Substitution into Equation~\eqref{eq:population_bregman} gives
\begin{align*}
\mathrm{BD}_{g^{\mathrm{PU}}}(\alpha)
={}&-\widetilde C\,\bbE\sqb{\log(1-\alpha(X))}\\
&-\bbE\Bigp{m\bigp{W,\widetilde C\{\log(\alpha)-\log(1-\alpha)\}}}.
\end{align*}
We estimate $\alpha_0$ by minimizing the empirical analogue plus $\lambda J(\alpha)$.

PU learning is a classical problem. For example, \citet{Lancaster1996casecontrolstudies} study this problem under a stratified sampling scheme \citep{Wooldridge2001asymptoticproperties}. \citet{duPlessis2015convexformulation} rediscover this formulation and call it unbiased PU learning. \citet{Kato2019learningfrom} point out the relationship between PU learning and density ratio estimation, and \citet{Kato2021nonnegativebregman} show that PU learning is a special case of density ratio model fitting under a Bregman divergence. The formulation above extends these density-ratio results to Riesz representer estimation under the case-control PU sampling scheme, whereas censoring PU learning uses a different observation model \citep{Elkan2008learningclassifiers}. \citet{Kato2025puate} consider ATE estimation in a PU learning setup and apply our method in their study.

\begin{table*}[!t]
\caption{Correspondence between density-ratio estimation and Riesz representer estimation.}
\label{tbl:dre_rre}
\begin{center}
{\renewcommand{\arraystretch}{0.98}%
\resizebox{\linewidth}{!}{
\begin{tabular}{lll}
\hline
$g(\alpha)$ & \textbf{Density-ratio estimation} & \textbf{Riesz representer estimation} \\
\hline
\multirow{2}{*}{$(\alpha-C)^2$} & LSIF & SQ-Riesz regression \\
& \citep{Kanamori2009aleastsquares} & (\textbf{Ours}) \\
 & KuLSIF & Riesz regression (RieszNet and RieszForest) \\
& \citep{Kanamori2012statisticalanalysis} & \citep{Chernozhukov2021automaticdebiased,Chernozhukov2022riesznet} \\
& & RieszBoost \\
& & \citep{Lee2025rieszboost} \\
& & Kernel ridge Riesz regression \\
& & \citep{Singh2024kernelridge} \\
& & Nearest neighbor matching \\
& & \citep{Lin2023estimationbased} \\
 & & Causal tree and causal forest \\
& & \citep{Wager2018estimationinference} \\
\multicolumn{3}{c}{\textbf{Dual solution with a linear link function}} \\
 & Kernel mean matching & Sieve Riesz representer\\
& \citep{Gretton2009covariateshift} & \citep{Chen2015sievesemiparametric,Chen2015sievewald} \\
& & Stable balancing weights \\
& & \citep{Zubizarreta2015stableweights,BrunsSmith2025augmentedbalancing} \\
 & & Approximate Residual Balancing\\
 & & \citep{Athey2018approximateresidual} \\
 & & Covariate balancing by support vector machine \\
& & \citep{Tarr2025estimatingaverage} \\
 & & Distributional balancing \\
& & \citep{Santra2026distributionalbalancing} \\
\hline
\multirow{2}{*}{$\bigp{|\alpha| - C}\log \bigp{|\alpha| - C} - |\alpha|$} & UKL divergence minimization & UKL-Riesz regression\\
 & \citep{Nguyen2010estimatingdivergence} & (\textbf{Ours}) \\
 & & Tailored loss minimization ($\alpha=\beta=-1$)\\
 & & \citep{Zhao2019covariatebalancing} \\
 & & Calibrated estimation \\
 & & \citep{Tan2019regularizedcalbrated} \\
\multicolumn{3}{c}{\textbf{Dual solution with a logistic or log link function}} \\
& KLIEP & Entropy balancing weights \\
& \citep{Sugiyama2008directimportance} & \citep{Hainmueller2012entropybalancing} \\
\hline
$(|\alpha| - C)\log \bigp{|\alpha| - C}$ & BKL divergence minimization & BKL-Riesz regression \\
$\qquad - (|\alpha| + C)\log(|\alpha| + C)$ & \citep{Qin1998inferencesfor} & (\textbf{Ours}) \\
 & Telescoping ratio estimation & Logistic propensity-score plug-in \\
 & \citep{Rhodes2020telescopingdensityratio} & (Related baseline) \\
& & Tailored loss minimization ($\alpha=\beta=0$)\\
& & \citep{Zhao2019covariatebalancing}\\
\hline
$\frac{\bigp{|\alpha| - C}^{1 + \omega} - \bigp{|\alpha| - C}}{\omega} - |\alpha|$ & BP divergence minimization & BP-Riesz regression\\
for some $\omega \in (0, \infty)$ & \citep{Sugiyama2011densityratio} & (\textbf{Ours}) \\
\hline
$\widetilde C\log\left(1-\alpha\right)$ & PU learning & PU-Riesz regression\\
$\ \ \ \ + \widetilde C\alpha\left(\log\left(\alpha\right)-\log\left(1-\alpha\right)\right)$ & \citep{duPlessis2015convexformulation} & (\textbf{Ours})\\
$\ \ \ \ $ for $\alpha \in (0, 1)$ & Nonnegative PU learning & \\
 & \citep{Kiryo2017positiveunlabeledlearning} & \\
\hline
General formulation by Bregman & Density-ratio matching & Generalized Riesz regression \\
 divergence minimization & \citep{Sugiyama2011densityratio} & (\textbf{Ours}) \\
 & Deep direct density-ratio estimation & \\
 & \citep{Kato2021nonnegativebregman} & \\
\hline
\end{tabular}}}
\vspace{-5mm}
\end{center}
\end{table*}

Entries in the same row of Table~\ref{tbl:dre_rre} record different types of relationships, including identical objectives, common generators over different model classes, dual moment equations, and related methods. The logistic propensity-score plug-in estimates the propensity score by Bernoulli likelihood and then forms inverse weights. It is related to binary-loss density-ratio estimation, but it is not the direct BKL-Riesz objective defined above. Appendix~\ref{app:relations_dre_rre} gives the detailed relationships, and Appendix~\ref{app:bkl_family} gives the distinction for the binary Kullback--Leibler row. Figure~\ref{fig:bregman_concept} in the Online Appendix gives a schematic overview of these connections.

\section{Automatic Regressor Balancing}
\label{sec:automaticcovariatebalancing}
This section identifies the regression directions protected by the first-order conditions. Any smooth pair yields empirical Riesz equations in model-dependent tangent directions, while compatibility aligns those directions with regressors selected before estimation and thereby gives automatic regressor balancing.

\subsection{Generalized Linear Models}
\label{subsec:lin_score_models}
Let $\bmphi=(\phi_1,\ldots,\phi_p)^\top$ be regressor functions on $\calX$. We consider
\begin{align}
\label{eq:param_riesz_model}
\alpha_{\bmbeta}(x)=\zeta^{-1}(x,\bmphi(x)^\top\bmbeta),
\qquad \bmbeta\in\bbR^p,
\end{align}
where $\zeta^{-1}$ is an inverse link function. A linear model, $\alpha_{\bmbeta}(x)=C+\bmphi(x)^\top\bmbeta$, is available when the representer is modeled on its original scale.

Additional structure can be incorporated through the link. For the ATE, suppose that the propensity score is modeled as
\[
e_{\bmbeta}(Z)=\frac{1}{1+\exp(-\bmphi(Z)^\top\bmbeta)}.
\]
Substitution into the ATE representer gives
\begin{align}
\label{eq:riesz_model1}
\alpha^{\mathrm{ATE}}_{\bmbeta}(D,Z)
={}&D\Bigp{1+\exp(-\bmphi(Z)^\top\bmbeta)}\\
&-(1-D)\Bigp{1+\exp(\bmphi(Z)^\top\bmbeta)}.
\end{align}
The treatment indicator is part of the representer model even though the regressor functions in the propensity index depend only on $Z$. If the regression function contains treatment-specific components that are not well approximated by functions of $Z$ alone, the representer model may instead use regressor functions of $(D,Z)$. Under covariate shift, a common positive model is $\alpha_{\bmbeta}(X)=\exp(-\bmphi(X)^\top\bmbeta)$.

\subsection{Key Structural Requirement: Linearity in Dual Coordinates}
\label{subsec:loss_link_pairs}
Define the dual coordinate $u_{\bmbeta}=\partial g\circ\alpha_{\bmbeta}$. Automatic regressor balancing follows when this coordinate is linear in the coefficients. After an affine reparameterization of the index, we write
\begin{align}
\label{eq:score_linearization}
u_{\bmbeta}(x)=\sum_{j=1}^p\beta_j\widetilde\phi_j(x).
\end{align}
An intercept can be included among the functions $\widetilde\phi_j$. This convention leaves the representer model unchanged and gives the dual program below without an additional fixed linear term.

\begin{definition}[Compatible loss--link pair]
\label{def:compatible_pair}
Let $\ell\colon J\to A$ be a continuously differentiable and strictly increasing bijection between open intervals. A differentiable and strictly convex function $g\colon A\to\bbR$ and the link $\ell$ form a compatible loss--link pair if there are constants $a>0$ and $b\in\bbR$ such that $\partial g(\ell(t))=at+b$ for every $t\in J$.
\end{definition}

\begin{proposition}[Characterization of compatible loss--link pairs]
\label{prop:compatible_characterization}
Under Definition~\ref{def:compatible_pair}, $g$ and $\ell$ are compatible if and only if, for some $a>0$, $b,c\in\bbR$, and fixed $\alpha_\circ\in A$,
\begin{align}
\label{eq:compatible_characterization}
g(\alpha)=a\int_{\alpha_\circ}^{\alpha}\ell^{-1}(v)\,\rmd v+b\alpha+c.
\end{align}
Equivalently, for fixed $g$, every compatible link has the form $\ell(t)=(\partial g)^{-1}(at+b)$. Thus, a fixed link determines the compatible Bregman divergence up to a positive scale and affine terms.
\end{proposition}

\begin{proof}
Compatibility gives $\partial g(\alpha)=a\ell^{-1}(\alpha)+b$, and integration yields Equation~\eqref{eq:compatible_characterization}. Conversely, differentiation proves compatibility, while strict convexity makes $\partial g$ invertible on $A$.
\end{proof}

Proposition~\ref{prop:compatible_characterization} characterizes the derivative of a Riesz objective. Results on proper binary losses instead characterize probability calibration and transformations between a class probability and a density ratio \citep{Menon2016linkinglosses}. The questions coincide in some binary density-ratio models, while the Riesz formulation extends to signed representers and linear functionals outside binary classification.

\subsection{Automatic Regressor Balancing as KKT Conditions}
\label{subsec:acb_kkt}
For an interior minimizer, the Karush--Kuhn--Tucker (KKT) conditions take the form below. For Equation~\eqref{eq:score_linearization}, the empirical objective is
\begin{align}
\label{eq:bd_emp_beta}
\widehat{\mathrm{BD}}_g(\alpha_{\bmbeta})
=\frac1n\sum_{i=1}^n\Bigp{-g(\alpha_{\bmbeta}(X_i))+\alpha_{\bmbeta}(X_i)u_{\bmbeta}(X_i)-m(W_i,u_{\bmbeta})}.
\end{align}
Let $\calB\subseteq\bbR^p$ be the set on which the representer model remains in the domain of $g$. We estimate $\bmbeta$ by
\begin{align}
\label{eq:beta_erm}
\widehat\bmbeta\in\argmin_{\bmbeta\in\calB}
\left\{\widehat{\mathrm{BD}}_g(\alpha_{\bmbeta})+\frac{\lambda}{q}\|\bmbeta\|_q^q\right\},
\qquad q\geq1,
\end{align}
and set $\widehat\alpha=\alpha_{\widehat\bmbeta}$. For a function $f$, define
\begin{align}
\label{eq:imbalance_def_rewrite}
\widehat\Delta(\alpha,f)
\coloneqq\frac1n\sum_{i=1}^n\left(\alpha(X_i)f(X_i)-m(W_i,f)\right),
\end{align}
and let $\widehat\Delta_j(\alpha)=\widehat\Delta(\alpha,\widetilde\phi_j)$.

\begin{theorem}[Automatic regressor balancing]
\label{thm:autocovariance}
Suppose that $g$ is strictly convex and differentiable on an open domain, Equation~\eqref{eq:score_linearization} holds, and $\widehat\bmbeta$ is a minimizer of Equation~\eqref{eq:beta_erm} that lies in the interior of $\calB$. Then, there are subgradients $s_j$ such that
\begin{align}
\label{eq:kkt_identity}
\widehat\Delta_j(\widehat\alpha)+\lambda s_j=0,
\qquad j=1,\ldots,p.
\end{align}
If $q=1$, then $|\widehat\Delta_j(\widehat\alpha)|\leq\lambda$. If $q>1$, then it holds that
\begin{align}
\label{eq:kkt_bound_lp}
|\widehat\Delta_j(\widehat\alpha)|=\lambda|\widehat\beta_j|^{q-1}.
\end{align}
If $\lambda=0$ and a minimizer exists, exact balance holds for the sample used to solve Equation~\eqref{eq:beta_erm}.
\end{theorem}

\begin{proof}
Differentiating Equation~\eqref{eq:bd_emp_beta} and using $u_{\bmbeta}=\partial g\circ\alpha_{\bmbeta}$ gives $\partial_{\beta_j}\widehat{\mathrm{BD}}_g(\alpha_{\bmbeta})=\widehat\Delta_j(\alpha_{\bmbeta})$, and the coordinatewise KKT condition then gives Equation~\eqref{eq:kkt_identity}. The two bounds follow from the subgradient of $|\beta_j|^q/q$.
\end{proof}

\paragraph{Active parameter constraints.}
If the objective in Equation~\eqref{eq:beta_erm} is convex and $\calB$ is closed and convex, then, at any minimizer, there are $\bms\in\partial(\|\bmbeta\|_q^q/q)|_{\widehat\bmbeta}$ and $\bmv\in N_{\calB}(\widehat\bmbeta)$ such that
\begin{align}
\label{eq:kkt_normal_cone}
\widehat\Delta_j(\widehat\alpha)+\lambda s_j+v_j=0,
\qquad j=1,\ldots,p.
\end{align}
Here, $N_{\calB}(\widehat\bmbeta)=\cb{\bmv\in\bbR^p:\bmv^\top(\bmbeta-\widehat\bmbeta)\leq0\text{ for every }\bmbeta\in\calB}$. Thus, $|\widehat\Delta_j(\widehat\alpha)+v_j|\leq\lambda$ for $q=1$, while $|\widehat\Delta_j(\widehat\alpha)+v_j|=\lambda|\widehat\beta_j|^{q-1}$ for $q>1$. At an interior minimizer, $\bmv=\bmzero$ and Theorem~\ref{thm:autocovariance} follows.

\paragraph{Interpretation and scope.}
Theorem~\ref{thm:autocovariance} applies to the observations used in Equation~\eqref{eq:beta_erm}, and Equation~\eqref{eq:kkt_normal_cone} attributes remaining imbalance to regularization and active parameter constraints. On different observations, the empirical imbalance records how well the Riesz equations carry to those observations; approximate optimization adds an optimization residual to Equation~\eqref{eq:kkt_identity}. Reporting these quantities separately identifies the roles of regularization, constraints, optimization, and evaluation on different observations.

When the empirical objective is convex and the dual coordinate is linear, the constrained balancing formulation is a genuine Fenchel or Lagrange dual under the usual constraint qualification. For nonconvex models such as neural networks, the same equations serve as first-order diagnostics, and their residuals record the accuracy of the attained stationary point.

\begin{corollary}[Balancing of the original basis functions]
\label{cor:automaticcovariate}
If $\widetilde\phi_j=\phi_j$ for every $j$, Theorem~\ref{thm:autocovariance} applies to the basis functions in the representer model. In particular, an $\ell_1$ penalty gives
\[
\left|\frac1n\sum_{i=1}^n\left(\widehat\alpha(X_i)\phi_j(X_i)-m(W_i,\phi_j)\right)\right|\leq\lambda.
\]
\end{corollary}

\begin{proposition}[Balancing-weight program for an $\ell_1$ penalty]
\label{prop:dual_balancing_program}
Suppose that the empirical objective is convex, the dual coordinate is linear, and strong duality holds. Then, Equation~\eqref{eq:beta_erm} with $q=1$ is equivalent to
\begin{align}
\label{eq:dual_balancing_program}
\min_{\alpha_i\in\calA_i}\quad &\frac1n\sum_{i=1}^n g(\alpha_i)\\
\text{subject to}\quad &
\left|\frac1n\sum_{i=1}^n\left(\alpha_i\widetilde\phi_j(X_i)-m(W_i,\widetilde\phi_j)\right)\right|\leq\lambda,
\quad j=1,\ldots,p,
\end{align}
where $\calA_i$ is the part of the loss domain allowed by the representer model at observation $i$.
\end{proposition}

\begin{proof}
The result follows from Fenchel duality. The conjugate of the $\ell_1$ penalty is the indicator of the $\ell_\infty$ ball, which gives the constraints in Equation~\eqref{eq:dual_balancing_program}.
\end{proof}

\begin{proposition}[First-order conditions under an arbitrary pair]
\label{prop:arbitrary_pair_foc}
Suppose that $g$ is twice differentiable on the relevant open domain and that $\alpha_{\bmbeta}$ is differentiable in $\bmbeta$. Define
\[
\psi_{j,\bmbeta}(x)
\coloneqq g''(\alpha_{\bmbeta}(x))\,\partial_{\beta_j}\alpha_{\bmbeta}(x).
\]
Then,
\begin{align}
\label{eq:arbitrary_pair_gradient}
\partial_{\beta_j}\widehat{\mathrm{BD}}_g(\alpha_{\bmbeta})
=\widehat\Delta(\alpha_{\bmbeta},\psi_{j,\bmbeta}).
\end{align}
Consequently, an interior stationary point of Equation~\eqref{eq:beta_erm} satisfies
\[
\widehat\Delta(\widehat\alpha,\psi_{j,\widehat\bmbeta})+\lambda s_j=0.
\]
For a generalized linear model $\alpha_{\bmbeta}(x)=\ell(\bmphi(x)^\top\bmbeta)$, suppose that the scalar index ranges over an open interval $J$. Then, there is a common constant $a>0$ such that $\psi_{j,\bmbeta}(x)=a\phi_j(x)$ for every $j$, $x$, and $\bmbeta$ if and only if $(g,\ell)$ is a compatible loss--link pair on $J$.
\end{proposition}

Appendix~\ref{appdx:proof_arbitrary_pair_foc} proves Proposition~\ref{prop:arbitrary_pair_foc}. Thus, optimization always imposes regularization-adjusted empirical Riesz equations in tangent directions, and compatibility makes those directions coincide with the regressors selected for bias control. Without compatibility, the first-order conditions can hold while imbalance remains in the intended regressors. Appendix~\ref{appdx:compatibility_figure} gives a numerical illustration.

Theorem~\ref{thm:autocovariance} states a property of the sample on which the representer is estimated. Regularization, numerical error, and evaluation on a different sample can all make the observed imbalance nonzero. These sources of inexact balance are distinct and should be reported separately.

\subsection{Connection to Balancing Weights}
Proposition~\ref{prop:dual_balancing_program} shows that generalized Riesz regression selects minimum-loss weights among approximately balanced solutions. Squared loss gives a quadratic criterion related to stable balancing weights \citep{Zubizarreta2015stableweights,BrunsSmith2025augmentedbalancing}, whereas the unnormalized Kullback--Leibler loss gives an entropy criterion related to entropy balancing and tailored loss minimization \citep{Hainmueller2012entropybalancing,Zhao2019covariatebalancing}. Kernel mean matching and sieve Riesz representers impose analogous equations on their respective linear spaces \citep{Gretton2009covariateshift,Chen2015sievesemiparametric,Chen2015sievewald}.

If exact balance holds for a fixed collection of regressor functions, the identity in Equation~\eqref{eq:rw_equals_aipw_on_working_space} depends on those equations rather than on the strictly convex loss used to select a feasible representer. The loss becomes consequential when the equations are inexact or admit several solutions whose behavior outside the specified linear space differs. Exact balance fixes the specified moments of the representer, so distinct feasible representers can agree on the balanced space and differ elsewhere.

\subsection{Feasibility and the Role of Regularization}
Exact balance can fail when the number of regressor functions is large, the empirical equations are incompatible with the allowed range of the representer, or the objective has no finite minimizer. Without range restrictions, exact balance is a system of $p$ linear equations in $n$ weights and is generally solvable when $p\leq n$ and the relevant design matrix has full rank. Positivity or sign restrictions add geometric constraints, so infeasibility may remain even when $p\leq n$.

An $\ell_1$ penalty replaces exact equations by the inequalities in Equation~\eqref{eq:dual_balancing_program}. The parameter $\lambda$ therefore controls both balance and feasibility. With a differentiable $\ell_q$ penalty, the discrepancy also depends on the fitted coefficient through Equation~\eqref{eq:kkt_bound_lp}. If the unregularized problem approaches the boundary of the allowed range, the fitted representer may become extreme before a minimizer is attained.

Regularization also affects the variability of the weights: a small value of $\lambda$ can produce extreme weights when overlap is weak, whereas a large value permits more imbalance. Thus, $\lambda$ governs a trade-off between the empirical Riesz equations and the dispersion of the fitted representer, in addition to its role in coefficient sparsity.

\subsection{Loss--Link Pairs for Automatic Regressor Balancing}
\label{rem:automaticbklriesz}
The compatible link is the inverse derivative of the generator, up to an affine transformation. Squared loss gives an affine link, whereas the unnormalized Kullback--Leibler loss gives an exponential link on each sign-specific part of the domain. Basu's power loss gives a power link, and a binary Kullback--Leibler generator gives a log-odds link. The usual Bernoulli likelihood with a sigmoid propensity-score model targets different first-order equations for the ATE; the compatible link in Appendix~\ref{appdx:loss_link_recipes} gives the empirical Riesz equations for the same generator.

\subsection{Practical Interpretation of the Regularization Parameter}
For an $\ell_1$ penalty, $\lambda$ is an upper bound on each empirical imbalance. For $q>1$, Equation~\eqref{eq:kkt_bound_lp} shows that the bound is coefficient dependent. Consequently, the same numerical value of $\lambda$ has different meanings across penalties and across rescalings of the regressor functions, so we report the resulting imbalance and weight dispersion rather than interpreting $\lambda$ in isolation.

\subsection{Regressor Balancing as Moment Matching Under Bregman Projection}
At the population level, generalized Riesz regression minimizes a Bregman divergence from the true representer to the model class, and a linear dual coordinate makes its first-order condition the Riesz equation on the linear space generated by the regressor functions. Theorem~\ref{thm:autocovariance} gives the corresponding empirical equations. When those equations do not determine a unique solution, or when regularization makes balance inexact, the loss determines which representer is selected. Appendix~\ref{appdx:kkt_riesz_linear_equation} gives the Hilbert-space formulation.

\subsection{Covariate Balancing}
For the ATE, the representer model may use functions $\bmphi(Z)$ together with the treatment indicator in Equation~\eqref{eq:riesz_model1}, in which case the first-order conditions compare weighted moments across the treated and control groups. Classical covariate balancing is therefore a special case of automatic regressor balancing. By contrast, a model of the form $\alpha(D,Z)=\bmphi(Z)^\top\bmbeta$ without treatment-specific terms cannot represent the signed ATE Riesz representer because $m(W,\phi_j(Z))=0$ for ATE functionals. Appendix~\ref{appdx:automatic_covariate_balancing_supp} gives the detailed specialization.

\section{Automatic Neyman Orthogonalization and Automatic Neyman Error Minimization}
\label{sec:automaticneyman}
Automatic regressor balancing has two consequences for the orthogonal score: the balance equations control the component of the Neyman error generated by regression functions in the specified linear space, while exact balance makes the Riesz weighted estimator equal to an estimator based on the orthogonal score for every regression function in that space. These finite-sample statements concern the observations on which the imbalance is evaluated and do not depend on a particular sample-splitting scheme.

Regularization, numerical error, and evaluation on observations different from those used to estimate the representer can all produce inexact balance. We therefore state the results for a general empirical imbalance and separate those algebraic statements from the conditions used for asymptotic normality.

\subsection{Recap of Estimators, the Imbalance Gap, and the Neyman Error}
For observations indexed by $\calI$, let $P_{\calI}f\coloneqq|\calI|^{-1}\sum_{i\in\calI}f(W_i)$ and define
\begin{align}
\label{eq:imbalance_any_sample}
\Delta_{\calI}(\alpha,f)
\coloneqq P_{\calI}\bigp{\alpha(X)f(X)-m(W,f)}.
\end{align}
If $\calI$ is the sample used in Equation~\eqref{eq:beta_erm}, Theorem~\ref{thm:autocovariance} controls $\Delta_{\calI}(\widehat\alpha,\widetilde\phi_j)$. For another sample, Equation~\eqref{eq:imbalance_any_sample} is still the relevant empirical quantity, but the KKT identity does not set it to zero.

Let $\varepsilon=Y-\gamma_0(X)$. The difference between the estimated score and the score at the true nuisance parameters satisfies
\begin{align}
\label{eq:neyman_error_decomp}
P_{\calI}\bigp{\psi(W;\widehat\eta,\theta_0)-\psi(W;\eta_0,\theta_0)}
=-\Delta_{\calI}(\widehat\alpha,\widehat\gamma-\gamma_0)
+P_{\calI}\bigp{(\widehat\alpha-\alpha_0)\varepsilon}.
\end{align}
The first term is determined by the empirical Riesz equations and approximation of the regression error, whereas empirical-process conditions or sample splitting control the second.

\subsection{Automatic Regressor Balancing Implies Automatic Neyman Error Minimization}
For $R>0$, define
\[
\Gamma_1(R)=\left\{\gamma=\sum_{j=1}^p\rho_j\widetilde\phi_j:\|\bmrho\|_1\leq R\right\}.
\]
Linearity of $m$ gives $\Delta_{\calI}(\alpha,\gamma)=\sum_j\rho_j\Delta_{\calI}(\alpha,\widetilde\phi_j)$ for every $\gamma$ in this space. Hence, we have
\begin{align}
\label{eq:automatic_neyman_min}
\sup_{\gamma\in\Gamma_1(R)}|\Delta_{\calI}(\widehat\alpha,\gamma)|
=R\max_{1\leq j\leq p}|\Delta_{\calI}(\widehat\alpha,\widetilde\phi_j)|.
\end{align}
We call Equation~\eqref{eq:automatic_neyman_min} \emph{automatic Neyman error minimization}. The name refers to the sharp dual-norm value of the systematic score error on $\Gamma_1(R)$. If $\calI$ is the estimation sample and an $\ell_1$ penalty is used, Theorem~\ref{thm:autocovariance} gives
\[
\sup_{\gamma\in\Gamma_1(R)}|\Delta_{\calI}(\widehat\alpha,\gamma)|\leq R\lambda.
\]
For other penalties, the bound follows from Equation~\eqref{eq:kkt_bound_lp}.

\subsection{Exact Balance Implies Automatic Neyman Orthogonalization}
Let $\Gamma_\phi=\cb{\gamma(x)=\sum_{j=1}^p\rho_j\widetilde\phi_j(x):\bmrho\in\bbR^p}$. If exact balance holds on $\calI$, then it holds that
\begin{align}
\label{eq:exact_balance_implies_riesz_identity}
P_{\calI}[\widehat\alpha(X)\gamma(X)]=P_{\calI}[m(W,\gamma)]
\qquad\text{for every }\gamma\in\Gamma_\phi.
\end{align}
Adding $P_{\calI}[\widehat\alpha(X)(Y-\gamma(X))]$ to both sides gives
\begin{align}
\label{eq:rw_equals_aipw_on_working_space}
P_{\calI}[\widehat\alpha(X)Y]
=P_{\calI}\bigp{m(W,\gamma)+\widehat\alpha(X)(Y-\gamma(X))}.
\end{align}

\begin{theorem}[Automatic Neyman orthogonalization]
\label{thm:automaticneymanorthogonalization}
Suppose that exact balance holds on $\Gamma_\phi$. If $\gamma_0\in\Gamma_\phi$, then the Riesz weighted estimator on the same sample equals the estimator based on the Neyman orthogonal score with $\gamma_0$:
\begin{align}
\label{eq:rw_equals_score_gamma0}
P_{\calI}[\widehat\alpha(X)Y]
=P_{\calI}\bigp{m(W,\gamma_0)+\widehat\alpha(X)(Y-\gamma_0(X))}.
\end{align}
\end{theorem}

\begin{proof}
Equation~\eqref{eq:exact_balance_implies_riesz_identity} applied to $\gamma_0$ gives $P_{\calI}[\widehat\alpha(X)\gamma_0(X)]=P_{\calI}[m(W,\gamma_0)]$. Adding $P_{\calI}[\widehat\alpha(X)(Y-\gamma_0(X))]$ to both sides proves Equation~\eqref{eq:rw_equals_score_gamma0}.
\end{proof}

The theorem is a finite-sample identity on the sample where exact balance holds and does not require an estimator of $\gamma_0$.

\begin{corollary}[Efficiency under exact balance]
\label{cor:rw_efficiency_exact_balance}
Suppose that the conditions of Theorem~\ref{thm:automaticneymanorthogonalization} hold with $\calI=\{1,\ldots,n\}$. Suppose that $\widehat\alpha$ belongs with probability approaching one to a fixed class $\calA_0$ for which
\[
\calF_0=\left\{(\alpha(X)-\alpha_0(X))(Y-\gamma_0(X)): \alpha\in\calA_0\right\}
\]
is $P_0$-Donsker with a square-integrable envelope. If $\|(\widehat\alpha-\alpha_0)(Y-\gamma_0)\|_{L_2(P_0)}=o_P(1)$, then it holds that
\[
\sqrt n(\widehat\theta^{\mathrm{RW}}-\theta_0)
=\frac1{\sqrt n}\sum_{i=1}^n\psi(W_i;\eta_0,\theta_0)+o_P(1).
\]
If $\psi(W;\eta_0,\theta_0)$ is the efficient influence function, the Riesz weighted estimator attains the semiparametric efficiency bound.
\end{corollary}

\begin{proof}
Theorem~\ref{thm:automaticneymanorthogonalization} gives
\[
\widehat\theta^{\mathrm{RW}}-\theta_0
=(P_n-P_0)\psi(\cdot;\eta_0,\theta_0)
+(P_n-P_0)\left[(\widehat\alpha-\alpha_0)(Y-\gamma_0)\right].
\]
For every fixed $\alpha$, $P_0[(\alpha-\alpha_0)(Y-\gamma_0)]=0$ because $\bbE[Y-\gamma_0(X)\mid X]=0$. Stochastic equicontinuity of $\calF_0$ and the stated $L_2(P_0)$ convergence make the second empirical-process term $o_P(n^{-1/2})$. The central limit theorem proves the result.
\end{proof}

\subsection{Inexact Balance Implies Approximate Neyman Orthogonality With Bias Control}
Suppose that the regression error admits the decomposition
\[
\widehat\gamma-\gamma_0=\sum_{j=1}^p\rho_j\widetilde\phi_j+r.
\]
Then, we have
\begin{align}
\label{eq:inexact_balance_bound}
|\Delta_{\calI}(\widehat\alpha,\widehat\gamma-\gamma_0)|
\leq\|\bmrho\|_1\max_j|\Delta_{\calI}(\widehat\alpha,\widetilde\phi_j)|
+|\Delta_{\calI}(\widehat\alpha,r)|.
\end{align}
Equation~\eqref{eq:inexact_balance_bound} separates the component controlled by the empirical Riesz equations from the approximation residual. Regularization, numerical error, and evaluation on a separate sample enter through the observed imbalance, while the chosen span determines the residual $r$.

At the population level, Equation~\eqref{eq:orth_score_bias_identity} yields the usual second-order bound
\[
|\bbE[\psi(W;\widehat\eta,\theta_0)]|
\leq\|\widehat\alpha-\alpha_0\|_{L_2(P_X)}\|\widehat\gamma-\gamma_0\|_{L_2(P_X)}.
\]
The empirical and population bounds answer different questions. Equation~\eqref{eq:inexact_balance_bound} describes which regression directions receive direct protection from balance, whereas the population bound supplies the convergence rate condition used for inference. Section~\ref{subsec:rw_approx_balance} gives an efficiency result when the number of regressor functions increases with $n$.

\subsection{Special Case: Linear Riesz Models With Linear Regression Models}
Let $\Phi$ be the $n\times p$ matrix whose $i$th row is $\bmphi(X_i)^\top$, and let $\widehat b=n^{-1}\sum_i m(W_i,\bmphi)$. Suppose that $\widehat\alpha(X)=\bmphi(X)^\top\widehat\bmbeta$ and exact balance holds. Then, we have $n^{-1}\Phi^\top\Phi\widehat\bmbeta=\widehat b$. Let $\widehat\bmrho^{\mathrm{OLS}}=(\Phi^\top\Phi)^\dagger\Phi^\top Y$ be the ordinary least-squares (OLS) coefficient vector. The two normal equations imply
\begin{align}
\label{eq:rw_equals_ols}
\widehat\theta^{\mathrm{RW}}=\widehat b^\top\widehat\bmrho^{\mathrm{OLS}}.
\end{align}
Thus, weighting and regression solve the same normal equations in this special case. Ridge versions give an analogous identity with an effective regression penalty \citep{Robins2007commentperformance,BrunsSmith2025augmentedbalancing,Singh2024kernelridge}.

\subsection{Empirical Process Conditions}
The identities above hold for any index set. If the nuisance parameters and the score are evaluated on the same observations, a Donsker condition can control the empirical-process term. Alternatively, nuisance parameters can be estimated on observations separate from those used to evaluate each score summand, and cross-fitting rotates these sample splits and averages the score summands \citep{Chernozhukov2018doubledebiased}. Under this construction, the empirical Riesz equations hold on the observations used to estimate the representer, while the inexact-balance result applies to the observations used to evaluate the score. Section~\ref{sec:convanalysis} establishes asymptotic normality under either route.

\subsection{Estimation Equation and TMLE Approaches}
\label{sec:comptmle}
TMLE updates an initial estimator of $\gamma_0$ so that an empirical score equation holds \citep{vanderLaan2006targetedmaximum,vanderLaan2018targetedlearning_book2,Schuler2024introductionmodern}. For a continuous outcome, a linear fluctuation takes the form $\widehat\gamma^{(1)}(x)=\widehat\gamma(x)+\widehat\epsilon\widehat\alpha(x)$, where $\widehat\epsilon$ is chosen so that $P_n[\widehat\alpha(X)(Y-\widehat\gamma^{(1)}(X))]=0$. Generalized Riesz regression instead estimates $\alpha_0$ so that empirical Riesz equations hold for specified regressor functions. Because the two operations concern different nuisance parameters, they can be combined, and Appendix~\ref{appdx:tmle_details} gives the corresponding equations for bounded outcomes.

\subsection{Regression and Representer Models}
The regressor functions in the representer model determine the linear space controlled by balance. To use Equation~\eqref{eq:inexact_balance_bound}, the same functions should approximate the components of $\widehat\gamma-\gamma_0$ that matter for the orthogonal score. In ATE estimation with heterogeneous effects, functions of $Z$ alone may balance conventional covariate moments but fail to approximate treatment-specific regression components. Terms such as $D\phi(Z)$ and $(1-D)\phi(Z)$ then enlarge the protected linear space. Proposition~\ref{prop:arbitrary_pair_foc} adds a second requirement: under an incompatible pair, the first-order conditions protect the realized span of $\psi_{j,\widehat\bmbeta}$ rather than the intended regressor span. Reporting imbalance in the intended regressors detects this discrepancy. Neither the loss nor optimization accuracy can compensate for an omitted regression direction.

Calibration based on level-set indicators or other finite-dimensional functions of a fitted summary $S=s(X)$ imposes empirical Riesz equations on subspaces of $L_2(\sigma(S))$. Conditional on any data used to construct $S$, Proposition~\ref{prop:summary_space_projection} shows that the fixed regressor span and $L_2(\sigma(S))$ need not contain one another, while their closed sum weakly improves approximation. These functions of $S$ can be combined with the fixed regressors in Proposition~\ref{prop:dual_balancing_program}; the optimization remains convex, and Equation~\eqref{eq:inexact_balance_bound} applies to the combined span.

\section{Practical Implications for Choice of Regression Models, Link, and Loss Functions}
\label{sec:loss_model_bias_correction}
The preceding results separate four specification choices. The target functional determines the sign and range of the representer. The generator and representer model determine the tangent directions and Bregman geometry, compatibility aligns those directions with regressors selected before estimation, the regressors define the component of regression error protected by balance, and regularization governs the trade-off between imbalance and weight dispersion.

\subsection{Summary}
A specification should first respect the representer's range and sign. Compatibility then places the intended regressors in the first-order conditions, and those regressors should approximate the components of regression error entering the orthogonal score. Regularization controls the empirical discrepancy and weight dispersion. Neither a different loss nor tighter optimization can recover an omitted regression direction or information lost through weak overlap.

\subsection{Loss Choice Selects a Bregman Projection}
If the representer model contains $\alpha_0$, every strictly convex Bregman divergence has the same population minimizer. Under misspecification, the minimizer is the Bregman projection of $\alpha_0$ onto the representer model and therefore depends on $g$. The loss also determines the curvature that maps an error in the linear index into an error in the representer. These two facts explain why loss choice matters under misspecification even when two estimators use the same regressor functions.

The loss also determines how the dual problem selects among approximately balanced solutions. Squared loss penalizes large weights quadratically, whereas an entropy loss favors weights that remain close in relative scale to the reference measure. Basu power losses interpolate between these geometries, which govern the trade-off between imbalance and weight concentration after the representer model and regressor functions are fixed.

Under correct specification, every strictly convex generator has the true representer as its population minimizer, although curvature and admissible ranges still produce different finite-sample behavior. Under misspecification, each generator defines a different projection, so comparisons across losses hold the representer model and regressor functions fixed.

Proposition~\ref{prop:generator_condition_numbers} quantifies curvature on a common admissible sign-specific range. Under ATE overlap, $C=1$, and $0<\omega\leq1$, the condition numbers of SQ, BP, UKL, and BKL have orders $1$, $\eta^{-2(1-\omega)}$, $\eta^{-2}$, and $\eta^{-3}$. The resulting condition numbers quantify inverse-link sensitivity on the common range, while the dual multipliers and design matrix determine the realized weights. When a representer identity does not fix $C$, sensitivity analysis compares its admissible values.

\subsection{Representative Loss--Link Pairs in Generalized Riesz Regression}
For squared loss, $g^{\mathrm{SQ}}(\alpha)=\frac12(\alpha-C)^2$ and the compatible representer model is
\[
\alpha_{\bmbeta}(x)=C+\bmphi(x)^\top\bmbeta.
\]
This specification is natural when the representer is well approximated on its original scale, and its constant curvature often limits the amplification of index errors.

For UKL loss, the compatible model is exponential on each sign-specific part of the domain. If $s(x)\in\{-1,1\}$ is fixed by the estimand, the model
\[
\alpha_{\bmbeta}(x)=s(x)\Bigp{C+\exp(s(x)\bmphi(x)^\top\bmbeta)}
\]
satisfies $\partial g^{\mathrm{UKL}}(\alpha_{\bmbeta}(x))=\bmphi(x)^\top\bmbeta$. In ATE estimation, treatment status fixes $s(x)$, and a sigmoid propensity-score model induces this logarithmic representer scale \citep{Zhao2019covariatebalancing}.

Basu power links interpolate between the affine and exponential models on their admissible domains. BKL gives a log-odds link and is natural for binary classification, with the resulting target determined by the relation between the propensity model and the representer. If the representer can cross zero at unknown values, as in many AME problems, a full-domain squared loss avoids imposing an unsupported sign restriction.

\subsection{Loss Choice and Its Impact on Riesz Estimation Under Inexact Balance}
With an $\ell_1$ penalty, $\lambda$ directly bounds the maximum empirical imbalance on the estimation sample, whereas the bound also depends on the fitted coefficients under a differentiable $\ell_q$ penalty. Increasing regularization can improve numerical stability and relax infeasible equations, but it increases the bound in Equation~\eqref{eq:automatic_neyman_min}. Therefore, the scale of the regressor functions matters when $\lambda$ is interpreted.

When the nuisance parameters are estimated on one sample and the score is evaluated on another, the balance equations should be examined on both samples. Estimation-sample imbalance measures optimization fit, whereas evaluation-sample imbalance measures how well the Riesz equations transfer. Equation~\eqref{eq:inexact_balance_bound} then isolates the residual determined by approximation of the regression error with the selected functions.

\subsection{Specification of the Representer and Regression Models}
\label{subsec:specification_reporting}
For ATE estimation, a logarithmic representer model is suitable when a propensity-score index is expected to be approximately linear. A squared-loss model can be preferable when the inverse propensity score is more nearly linear on its original scale or when exponential weights are unstable. For AME estimation, the score can change sign at unknown values, so an identity link provides a full-domain starting point unless additional structure determines the sign. For covariate shift, the identity, logarithmic, and power scales correspond to different approximation assumptions for the density ratio.

After the representer model and loss have been specified, we report empirical imbalance, maximum absolute weight, effective sample size, optimization failures, and domain violations. For balance functions constructed from a summary $S$, we also report $R_j(S)=\bbE[\Var(\widetilde\phi_j(X)\mid S)]/\Var(\widetilde\phi_j(X))$ for important nonconstant fixed regressors, estimated by the ratio of within-cell to total sums of squares. A value near one indicates poor representation by functions of $S$. This quantity is a function-space diagnostic rather than a direct estimate of remaining bias. Appendix~\ref{appdx:summary_space_diagnostic} gives an analytic example.

Inference combines these diagnostics of balance, function-space coverage, and weight concentration with the nuisance-rate conditions in Section~\ref{sec:convanalysis}.

Raw Bregman objective values are not comparable across generators. Multiplying $g$ by a positive constant changes the objective without changing the unregularized population minimizer, and affine additions leave the Bregman divergence unchanged. Comparisons across losses should use quantities on the scale of the target parameter together with the balance and weight summaries above. If a separate sample is used to compare specifications, the same observations should not also be used to report an unadjusted estimate of sampling uncertainty.

\section{Convergence Rate Analysis}
\label{sec:convanalysis}
We first derive rates for RKHS and neural-network models, then state general conditions for inference, and finally specialize the dual-linear setup to sparse high-dimensional models. For the nonlinear models, write $\alpha_f(x)=\zeta^{-1}(x,f(x))$, where $\zeta^{-1}$ is Lipschitz in its second argument and $f$ need not be linear. For $\alpha$ in either nonlinear class, let $\ell_\alpha(W)=\partial g(\alpha(X))\alpha(X)-g(\alpha(X))-m(W,\partial g\circ\alpha)$ denote the per-observation objective. The assumptions below are stronger than the continuity condition used to define the Riesz representer because they control this objective.

\begin{assumption}[Range and curvature]
\label{asm:compact_curvature}
There are compact intervals $\calI_1,\ldots,\calI_K$, each contained in the interior of a connected component of $\calA$, a known measurable map $c\colon\calX\to\cb{1,\ldots,K}$, and constants $0<\mu\leq C_g<\infty$ such that $\mu\leq g''(t)\leq C_g$ for every $t\in\bigcup_{k=1}^K\calI_k$. The true representer and every representer in the model classes below satisfy $\alpha(x)\in\calI_{c(x)}$ for every $x$.
\end{assumption}

\begin{assumption}[Regularity of the functional]
\label{asm:m_reg}
The map $h\mapsto m(W,h)$ is linear almost surely. There is a constant $C_m<\infty$ such that $\bbE[m(W,h)^2]\leq C_m^2\bbE[h(X)^2]$ for every square-integrable $h$ in the classes used below. There is also a constant $B_m<\infty$ such that $|m(W,h)|\leq B_m\|h\|_\infty$ almost surely for every bounded $h$.
\end{assumption}

Assumption~\ref{asm:m_reg} holds for evaluation functionals such as the ATE under strict overlap. It does not hold in this form for derivative functionals such as the AME. Those functionals require a norm that controls derivatives.

\subsection{RKHS}
Let $\calF_M^{\mathrm{RKHS}}$ be a fixed RKHS ball that contains $f_0$, and let $\calG_M=\{\ell_{\alpha_f}:f\in\calF_M^{\mathrm{RKHS}}\}$ be the induced class of per-observation objectives. Suppose that its bracketing entropy satisfies
\[
H_B(\delta,\calG_M,P_0)\leq A(M/\delta)^\tau,
\qquad 0<\tau<2.
\]
Let $\widehat f^{\mathrm{RKHS}}$ minimize the empirical generalized Riesz objective plus $\lambda\|f\|_{\calF}^2$ over $\calF_M^{\mathrm{RKHS}}$, and set $\widehat\alpha^{\mathrm{RKHS}}=\alpha_{\widehat f^{\mathrm{RKHS}}}$.

\begin{theorem}[$L_2$ estimation error for an RKHS model]
\label{thm:l2norm}
Suppose that Assumptions~\ref{asm:compact_curvature} and~\ref{asm:m_reg} hold, the kernel is bounded, and $\alpha_0=\alpha_{f_0}$ for some $f_0\in\calF_M^{\mathrm{RKHS}}$. If $\lambda\to0$, then it holds that
\[
\|\widehat\alpha^{\mathrm{RKHS}}-\alpha_0\|_{L_2(P_X)}
=O_P\left(\max\{n^{-1/(2+\tau)},\lambda^{1/2}\}\right).
\]
In particular, $\lambda\asymp n^{-2/(2+\tau)}$ gives the rate $n^{-1/(2+\tau)}$.
\end{theorem}

The proof applies the bracketing bound to the induced loss class, rather than to the RKHS alone, and follows the argument used for kernel least-squares density-ratio estimation by \citet{Kanamori2012statisticalanalysis}.

\subsection{Neural Networks}
The neural-network result additionally requires the local Rademacher average of the induced class $\{m(\cdot,\partial g\circ\alpha_f)-m(\cdot,\partial g\circ\alpha_0)\}$ to be bounded by a constant multiple of the local Rademacher average of $\{\alpha_f-\alpha_0\}$. We refer to this requirement as the induced-class local-complexity condition and impose it separately from Assumption~\ref{asm:m_reg}.

Let $\calF_n^{\mathrm{FNN}}$ be a class of bounded feedforward ReLU neural networks and let $p_n=\operatorname{Pdim}(\calF_n^{\mathrm{FNN}})$, where $\operatorname{Pdim}$ denotes pseudodimension. Let $\widehat f^{\mathrm{FNN}}\in\calF_n^{\mathrm{FNN}}$ and let $\delta_n\geq0$ satisfy
\begin{align}
\label{eq:nn_optimization_error}
P_n\ell_{\alpha_{\widehat f^{\mathrm{FNN}}}}
\leq\inf_{f\in\calF_n^{\mathrm{FNN}}}P_n\ell_{\alpha_f}+\delta_n.
\end{align}
We set $\widehat\alpha^{\mathrm{FNN}}=\alpha_{\widehat f^{\mathrm{FNN}}}$, and $\delta_n$ measures the optimization error in the empirical generalized Riesz objective.

\begin{theorem}[Estimation error for neural networks]
\label{thm:est_error_nn}
Suppose that Assumptions~\ref{asm:compact_curvature} and~\ref{asm:m_reg} hold, the induced-class local-complexity condition stated above holds, and $\zeta^{-1}(x,\cdot)$ is nondecreasing and Lipschitz for every $x$. Then, for every $z>0$, with probability at least $1-\exp(-z)$,
\[
\|\widehat\alpha^{\mathrm{FNN}}-\alpha_0\|_{L_2(P_X)}^2
\leq C\left(\frac{p_n\log n}{n}
+\inf_{f\in\calF_n^{\mathrm{FNN}}}\|\alpha_f-\alpha_0\|_{L_2(P_X)}^2
+\delta_n
+\frac{z}{n}\right).
\]
If $f_0$ is $\nu$-H\"older smooth on $[0,1]^d$, the width and depth are chosen as in Appendix~\ref{appex:neural_proof}, and $\delta_n=O_P(n^{-2\nu/(d+2\nu)}\log^4n)$, then it holds that
\[
\|\widehat\alpha^{\mathrm{FNN}}-\alpha_0\|_{L_2(P_X)}^2
=O_P\left(n^{-2\nu/(d+2\nu)}\log^4 n\right).
\]
\end{theorem}

The rate agrees with the classical H\"older regression rate up to logarithmic factors \citep{Stone1982optimalglobal,Anthony1999neuralnetwork,SchmidtHieber2020nonparametricregression,Zheng2022anerror}. \citet{SchmidtHieber2020nonparametricregression} likewise separates approximation and stochastic error from the discrepancy between the attained empirical risk and its infimum. The term $\delta_n$ has the analogous role for the generalized Riesz objective, for which Theorem~\ref{thm:est_error_nn} establishes the corresponding upper rate.

\subsection{Asymptotic Normality}
We combine convergence of the representer and regression estimators with either a Donsker condition or cross-fitting to obtain asymptotic normality.
\begin{assumption}[Convergence rate conditions]
\label{asm:conv_rate}
It holds that $\|\widehat\alpha-\alpha_0\|_{L_2(P_X)}=o_P(1)$, $\|\widehat\gamma-\gamma_0\|_{L_2(P_X)}=o_P(1)$,
\[
\|\widehat\alpha-\alpha_0\|_{L_2(P_X)}\|\widehat\gamma-\gamma_0\|_{L_2(P_X)}=o_P(n^{-1/2}),
\]
and
\[
\|(\widehat\alpha-\alpha_0)(\widehat\gamma-\gamma_0)\|_{L_2(P_X)}=o_P(1).
\]
Under cross-fitting, these conditions hold uniformly over the fixed number of folds.
\end{assumption}

\begin{assumption}[Donsker condition or cross-fitting]
\label{asm:donsker}
At least one of the following alternatives holds. Under the first alternative, the estimated nuisance parameters lie with probability approaching one in fixed, uniformly bounded $P_0$-Donsker classes, and the induced score differences converge to zero in $L_2(P_0)$. Under the second alternative, the nuisance parameters are estimated via cross-fitting with a fixed number of folds. Each score summand uses nuisance parameter estimates constructed without observations in the fold containing that summand, and Assumption~\ref{asm:conv_rate} holds uniformly over the folds.
\end{assumption}

\begin{assumption}[Moments and variance]
\label{asm:score_moments}
The Riesz representer $\alpha_0$ is essentially bounded, and the conditional variance of $Y-\gamma_0(X)$ is uniformly bounded. The efficient influence function has a finite $2+\delta$ moment for some $\delta>0$, and its variance $V^*=\bbE[\psi(W;\eta_0,\theta_0)^2]$ is positive and finite.
\end{assumption}

\begin{theorem}[Asymptotic normality]
\label{thm:asymp_normality}
Suppose that Assumptions~\ref{asm:m_reg} and~\ref{asm:conv_rate}--\ref{asm:score_moments} hold. Then, it holds that
\[
\sqrt n(\widehat\theta^{\mathrm{ARW}}-\theta_0)\xrightarrow{\rmd}\calN(0,V^*).
\]
The sample variance of the estimated influence functions is consistent under the same conditions.
\end{theorem}

Appendix~\ref{appdx:proof_asymp_normality} gives the proof.

The Donsker condition and cross-fitting are alternative ways to control the empirical-process term. The Donsker condition makes cross-fitting unnecessary.

\subsection{Approximate Balance and the Riesz Weighted Estimator}
\label{subsec:rw_approx_balance}
Let $p_n$ be allowed to increase with $n$, let $\widetilde\phi_{n,1},\ldots,\widetilde\phi_{n,p_n}$ be regressor functions, let $\bmrho_n\in\bbR^{p_n}$ be nonrandom, and write $\gamma_n=\sum_{j=1}^{p_n}\rho_{n,j}\widetilde\phi_{n,j}$ and $r_n=\gamma_0-\gamma_n$. The separation of $\gamma_n$ from the residual follows the standard series analysis used by \citet{Belloni2015somenew}. The result below applies this separation to the Riesz weighted estimator through the empirical imbalance.

\begin{theorem}[Asymptotic normality under approximate balance]
\label{thm:rw_efficiency_approx_balance}
Suppose that $V^*=\bbE[\psi(W;\eta_0,\theta_0)^2]$ is positive and finite and that
\begin{align}
\label{eq:rw_approx_balance_conditions}
\sqrt n\,|\widehat\Delta(\widehat\alpha,\gamma_n)|&=o_P(1),\nonumber\\
\sqrt n\,\|\widehat\alpha-\alpha_0\|_{L_2(P_X)}\|r_n\|_{L_2(P_X)}&=o_P(1),\nonumber\\
\sqrt n(P_n-P_0)[(\widehat\alpha-\alpha_0)(Y-\gamma_0)]&=o_P(1),\nonumber\\
\sqrt n(P_n-P_0)[\widehat\alpha r_n-m(W,r_n)]&=o_P(1).
\end{align}
Then, it holds that $\sqrt n(\widehat\theta^{\mathrm{RW}}-\theta_0)=n^{-1/2}\sum_{i=1}^n\psi(W_i;\eta_0,\theta_0)+o_P(1)$. If $\psi(W;\eta_0,\theta_0)$ is the efficient influence function, the Riesz weighted estimator attains the semiparametric efficiency bound.
\end{theorem}

Appendix~\ref{appdx:proof_rw_approx_balance} gives the proof.

Automatic Neyman error minimization shows that the first condition follows from $\sqrt n\,\|\bmrho_n\|_1\max_{1\leq j\leq p_n}|\widehat\Delta(\widehat\alpha,\widetilde\phi_{n,j})|=o_P(1)$. Under the interior conditions of Theorem~\ref{thm:autocovariance}, an $\ell_1$ penalty gives the sufficient condition $\sqrt n\,\|\bmrho_n\|_1\lambda_n\to0$. Exact balance makes this term zero and recovers Corollary~\ref{cor:rw_efficiency_exact_balance} when $r_n=0$. Stochastic equicontinuity under suitable Donsker conditions implies the last two conditions in Equation~\eqref{eq:rw_approx_balance_conditions}, while estimation of the nuisance parameters via cross-fitting gives the same conclusion when the corresponding conditional second moments vanish.

For an $s$-smooth regression approximated by a sufficiently high-order series of splines, wavelets, or local polynomials, \citet{Belloni2015somenew} give $\|r_n\|_{L_2(P_X)}=O(p_n^{-s/d})$. If $\|\widehat\alpha-\alpha_0\|_{L_2(P_X)}=O_P(a_n)$, the second condition in Equation~\eqref{eq:rw_approx_balance_conditions} holds when $\sqrt n\,a_n p_n^{-s/d}\to0$. No outcome regression estimate enters $\widehat\theta^{\mathrm{RW}}$.

\subsection{Sparse Models Linear in Dual Coordinates}
\label{subsec:sparse_dual_models}
We now connect the empirical Riesz equations in Section~\ref{sec:automaticcovariatebalancing} to a high-dimensional rate for models whose dual coordinate is linear. The regressors $\widetilde\phi_1,\ldots,\widetilde\phi_p$ are treated as nonrandom; the same result holds conditionally when they are constructed on an independent auxiliary sample. Let $\widehat\bmbeta$ solve Equation~\eqref{eq:beta_erm} with $q=1$ over $\calB_{R_1}=\{\bmbeta:\|\bmbeta\|_1\leq R_1\}$, and let $\bmbeta^*$ be an interior population minimizer with support size $s$. Appendix~\ref{appdx:sparse_dual_proof} states the bounded-regressor, compact-curvature, and restricted-eigenvalue conditions. Write $B_\phi$ for the regressor bound and $\kappa$ for the restricted-eigenvalue constant. Let $A_\alpha$ be the largest absolute representer value on the induced range, set $B_\xi=B_\phi(A_\alpha+B_m)$, and define $\kappa_g=C_g/\mu$.

\begin{theorem}[Sparse models linear in dual coordinates]
\label{thm:sparse_dual_rate}
Suppose that the conditions in Appendix~\ref{appdx:sparse_dual_proof} hold and
\[
16sB_\phi^2\sqrt{\frac{2\log(2p^2/\varepsilon)}{n}}\leq\frac{\kappa^2}{2}.
\]
For $\lambda=2B_\xi\sqrt{2\log(2p/\varepsilon)/n}$, it holds with probability at least $1-2\varepsilon$ that
\[
\|\widehat\alpha-\alpha_{\bmbeta^*}\|_{L_2(P_X)}
\leq\frac{6\kappa_g\sqrt{s}\lambda}{\kappa}.
\]
The appendix also gives the corresponding $\ell_2$ and $\ell_1$ coefficient rates. If the $\ell_1$-ball constraint is inactive at $\widehat\bmbeta$, then
\[
\max_{j\leq p}|\widehat\Delta_j(\widehat\alpha)|\leq\lambda,
\qquad
\sup_{\gamma\in\Gamma_1(R)}|\widehat\Delta(\widehat\alpha,\gamma)|\leq R\lambda.
\]
\end{theorem}

Under correct specification, applying Theorem~\ref{thm:sparse_dual_rate} on each fold-specific training sample verifies the representer-rate condition in Theorem~\ref{thm:asymp_normality}. Let $\varepsilon=\varepsilon_n\to0$ with $\log(1/\varepsilon_n)=O(\log(p\vee n))$, and suppose that the sample restricted-eigenvalue event in the theorem holds with probability approaching one. The displayed transfer condition is sufficient; when $B_\phi$ and $\kappa$ remain bounded, it follows from $s\sqrt{\log(p\vee n)/n}\to0$. If
\[
\frac{\kappa_g B_\xi}{\kappa}\sqrt{\frac{s\log(p\vee n)}{n}}\,\|\widehat\gamma-\gamma_0\|_{L_2(P_X)}=o_P(n^{-1/2}),
\]
then the ARW estimator is asymptotically normal under the remaining conditions of Theorem~\ref{thm:asymp_normality}. If $\|\widehat\gamma-\gamma_0\|_{L_2(P_X)}=O_P(\sqrt{s_\gamma\log(p\vee n)/n})$ and the displayed constants remain bounded, the conditions $s^2\log(p\vee n)=o(n)$ and $ss_\gamma\log^2(p\vee n)=o(n)$ are sufficient. Appendix~\ref{appdx:sparse_dual_proof} proves these statements. For the squared generator, the theorem gives the standard sparse Riesz rate; for UKL in ATE models, it complements high-dimensional calibrated propensity-score analysis. The result covers general linear functionals and every compatible loss--link pair, with generator curvature entering through $\kappa_g$.

\section{Applications}
\label{sec:application}
This section applies generalized Riesz regression to ATE, AME, and covariate shift, while Appendix~\ref{sec:additionalapplications} treats additional parameters.

\subsection{ATE Estimation}
\label{appdx:ate_estimation}
Let $X=(D,Z)$, where $D\in\{0,1\}$. We use the potential-outcome formulation of \citet{Neyman1923surapplications} and \citet{Rubin1974estimatingcausal}. Under unconfoundedness and overlap, the ATE is $\theta_0^{\mathrm{ATE}}=\bbE[\gamma_0(1,Z)-\gamma_0(0,Z)]$, and
\[
m^{\mathrm{ATE}}(W,\gamma)=\gamma(1,Z)-\gamma(0,Z),
\qquad
\alpha_0^{\mathrm{ATE}}(D,Z)=\frac{D}{e_0(Z)}-\frac{1-D}{1-e_0(Z)}.
\]
A sigmoid propensity-score model induces Equation~\eqref{eq:riesz_model1}. The compatible UKL loss then gives tailored loss minimization and the entropy-balancing dual, whereas SQ-Riesz models the signed inverse propensity weight on its original scale. BKL and BP give other loss--link combinations, provided the representer model remains in their domains.

For inference, the ARW estimator is the usual augmented inverse probability weighted estimator, and the RW estimator is the inverse probability weighted estimator \citep{Horvitz1952generalization}. Covariate-balancing estimators provide related direct weighting methods \citep{Lee2010improvingpropensity,Kallus2020generalizedoptimal,Kong2023covariatebalancing}. If the outcome regression belongs to the balanced linear space, exact balance gives the identity in Theorem~\ref{thm:automaticneymanorthogonalization}. Appendix~\ref{appdx:att_estimation} gives the explicit objectives for the average treatment effect on the treated (ATT), and Appendix~\ref{appdx:att_correction} gives the variance correction used below.

\subsection{AME Estimation}
Let $D$ be a continuously distributed treatment and suppose that integration by parts is valid. For $m^{\mathrm{AME}}(W,\gamma)=\partial_d\gamma(D,Z)$, the Riesz representer is
\[
\alpha_0^{\mathrm{AME}}(D,Z)=-\partial_d\log f_0(D,Z).
\]
The squared generalized Riesz objective coincides with Hyv\"arinen score matching \citep{Hyvarinen2005estimationof}. Because the score may cross zero at unknown values, a full-domain model is suitable unless additional information determines the sign. The $L_2$ pathwise bound used for evaluation functionals does not apply directly to derivatives. Appendix~\ref{appdx:ame_estimation} states the corresponding smoothness conditions.

\subsection{Covariate Shift Adaptation}
\label{appdx:covariate_shift}
Let $P_{X,1}\ll P_{X,0}$ and let $r_0=\rmd P_{X,1}/\rmd P_{X,0}$. The target is the mean of $\gamma_0$ under $P_{X,1}$, and the Riesz representer relative to $L_2(P_{X,0})$ is $r_0$. The resulting augmented estimator is the covariate-shift counterpart of doubly robust estimation \citep{Reddi2015doublyrobust}. With labeled observations $(X_i,Y_i)$ from the source law and covariates $\widetilde X_j$ from the target law, the augmented estimator is
\begin{align}
\label{eq:covariate_shift_arw_main}
\widehat\theta^{\mathrm{CS}}
=\frac1m\sum_{j=1}^m\widehat\gamma(\widetilde X_j)
+\frac1n\sum_{i=1}^n\widehat r(X_i)(Y_i-\widehat\gamma(X_i)).
\end{align}
Squared, logarithmic, power, and classification-based ratio estimators correspond to the losses in Table~\ref{tbl:dre_rre}. Direct density-ratio estimation avoids separately estimating two high-dimensional probability laws \citep{Sugiyama2008directimportance,Sugiyama2012densityratio}. With separate source and target samples, the two expectations in the population objective are replaced by their corresponding empirical averages. Differentiating that objective gives the same Riesz-moment gradient with source and target averages in their respective terms. Appendix~\ref{supp:covariate_shift_details} gives the explicit two-sample objectives. The construction also applies to probability mass functions because $r_0$ is a Radon--Nikodym derivative.

\section{Experiments}
\label{sec:experiments}
The experiments study approximation error, weight concentration, and numerical failure, while the Online Appendix reports the remaining estimators and figures.

\subsection{Simulation Study}
We generate $n=2000$ observations with three covariates. The treatment probability follows a logistic model whose index contains the centered quadratic term $Z_2^2-1$, and the outcome regression contains the same term. We compare regressor sets that include or omit $Z_2^2-1$, use coefficients $0.5$ and $2.5$ for stronger and weaker overlap, and run 400 replications in each setting. A fit is invalid when the optimizer does not converge, the exact link is undefined, the KKT residual exceeds the stated tolerance, or the resulting score is not finite. In Table~\ref{tab:coverage_main}, MC SD is the Monte Carlo standard deviation, Mean SE is the average estimated standard error, CR is the coverage rate of the nominal 95 percent interval, and Valid is the number of valid fits. For each replication, $\widehat B$ is the largest absolute empirical imbalance among the regressor functions used in the representer model.

\begin{table}[!t]
\centering
\caption{ATE estimates when the regressor functions include or omit $Z_2^2-1$.}
\label{tab:coverage_main}
\setlength{\tabcolsep}{2.2pt}
\resizebox{\linewidth}{!}{
\begin{tabular}{lllrrrrrrr}
\toprule
Overlap & Regressors & Loss & Bias & MC SD & Mean SE & CR & $\widehat B$ & Max $|\widehat\alpha|$ & Valid \\
\midrule
\multirow{8}{*}{Stronger}
& \multirow{4}{*}{Include term}
& BKL & 0.007 & 0.062 & 0.062 & 0.963 & 0.469 & 48.43 & 400 \\
& & BP(0.5) & -- & -- & -- & -- & -- & -- & 0 \\
& & SQ & 0.005 & 0.046 & 0.046 & 0.940 & 0.112 & 6.25 & 400 \\
& & UKL & 0.005 & 0.049 & 0.050 & 0.965 & 0.234 & 22.91 & 400 \\
\cmidrule(lr){2-10}
& \multirow{4}{*}{Omit term}
& BKL & 0.269 & 0.067 & 0.070 & 0.033 & 0.140 & 28.07 & 400 \\
& & BP(0.5) & 0.261 & 0.053 & 0.057 & 0.000 & 0.063 & 4.50 & 165 \\
& & SQ & 0.262 & 0.051 & 0.055 & 0.000 & 0.071 & 3.72 & 400 \\
& & UKL & 0.264 & 0.052 & 0.059 & 0.003 & 0.075 & 6.91 & 400 \\
\midrule
\multirow{8}{*}{Weaker}
& \multirow{4}{*}{Include term}
& BKL & 0.001 & 0.158 & 0.150 & 0.943 & 0.481 & 50.00 & 400 \\
& & BP(0.5) & -- & -- & -- & -- & -- & -- & 0 \\
& & SQ & 0.005 & 0.066 & 0.059 & 0.925 & 0.204 & 10.78 & 400 \\
& & UKL & $-0.012$ & 0.174 & 0.161 & 0.940 & 0.622 & 99.97 & 400 \\
\cmidrule(lr){2-10}
& \multirow{4}{*}{Omit term}
& BKL & 0.919 & 0.159 & 0.162 & 0.000 & 0.386 & 50.00 & 400 \\
& & BP(0.5) & -- & -- & -- & -- & -- & -- & 0 \\
& & SQ & 0.743 & 0.077 & 0.066 & 0.000 & 0.125 & 7.47 & 400 \\
& & UKL & 0.899 & 0.144 & 0.141 & 0.000 & 0.330 & 81.76 & 400 \\
\bottomrule
\end{tabular}}
\end{table}

When the regressor functions include the quadratic term, the valid SQ, UKL, and BKL estimators have small bias and coverage close to the nominal level. Omitting the term produces substantial bias even when the fitted weights are moderate. The Valid column also records the absence of a valid exact BP fit in several settings.

\subsection{IHDP Semi-Synthetic Data}
The Infant Health and Development Program (IHDP) data contain 747 observations and 25 covariates, with simulated potential outcomes \citep{Hill2011bayesiannonparametric}. The Online Appendix reports ATE and ATT results for ARW, RA, RW, and TMLE. In this experiment, RW is substantially more variable than the estimators based on the orthogonal score.

\subsection{Lalonde Job-Training Data}
We use the Lalonde job-training data to describe overlap, balance, and weight geometry for the ATT of the National Supported Work (NSW) treated group. Because the true effect for the observational comparison is unobserved, Table~\ref{tab:lalonde_main} reports estimates together with standard errors and weight summaries. The ATT standard errors use the recentered influence function for an estimated treated proportion. Max SMD is the largest absolute standardized mean difference, and ESS C is the effective sample size among the controls.

\begin{table}[!t]
\centering
\caption{ATT estimates in the Lalonde data.}
\label{tab:lalonde_main}
\setlength{\tabcolsep}{4pt}
\begin{tabular}{llrrrrr}
\toprule
Loss & Estimator & Estimate & SE & Max $|\widehat\alpha|$ & Max SMD & ESS C \\
\midrule
\multirow{2}{*}{BKL} & ARW & 1517.3 & 846.8 & 50.00 & 0.135 & 46.33 \\
& TMLE & 1647.0 & 853.0 & 50.00 & 0.135 & 46.33 \\
\multirow{2}{*}{BP(0.5)} & ARW & 1781.1 & 656.3 & 4.080 & 1.150 & 304.0 \\
& TMLE & 1775.3 & 656.3 & 4.080 & 1.150 & 304.0 \\
\multirow{2}{*}{SQ} & ARW & 1805.8 & 635.8 & 3.487 & 1.308 & 334.4 \\
& TMLE & 1803.9 & 635.8 & 3.487 & 1.308 & 334.4 \\
\multirow{2}{*}{UKL} & ARW & 1764.9 & 682.5 & 6.330 & 0.869 & 227.2 \\
& TMLE & 1757.2 & 682.5 & 6.330 & 0.869 & 227.2 \\
\bottomrule
\end{tabular}
\end{table}

The ARW and TMLE estimates are similar for SQ, UKL, and BP. BKL reaches the prespecified bound of 50, and several specifications leave large standardized mean differences. These estimates and diagnostics document how overlap, balance, and weight concentration differ across the fitted specifications.

\section{Discussion, Remarks, and Extensions}
\subsection{Instability of Density-Ratio Estimators}
Density-ratio estimation can overfit when the source and target probability laws have little common support. In treatment-effect estimation, trimming and overlap weights address the related problem by changing the target population \citep{Li2018addressingextreme}. Nonnegative Bregman objectives instead modify the empirical objective \citep{Kato2021nonnegativebregman}, whereas telescoping density-ratio estimation inserts intermediate probability laws and estimates a product of local ratios \citep{Rhodes2020telescopingdensityratio}.

Proposition~\ref{prop:generator_condition_numbers} quantifies the curvature channel. Under ATE overlap $e_0(Z)\in[\eta,1-\eta]$, the value $C=1$ used by the sigmoid ATE specifications, and $0<\omega\leq1$, the condition numbers of SQ, BP, UKL, and BKL have orders $1$, $\eta^{-2(1-\omega)}$, $\eta^{-2}$, and $\eta^{-3}$, respectively. The observed maximum weights in Tables~\ref{tab:coverage_main} and~\ref{tab:lalonde_main} follow the same qualitative ordering among valid fits, with the KL-type generators most sensitive under weak overlap. The condition number quantifies inverse-link sensitivity on a common range, while the dual multipliers and design determine realized weights. Approximation error remains governed by the selected regressor functions.

\subsection{Relation to Sieve Riesz Representers}
\label{sec:sieve_discussion}
The population and sieve Riesz representers are defined on different spaces. The population representer in Equation~\eqref{eq:riesz_identity} exists when the functional is bounded on $L_2(P_X)$, whereas a sieve representer is defined after restricting the functional to a finite-dimensional approximation space. Because the restricted functional is bounded on each such space, the sieve representer remains well-defined even when the population functional is irregular \citep{Chen2014sieveinference,Chen2015sievewald}. The behavior of its norm as the sieve dimension increases distinguishes regular from irregular functionals \citep{Chen2014sieveinference}, and \citet{Chen2015sievesemiparametric} give finite-dimensional formulas that make the sieve representer directly computable.

Generalized Riesz regression recovers the empirical sieve Riesz equations when the dual coordinate is linear in the chosen basis functions and $\lambda=0$; regularization yields their approximate form. The generator selects the Bregman projection when the representer model is restricted or exact balance is not attained. Appendix~\ref{appdx:sieve_relation} gives the detailed comparison. After our series of works, \citet{Brunssmith2026twoapproaches} also discusses that the automatic debiased machine learning and sieve formulations are numerically equivalent for unregularized and ridge-regularized linear, sieve, and RKHS models, although they can differ under lasso regularization or general nonparametric function classes. This distinction separates the common empirical Riesz equations from the objective and regularization used to select a representer.

Appendix~\ref{app:bkl_family} discusses the relation to binary probability losses, whereas Appendix~\ref{app:genriesz} describes the software implementation.

\section{Conclusion}
Generalized Riesz regression fits the Riesz representer under an observable Bregman objective. Its first-order conditions impose empirical Riesz equations in model-dependent tangent directions. Compatibility aligns those directions with regressors selected before estimation, so the resulting empirical Riesz equations protect the intended component of regression error. They provide sharp control of the systematic Neyman error and an orthogonal-score identity under exact balance.

The generator determines the Bregman projection and curvature, the representer model and generator jointly determine the tangent directions, the chosen regressors determine the protected function space, and regularization governs imbalance and weight dispersion. Sparse, RKHS, and neural-network rates connect these finite-sample equations to inference. The experiments confirm the two main design implications: omitted directions generate bias, and weak overlap amplifies curvature-sensitive generators. The framework therefore links representer fitting, balancing, and orthogonal-score construction through the directions selected by optimization.

%% file: Supplementary.tex
\section*{Appendix}
\appendix

\begin{figure}[!t]
    \centering
    \includegraphics[width=0.94\linewidth]{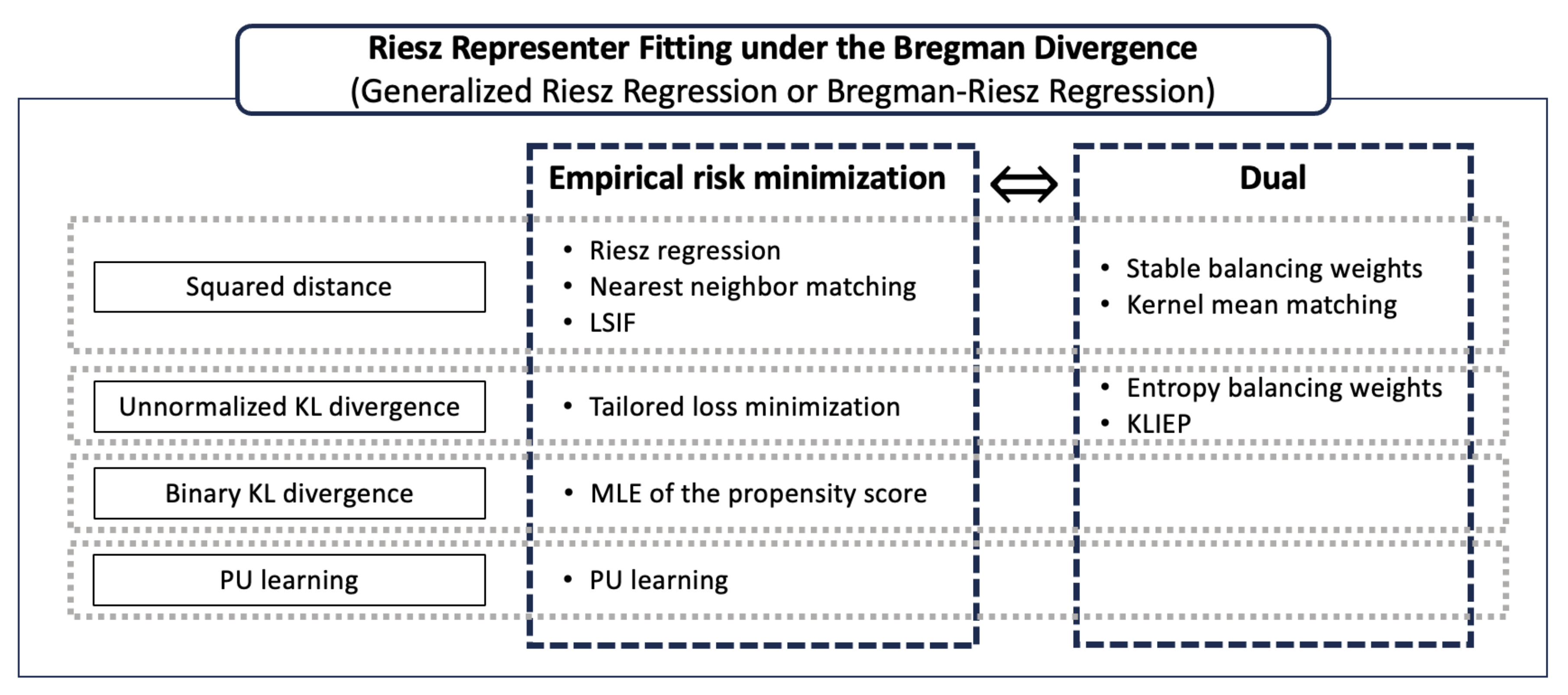}
    \caption{Generalized Riesz regression and selected dual balancing interpretations. The diagram is schematic; Table~\ref{tbl:dre_rre} and Appendix~\ref{app:relations_dre_rre} state the precise relationship represented by each entry.}
    \label{fig:bregman_concept}
\end{figure}

\section{ATT Variance With an Estimated Treatment Share}
\label{appdx:att_correction}
Let $\pi_1=P(D=1)$. If the ATT score is written with $\pi_1$ fixed and $\pi_1$ is replaced by $\widehat\pi_1=n^{-1}\sum_iD_i$, the point estimator has the efficient first-order expansion, but the variance based on the unrecentered score contains an additional term. If $\widehat\psi_i^{\mathrm{ATT}}$ is the score evaluated with $\widehat\pi_1$, a consistent variance estimator uses
\[
\widehat\psi_{i,\mathrm{rec}}^{\mathrm{ATT}}=\widehat\psi_i^{\mathrm{ATT}}-\frac{\widehat\theta^{\mathrm{ATT}}}{\widehat\pi_1}(D_i-\widehat\pi_1).
\]
The standard errors reported for the ATT use this recentered influence function \citep{Hahn1998ontherole}.

\section{Proof of the Automatic Covariate Balancing Property}
\label{appdx:proof:autocovariance}

We prove Theorem~\ref{thm:autocovariance} and Corollary~\ref{cor:automaticcovariate}, and give a dual derivation for the $\ell_1$ case that establishes Proposition~\ref{prop:dual_balancing_program}.

\subsection{Proof of Theorem~\ref{thm:autocovariance}}
Fix basis functions $\{\widetilde\phi_j\}_{j=1}^p$ and define the linear score index
$v_{\bmbeta}(x)=\sum_{j=1}^p\beta_j\widetilde\phi_j(x)$.
By assumption, the link is chosen so that
$\partial g(\alpha_{\bmbeta}(x))=v_{\bmbeta}(x)$.

\paragraph{Step 1: rewrite the empirical objective using convex conjugacy.}
For each observation $i$, let $\calA_i$ be the interval allowed by the representer model. Define $\bar g_i(\alpha)=g(\alpha)$ for $\alpha\in\calA_i$ and $\bar g_i(\alpha)=+\infty$ otherwise, and let $g_i^*$ be the convex conjugate of $\bar g_i$:
\[
g_i^*(v)\coloneqq \sup_{\alpha\in\calA_i}\{\alpha v-g(\alpha)\}.
\]
Because $\alpha_{\bmbeta}(X_i)$ lies in the interior of $\calA_i$ and $v_{\bmbeta}(X_i)=\partial g(\alpha_{\bmbeta}(X_i))$, strict convexity gives
\[
g_i^*\big(v_{\bmbeta}(X_i)\big)=\alpha_{\bmbeta}(X_i)v_{\bmbeta}(X_i)-g(\alpha_{\bmbeta}(X_i)).
\]
Thus, the empirical Bregman objective \eqref{eq:bd_emp_beta} becomes
\begin{align}
\label{eq:bd_as_conjugate}
\widehat{\mathrm{BD}}_g(\alpha_{\bmbeta})
=
\frac{1}{n}\sum_{i=1}^n
\Big\{
g_i^*\big(v_{\bmbeta}(X_i)\big)
-
m\big(W_i,v_{\bmbeta}\big)
\Big\}.
\end{align}
By linearity of $m(\cdot,\cdot)$ in its second argument and the linear expansion of $v_{\bmbeta}$,
\[
m(W_i,v_{\bmbeta})=\sum_{j=1}^p\beta_j\,m(W_i,\widetilde\phi_j).
\]

\paragraph{Step 2: compute the (sub)gradient in $\bmbeta$.}
Since $v_{\bmbeta}(X_i)$ is linear in $\bmbeta$ with
$\partial v_{\bmbeta}(X_i)/\partial \beta_j = \widetilde\phi_j(X_i)$,
we differentiate \eqref{eq:bd_as_conjugate} using the chain rule:
\begin{align*}
\frac{\partial}{\partial \beta_j}\widehat{\mathrm{BD}}_g(\alpha_{\bmbeta})
&=
\frac{1}{n}\sum_{i=1}^n
\Big\{
\partial g_i^*\big(v_{\bmbeta}(X_i)\big)\,\widetilde\phi_j(X_i)
-
m(W_i,\widetilde\phi_j)
\Big\}.
\end{align*}
On the range of $\partial g$ over $\calA_i$, we have $\partial g_i^*(v)=(\partial g)^{-1}(v)$. Since $\alpha_{\bmbeta}(x)=(\partial g)^{-1}(v_{\bmbeta}(x))$ by construction, it follows that $\partial g_i^*(v_{\bmbeta}(X_i))=\alpha_{\bmbeta}(X_i)$.
Therefore, we have
\begin{align}
\label{eq:grad_bd}
\frac{\partial}{\partial \beta_j}\widehat{\mathrm{BD}}_g(\alpha_{\bmbeta})
=
\frac{1}{n}\sum_{i=1}^n
\Big(
\alpha_{\bmbeta}(X_i)\widetilde\phi_j(X_i)-m(W_i,\widetilde\phi_j)
\Big)
=
\widehat\Delta_j(\alpha_{\bmbeta}).
\end{align}

\paragraph{Step 3: KKT conditions with $\ell_q$ penalty.}
Define the penalized objective
\[
Q_n(\bmbeta)\coloneqq \widehat{\mathrm{BD}}_g(\alpha_{\bmbeta})+\frac{\lambda}{q}\|\bmbeta\|_q^q.
\]
Let $\widehat{\bmbeta}$ be an interior minimizer relative to $\calB$. The normal-cone term is then zero, and the first-order optimality condition yields, for each $j$,
\[
0\in
\widehat\Delta_j(\widehat\alpha)
+
\lambda\,\partial\Big(\frac{|\beta_j|^q}{q}\Big)\Big|_{\beta_j=\widehat\beta_j}.
\]
Equivalently, there exists $s_j\in \partial(|\beta_j|^q/q)|_{\widehat\beta_j}$ such that
$\widehat\Delta_j(\widehat\alpha)+\lambda s_j=0$, proving \eqref{eq:kkt_identity}.

If $q=1$, then $s_j\in[-1,1]$, which implies $|\widehat\Delta_j(\widehat\alpha)|\le \lambda$. If $q>1$, then $s_j=\sign(\widehat\beta_j)|\widehat\beta_j|^{q-1}$, which implies $|\widehat\Delta_j(\widehat\alpha)|=\lambda|\widehat\beta_j|^{q-1}$. This completes the proof of Theorem~\ref{thm:autocovariance}.
\qed

\subsection{Proof of Corollary~\ref{cor:automaticcovariate}}
If $\widetilde\phi_j=\phi_j$ for all $j$, then the imbalance definition \eqref{eq:imbalance_def_rewrite} reduces to
\[
\widehat\Delta_j(\alpha)
=
\frac{1}{n}\sum_{i=1}^n\Big(\alpha(X_i)\phi_j(X_i)-m(W_i,\phi_j)\Big),
\]
and the bounds in Theorem~\ref{thm:autocovariance} apply directly.
\qed

\subsection{Dual Characterization for \texorpdfstring{$\ell_1$}{ell1} and Proof of Proposition~\ref{prop:dual_balancing_program}}
We give a standard Fenchel-duality derivation.

Let $q=1$. Define the design matrix $\widetilde\Phi\in\bbR^{n\times p}$ by
$\widetilde\Phi_{ij}\coloneqq \widetilde\phi_j(X_i)$,
and define $b\in\bbR^p$ by
$b_j\coloneqq \frac{1}{n}\sum_{i=1}^n m(W_i,\widetilde\phi_j)$.
Then, we have $v_{\bmbeta}(X_i)=(\widetilde\Phi\bmbeta)_i$, and Equation~\eqref{eq:bd_as_conjugate} gives
\[
\widehat{\mathrm{BD}}_g(\alpha_{\bmbeta})
=
\frac{1}{n}\sum_{i=1}^n g_i^*\big((\widetilde\Phi\bmbeta)_i\big) - b^\top\bmbeta.
\]
Hence, the primal problem in Equation~\eqref{eq:beta_erm} is
\begin{align}
\label{eq:primal_beta_for_dual}
\min_{\bmbeta\in\bbR^p}
\Big\{
\frac{1}{n}\sum_{i=1}^n g_i^*\big((\widetilde\Phi\bmbeta)_i\big)
- b^\top\bmbeta
+\lambda\|\bmbeta\|_1
\Big\}.
\end{align}

Using the conjugate representation $g_i^*(t)=\sup_{\alpha\in\calA_i}\{\alpha t-g(\alpha)\}$,
we may write
\[
\frac{1}{n}\sum_{i=1}^n g_i^*\big((\widetilde\Phi\bmbeta)_i\big)
=
\sup_{\alpha_i\in\calA_i,\ i=1,\ldots,n}
\Big\{
\frac{1}{n}\sum_{i=1}^n \alpha_i(\widetilde\Phi\bmbeta)_i
-\frac{1}{n}\sum_{i=1}^n g(\alpha_i)
\Big\}.
\]
Substituting into \eqref{eq:primal_beta_for_dual} and exchanging $\min$ and $\sup$ (justified by standard strong duality conditions), we obtain the dual
\[
\sup_{\alpha_i\in\calA_i,\ i=1,\ldots,n}
\inf_{\bmbeta\in\bbR^p}
\Big\{
\Big(\frac{1}{n}\widetilde\Phi^\top\bmalpha-b\Big)^\top\bmbeta
+\lambda\|\bmbeta\|_1
-\frac{1}{n}\sum_{i=1}^n g(\alpha_i)
\Big\}.
\]
The inner infimum is finite if and only if
$\big\|\frac{1}{n}\widetilde\Phi^\top\bmalpha-b\big\|_\infty\le \lambda$,
in which case it equals $0$ (this is the conjugacy between $\|\cdot\|_1$ and $\|\cdot\|_\infty$).
Therefore, the dual becomes
\[
\min_{\alpha_i\in\calA_i,\ i=1,\ldots,n}
\frac{1}{n}\sum_{i=1}^n g(\alpha_i)
\quad
\text{subject to}\quad
\Big\|\frac{1}{n}\widetilde\Phi^\top\bmalpha-b\Big\|_\infty\le \lambda.
\]
Expanding the $\ell_\infty$ constraint yields exactly \eqref{eq:dual_balancing_program}, proving Proposition~\ref{prop:dual_balancing_program}.
\qed

\subsection{Proof of Proposition~\ref{prop:arbitrary_pair_foc}}
\label{appdx:proof_arbitrary_pair_foc}
Differentiating the first two terms of the empirical objective gives
\[
\partial_{\beta_j}\{-g(\alpha_{\bmbeta})+\alpha_{\bmbeta}\partial g(\alpha_{\bmbeta})\}
=\alpha_{\bmbeta}g''(\alpha_{\bmbeta})\partial_{\beta_j}\alpha_{\bmbeta},
\]
because the two terms containing $\partial g(\alpha_{\bmbeta})\partial_{\beta_j}\alpha_{\bmbeta}$ cancel. Moreover,
\[
\partial_{\beta_j}(\partial g\circ\alpha_{\bmbeta})
=g''(\alpha_{\bmbeta})\partial_{\beta_j}\alpha_{\bmbeta}
=\psi_{j,\bmbeta}.
\]
Linearity of $m(W,\cdot)$ gives Equation~\eqref{eq:arbitrary_pair_gradient}, and the penalty adds the subgradient $\lambda s_j$. For $\alpha_{\bmbeta}(x)=\ell(\bmphi(x)^\top\bmbeta)$,
\[
\psi_{j,\bmbeta}(x)
=\left.\frac{\rmd}{\rmd t}\partial g(\ell(t))\right|_{t=\bmphi(x)^\top\bmbeta}\phi_j(x).
\]
The multiplier is a common positive constant on $J$ if and only if $\partial g(\ell(t))=at+b$ on $J$ for some $a>0$ and $b\in\bbR$, which is Definition~\ref{def:compatible_pair}.

\section{Preliminaries for the Convergence Rate Analysis}
\label{sec:prelim_convergence}
For $\alpha$ in a model class, define the per-observation objective
\[
\ell_\alpha(W)=\partial g(\alpha(X))\alpha(X)-g(\alpha(X))-m(W,\partial g\circ\alpha).
\]
The population excess objective satisfies
\begin{align}
\label{eq:excess_bregman_identity}
P_0(\ell_\alpha-\ell_{\alpha_0})
=\bbE[\mathrm{BD}^{\dagger}_g(\alpha_0(X)\mid\alpha(X))].
\end{align}
Assumption~\ref{asm:compact_curvature} therefore gives
\begin{align}
\label{eq:curvature_two_sided}
\frac\mu2\|\alpha-\alpha_0\|_{L_2(P_X)}^2
\leq P_0(\ell_\alpha-\ell_{\alpha_0})
\leq\frac{C_g}{2}\|\alpha-\alpha_0\|_{L_2(P_X)}^2.
\end{align}

Let $A(t)=t\partial g(t)-g(t)$. On $\bigcup_{k=1}^K\calI_k$, $A'(t)=t g''(t)$ is bounded. Assumption~\ref{asm:m_reg} and the bounded derivative of $g$ imply
\begin{align}
\label{eq:loss_l2_lipschitz}
\|\ell_\alpha-\ell_{\alpha_0}\|_{L_2(P_0)}
\leq L_\ell\|\alpha-\alpha_0\|_{L_2(P_X)}
\end{align}
for a finite constant $L_\ell$. The same assumptions give a uniform envelope on the model classes used below.

\subsection{Proof of Theorem~\ref{thm:sparse_dual_rate} and the Sparse-Model Inference Statement}
\label{appdx:sparse_dual_proof}
Let
\[
L_n(\bmbeta)=\frac1n\sum_{i=1}^n\left[g_i^*(\widetilde\bmphi(X_i)^\top\bmbeta)-m(W_i,u_{\bmbeta})\right],
\qquad u_{\bmbeta}=\widetilde\bmphi^\top\bmbeta,
\]
where $g_i^*$ is the convex conjugate of $g$ restricted to the sign component allowed at $X_i$. Let $L=P_0L_n$, let $\bmbeta^*$ be an interior minimizer of $L$ over $\calB_{R_1}$, and set $S=\operatorname{supp}(\bmbeta^*)$ and $s=|S|$. We impose $\max_{j\leq p}\|\widetilde\phi_j\|_\infty\leq B_\phi$. The range induced by $\calB_{R_1}$ lies in the intervals of Assumption~\ref{asm:compact_curvature}, and $A_\alpha$ bounds the absolute representer values on this range. Assumption~\ref{asm:m_reg} holds. Finally, with $\bmSigma=\bbE[\widetilde\bmphi(X)\widetilde\bmphi(X)^\top]$, suppose that
\[
\bmdelta^\top\bmSigma\bmdelta\geq\kappa^2\|\bmdelta\|_2^2
\quad\text{whenever}\quad\|\bmdelta_{S^c}\|_1\leq3\|\bmdelta_S\|_1.
\]
The constants below are explicit but not optimized.

\paragraph{Conjugate curvature.}
On the induced range, the inverse function theorem gives
\[
(g_i^*)'(v)=(\partial g|_{c(X_i)})^{-1}(v),
\qquad
(g_i^*)''(v)=\frac{1}{g''((g_i^*)'(v))}\in[1/C_g,1/\mu].
\]
Thus, $(g_i^*)'$ is $1/\mu$-Lipschitz.

\paragraph{Gradient concentration.}
The $j$th coordinate of the empirical gradient at $\bmbeta^*$ is
\[
\partial_{\beta_j}L_n(\bmbeta^*)=\frac1n\sum_{i=1}^n\xi_{ij},
\qquad\xi_{ij}=\alpha_{\bmbeta^*}(X_i)\widetilde\phi_j(X_i)-m(W_i,\widetilde\phi_j).
\]
Because $\bmbeta^*$ is an interior population minimizer, $\bbE[\xi_{ij}]=0$. Assumption~\ref{asm:m_reg} and the induced range give $|\xi_{ij}|\leq B_\xi=B_\phi(A_\alpha+B_m)$. Hoeffding's inequality and a union bound therefore imply
\begin{align}
\label{eq:sparse_gradient_event}
\|\nabla L_n(\bmbeta^*)\|_\infty\leq B_\xi\sqrt{\frac{2\log(2p/\varepsilon)}{n}}
\end{align}
with probability at least $1-\varepsilon$.

\paragraph{Sample restricted eigenvalue.}
Let $\widehat\bmSigma=n^{-1}\sum_i\widetilde\bmphi(X_i)\widetilde\bmphi(X_i)^\top$. Entrywise Hoeffding bounds give
\[
D_n=\max_{j,k}|\widehat\Sigma_{jk}-\Sigma_{jk}|\leq B_\phi^2\sqrt{\frac{2\log(2p^2/\varepsilon)}{n}}
\]
with probability at least $1-\varepsilon$. On the cone, $\|\bmdelta\|_1\leq4\sqrt{s}\|\bmdelta\|_2$. Under the sample-size condition in Theorem~\ref{thm:sparse_dual_rate},
\[
\bmdelta^\top\widehat\bmSigma\bmdelta\geq\frac{\kappa^2}{2}\|\bmdelta\|_2^2.
\]

\paragraph{Cone condition and coefficient rates.}
Work on the intersection of the two preceding events. The stated choice of $\lambda$ gives $\lambda\geq2\|\nabla L_n(\bmbeta^*)\|_\infty$. Optimality and convexity yield
\[
\|\widehat\bmbeta_{S^c}-\bmbeta^*_{S^c}\|_1\leq3\|\widehat\bmbeta_S-\bmbeta^*_S\|_1.
\]
Set $\bmdelta=\widehat\bmbeta-\bmbeta^*$. Conjugate curvature along the segment between the two coefficients gives
\[
L_n(\widehat\bmbeta)-L_n(\bmbeta^*)-\nabla L_n(\bmbeta^*)^\top\bmdelta
\geq\frac{1}{2C_g}\bmdelta^\top\widehat\bmSigma\bmdelta.
\]
Combining this bound with the basic inequality gives
\[
\frac{1}{2C_g}\bmdelta^\top\widehat\bmSigma\bmdelta\leq\frac{3\lambda}{2}\sqrt{s}\|\bmdelta\|_2.
\]
The sample restricted eigenvalue then implies
\[
\|\widehat\bmbeta-\bmbeta^*\|_2\leq\frac{6C_g\sqrt{s}\lambda}{\kappa^2},
\qquad
\|\widehat\bmbeta-\bmbeta^*\|_1\leq\frac{24C_gs\lambda}{\kappa^2},
\]
and
\[
\sqrt{\bmdelta^\top\widehat\bmSigma\bmdelta}\leq\frac{3\sqrt{2}C_g\sqrt{s}\lambda}{\kappa}.
\]

\paragraph{Representer rate.}
Because $(g_i^*)'$ is $1/\mu$-Lipschitz,
\[
|\widehat\alpha-\alpha_{\bmbeta^*}|\leq\mu^{-1}|\widetilde\bmphi^\top\bmdelta|.
\]
The entrywise deviation bound and the cone condition give $\bmdelta^\top\bmSigma\bmdelta\leq2\bmdelta^\top\widehat\bmSigma\bmdelta$. Consequently,
\[
\|\widehat\alpha-\alpha_{\bmbeta^*}\|_{L_2(P_X)}\leq\frac{6(C_g/\mu)\sqrt{s}\lambda}{\kappa},
\]
which proves Theorem~\ref{thm:sparse_dual_rate}. If the $\ell_1$-ball constraint is inactive, Theorem~\ref{thm:autocovariance} and Equation~\eqref{eq:automatic_neyman_min} give the two imbalance bounds.

Under correct specification, $\alpha_{\bmbeta^*}=\alpha_0$. The theorem then gives $\|\widehat\alpha-\alpha_0\|_{L_2(P_X)}=O_P(\sqrt{s\log(p\vee n)/n})$ when the displayed constants remain bounded. The rate condition in the main text verifies the product condition in Assumption~\ref{asm:conv_rate}. The compact induced range also gives
\[
\|(\widehat\alpha-\alpha_0)(\widehat\gamma-\gamma_0)\|_{L_2(P_X)}\leq2A_\alpha\|\widehat\gamma-\gamma_0\|_{L_2(P_X)}=o_P(1).
\]
The same statements hold uniformly over a fixed number of folds. Theorem~\ref{thm:asymp_normality} proves the sparse-model inference statement. This proof uses the standard decomposable-regularizer argument of \citet{Negahban2012aunified}.

\section{Proof of Theorem~\ref{thm:l2norm}}
\label{appex:kernel_proof}
Let $\widehat\alpha=\alpha_{\widehat f^{\mathrm{RKHS}}}$ and let $\alpha_0=\alpha_{f_0}$. Optimality gives
\begin{align}
\label{eq:rkhs_basic}
P_n\ell_{\widehat\alpha}+\lambda\|\widehat f^{\mathrm{RKHS}}\|_{\calF}^2
\leq P_n\ell_{\alpha_0}+\lambda\|f_0\|_{\calF}^2.
\end{align}
By the bracketing-entropy assumption for the induced loss class, Equation~\eqref{eq:loss_l2_lipschitz}, and the standard localized empirical-process bound used in \citet{VandeGeer2000empiricalprocesses} and \citet{Kanamori2012statisticalanalysis},
\begin{align}
\label{eq:rkhs_emp_process}
\left|(P_n-P_0)(\ell_\alpha-\ell_{\alpha_0})\right|
=O_P\left(\max\left\{\frac{\|\alpha-\alpha_0\|_{L_2(P_X)}^{1-\tau/2}}{\sqrt n},n^{-2/(2+\tau)}\right\}\right)
\end{align}
uniformly over the fixed RKHS ball.

Set $r_n=\|\widehat\alpha-\alpha_0\|_{L_2(P_X)}$. Combining Equations~\eqref{eq:curvature_two_sided}, \eqref{eq:rkhs_basic}, and~\eqref{eq:rkhs_emp_process} gives
\[
r_n^2
=O_P\left(\max\left\{\frac{r_n^{1-\tau/2}}{\sqrt n},n^{-2/(2+\tau)},\lambda\right\}\right).
\]
If the first term on the right is largest, division by $r_n^{1-\tau/2}$ gives $r_n=O_P(n^{-1/(2+\tau)})$. The other two cases give the same stochastic order or $r_n=O_P(\lambda^{1/2})$. Hence, we have
\[
\|\widehat\alpha^{\mathrm{RKHS}}-\alpha_0\|_{L_2(P_X)}
=O_P\left(\max\{n^{-1/(2+\tau)},\lambda^{1/2}\}\right).
\]
This proves Theorem~\ref{thm:l2norm}.

\section{Proof of Theorem~\ref{thm:est_error_nn}}
\label{appex:neural_proof}
Let $\calF=\calF_n^{\mathrm{FNN}}$, let $p_n=\operatorname{Pdim}(\calF)$, and write $D_f=\|\alpha_f-\alpha_0\|_{L_2(P_X)}$. For $h_f=\ell_{\alpha_f}-\ell_{\alpha_0}$, Equation~\eqref{eq:curvature_two_sided} gives $P_0h_f\asymp D_f^2$. Equation~\eqref{eq:loss_l2_lipschitz} gives $P_0h_f^2\leq\Lambda^2D_f^2$ for a constant $\Lambda$, and the compact range gives a fixed envelope.

Because $\zeta^{-1}(x,\cdot)$ is nondecreasing, the subgraphs of $\{\alpha_f:f\in\calF\}$ are reparameterizations of the subgraphs of $\calF$. Thus, their pseudodimension is at most $p_n$. The contraction inequality for $A\circ\alpha_f$ and the induced-class local-complexity condition stated in Section~\ref{sec:convanalysis} give, for $\rho$ above a constant multiple of $\sqrt{p_n\log n/n}$,
\begin{align}
\label{eq:nn_local_rad}
\bbE\left[\mathfrak{R}_n\{h_f:D_f\leq\rho\}\right]
\leq C\rho\sqrt{\frac{p_n\log n}{n}}.
\end{align}
The local Rademacher theorem of \citet{Bartlett2005localrademacher}, applied on dyadic shells and combined with the variance bound above, yields the following oracle bound for any $z>0$: with probability at least $1-\exp(-z)$,
\begin{align}
\label{eq:nn_oracle}
\|\widehat\alpha^{\mathrm{FNN}}-\alpha_0\|_{L_2(P_X)}^2
\leq C\left(\frac{p_n\log n}{n}
+\inf_{f\in\calF}\|\alpha_f-\alpha_0\|_{L_2(P_X)}^2
+\delta_n
+\frac{z}{n}\right).
\end{align}
Choose $f^\ast$ within a factor two of the approximation error. Equation~\eqref{eq:nn_optimization_error} gives $P_n\ell_{\widehat\alpha^{\mathrm{FNN}}}\leq P_n\ell_{\alpha_{f^\ast}}+\delta_n$. On every shell on which $P_0h_f$ is larger than a sufficiently large multiple of the approximation error, $\delta_n$, and the stochastic terms in Equation~\eqref{eq:nn_oracle}, Equation~\eqref{eq:nn_local_rad} and Bernstein's inequality make $(P_0-P_n)h_f$ smaller than $P_0h_f/2$. The approximate minimizing property then excludes that shell. A union bound over the shells proves Equation~\eqref{eq:nn_oracle}.

For the H\"older specialization, let $f_0\in\Sigma(\nu,M,[0,1]^d)$ and compose the networks with the range truncation $T_M(u)=\max\{-M,\min\{M,u\}\}$. The ReLU approximation result used by \citet{Zheng2022anerror} gives
\[
\inf_{f\in\calF}\|f-f_0\|_{L_2(P_X)}
\leq C n^{-\nu/(d+2\nu)}\log^{3/2}n
\]
for the width and depth stated there. The Lipschitz property of $\zeta^{-1}$ gives the same order for the induced representer approximation, while the corresponding pseudodimension satisfies $p_n\leq Cn^{d/(d+2\nu)}\log^3n$ \citep{Anthony1999neuralnetwork,Zheng2022anerror}. Consequently, if $\delta_n=O_P(n^{-2\nu/(d+2\nu)}\log^4n)$, substitution into Equation~\eqref{eq:nn_oracle} gives
\[
\|\widehat\alpha^{\mathrm{FNN}}-\alpha_0\|_{L_2(P_X)}^2
=O_P\left(n^{-2\nu/(d+2\nu)}\log^4n\right),
\]
which proves Theorem~\ref{thm:est_error_nn}.

\section{Proof of Theorem~\ref{thm:asymp_normality}}
\label{appdx:proof_asymp_normality}
Write $\varepsilon=Y-\gamma_0(X)$ and $\psi_0(W)=\psi(W;\eta_0,\theta_0)$. By linearity of $m$,
\begin{align}
\label{eq:asympnorm_score_decomp_main}
\psi(W;\eta,\theta_0)-\psi_0(W)
={}&m(W,\gamma-\gamma_0)-\alpha_0(X)(\gamma-\gamma_0)(X)\\
&+(\alpha-\alpha_0)(X)\varepsilon
-(\alpha-\alpha_0)(X)(\gamma-\gamma_0)(X).
\end{align}
The Riesz identity and $\bbE[\varepsilon\mid X]=0$ imply
\[
\bbE[\psi(W;\eta,\theta_0)-\psi_0(W)]
=-\bbE[(\alpha-\alpha_0)(X)(\gamma-\gamma_0)(X)].
\]
Assumption~\ref{asm:conv_rate} therefore makes the expectation of the estimated score difference $o_P(n^{-1/2})$.

Under the Donsker alternative in Assumption~\ref{asm:donsker}, stochastic equicontinuity gives
\[
\sqrt n(P_n-P_0)\{\psi(\cdot;\widehat\eta,\theta_0)-\psi_0\}=o_P(1).
\]
Under the cross-fitting alternative, condition on the observations used to estimate the nuisance parameters for a given fold. Assumptions~\ref{asm:m_reg},~\ref{asm:conv_rate}, and~\ref{asm:score_moments} imply that the conditional second moment of the score difference in Equation~\eqref{eq:asympnorm_score_decomp_main} is $o_P(1)$, uniformly over the fixed number of folds. Conditional Chebyshev's inequality then gives the same $o_P(1)$ empirical-process remainder after multiplication by $\sqrt n$.

In either case,
\[
\sqrt n(\widehat\theta^{\mathrm{ARW}}-\theta_0)
=\frac1{\sqrt n}\sum_{i=1}^n\psi_0(W_i)+o_P(1).
\]
The central limit theorem proves the stated limit. The same expansion and the finite $2+\delta$ moment imply consistency of the sample variance.

\subsection{Proof of Theorem~\ref{thm:rw_efficiency_approx_balance}}
\label{appdx:proof_rw_approx_balance}
Let $\varepsilon=Y-\gamma_0(X)$. Since $\gamma_0=\gamma_n+r_n$,
\[
\widehat\theta^{\mathrm{RW}}-\theta_0
=(P_n-P_0)\psi(\cdot;\eta_0,\theta_0)+P_n[(\widehat\alpha-\alpha_0)\varepsilon]
+\widehat\Delta(\widehat\alpha,\gamma_n)+\widehat\Delta(\widehat\alpha,r_n).
\]
For every realization of $\widehat\alpha$, $P_0[(\widehat\alpha-\alpha_0)\varepsilon]=0$ follows from $\bbE[\varepsilon\mid X]=0$. The Riesz identity gives
\[
\widehat\Delta(\widehat\alpha,r_n)=P_0[(\widehat\alpha-\alpha_0)r_n]+(P_n-P_0)[\widehat\alpha r_n-m(W,r_n)].
\]
Cauchy--Schwarz and Equation~\eqref{eq:rw_approx_balance_conditions} make the remainder $o_P(n^{-1/2})$, which proves the theorem.

\section{Related Work}
\label{sec:relatedwork}

\subsection{Asymptotically Efficient Estimation}
\paragraph{Early history.}
Constructing asymptotically efficient estimators is a classical problem in statistics, machine learning, economics, epidemiology, and related fields. In the semiparametric models considered here, the parameter of interest is low-dimensional, while the model also contains nuisance parameters. The objective is to obtain a $\sqrt{n}$-consistent and asymptotically normal estimator of the parameter of interest. The difficulty is that nuisance estimation errors typically decay more slowly than $n^{-1/2}$, so they can affect first-order inference. A large literature develops methods that reduce this effect, including \citet{Levit1976onthe}, \citet{Ibragimov1981statisticalestimation}, \citet{Pfanzagl1982contributionsto}, \citet{Klaassen1987consistentestimation}, \citet{Robinson1988rootnconsistent}, \citet{Newey1994theasymptotic}, \citet{VanderVaart1998asymptoticstatistics}, \citet{Bickel1998efficientadaptive}, \citet{Ai2012thesemiparametric}, and \citet{Chernozhukov2018doubledebiased}.

An asymptotically efficient estimator is $\sqrt{n}$-consistent and asymptotically normal, and its asymptotic variance attains the efficiency bound. Asymptotic efficiency bounds are called H\'ajek--Le Cam efficiency bounds, or semiparametric efficiency bounds when we consider semiparametric models \citep{Hajek1970acharacterization,LeCam1972theoryofstatisics,LeCam1986asymptoticmethods}. They share the same motivation as the Cram\'er--Rao lower bound. While the Cram\'er--Rao lower bound provides a lower bound for unbiased estimators, asymptotic efficiency bounds provide lower bounds for asymptotically unbiased estimators, called regular estimators. It is known that efficient estimators are regular and asymptotically linear (RAL) with the efficient influence function. Therefore, the construction of asymptotically efficient estimators is equivalent to the construction of RAL estimators \citep{VanderVaart1998asymptoticstatistics}.

The three main approaches to constructing efficient estimators are one-step bias correction, estimating equation methods, and TMLE \citep{Schuler2018comparisonmethods,VanderVaart2002semiparametricstatistics,vanderLaan2006targetedmaximum,vanderLaan2018targetedlearning_book2}. Although these approaches are asymptotically equivalent in many cases, their finite-sample behavior may differ. A related literature studies post-selection inference with high-dimensional control variables \citep{Belloni2011inference,Belloni2014highdimensionalmethods,Belloni2016postselection}.

\paragraph{Debiased machine learning and Riesz representer.}
Debiased and double machine learning (DML) provides a general recipe for constructing asymptotically linear and semiparametrically efficient estimators by combining flexible first-step learning with Neyman-orthogonal scores \citep{Chernozhukov2018doubledebiased}. In classical semiparametric theory, such orthogonalization is naturally expressed through the efficient influence function, obtained by projecting the pathwise derivative onto the nuisance tangent space \citep{Newey1994theasymptotic}. Related influence-function and projection-based bias corrections also appear in earlier semiparametric two-step and sieve inference work \citep{Chen2008semiparametricefficiency,Ackerberg2014asympototicefficiency}.

For many targets, the orthogonal score has an augmented form. In particular, for linear and local functionals of a regression-type nuisance, \citet{Chernozhukov2022automaticdebiased} make explicit that one can write an orthogonal score as a plug-in term plus a correction that multiplies the regression residual by the functional's Riesz representer. \citet{Chernozhukov2022automaticdebiased} treat the Riesz representer itself as an additional nuisance parameter and propose to estimate it from data via regularized \emph{Riesz representer regression}, combined with cross-fitting, yielding adaptive inference for regular and nonregular local functionals \citep{Chernozhukov2022automaticdebiased}. This places the adjustment term, often called the one-step bias-correction term or the clever covariate, into the same estimation procedure as the other nuisance parameters.

Closely related Riesz-representation-based characterizations and feasible approximations of this adjustment term have been developed in the sieve and semiparametric inference literature by Chen and coauthors. In semiparametric conditional moment restriction settings, \citet{Chen2015sievewald} characterize the pathwise derivative of a target functional as a linear functional on an infinite-dimensional Hilbert space and show that a \emph{population} Riesz representer exists if and only if this derivative is bounded. When the derivative is unbounded, the functional is irregular and the population representer may fail to exist \citep{Chen2015sievewald}. On any finite-dimensional sieve space the derivative is automatically bounded, so a \emph{sieve} Riesz representer is always well-defined. It can be used to construct implementable sieve influence functions and variance estimators \citep{Chen2015sievewald}. Building on this perspective, \citet{Chen2014sieveinference} emphasize that even when the population Riesz representer is difficult to compute, or does not exist on the infinite-dimensional space, the sieve Riesz representer can always be computed explicitly, enabling a unified treatment of regular and irregular functionals \citep{Chen2014sieveinference}. \citet{Chen2014sieveinference} also relate regularity to the behavior of the sieve Riesz representer norm as sieve dimension increases, providing a convenient diagnostic of whether root-$n$ inference is attainable \citep{Chen2014sieveinference}. Finally, \citet{Chen2015sievesemiparametric} show that while the population representer may not admit a closed-form solution, its sieve analogue often does and can be computed via finite-dimensional linear algebra, including generalized inverse formulas, yielding practical influence-function-based inference and variance estimation procedures \citep{Chen2015sievesemiparametric}. For the relationship between our generalized Riesz regression and the series Riesz representer, see Appendix~\ref{appdx:kkt_riesz_linear_equation}.

From this viewpoint, the Riesz representer regression terminology of \citet{Chernozhukov2022automaticdebiased} can be interpreted as a high-dimensional regularized analogue of the sieve Riesz representer constructions in \citet{Chen2015sievewald,Chen2014sieveinference,Chen2014sievem,Chen2015sievesemiparametric}. Both lines of work use Riesz representation to express and estimate the orthogonal-score adjustment term. However, \citet{Chernozhukov2022automaticdebiased} estimate the representer over large sets of basis functions through regularization and cross-fitting, complementing the closed-form series and sieve calculations emphasized in the sieve literature \citep{Chen2014sieveinference,Chen2015sievesemiparametric}.

Minimax methods provide a complementary route. Augmented minimax linear estimation controls imbalance over a function class while combining weighting with regression adjustment \citep{Hirshberg2021augmentedminimax}. Adversarial Riesz estimation estimates the representer through a min--max objective rather than a Bregman empirical-risk objective \citep{Chernozhukov2026adversarialestimation}. These methods are not special cases of generalized Riesz regression, although they target the same orthogonal-score adjustment term. The sparse result in Theorem~\ref{thm:sparse_dual_rate} instead extends the empirical-risk route to every compatible loss--link pair and makes the curvature dependence explicit.

\subsection{ATE Estimation}
Randomized controlled trials are the gold standard for causal inference, where treatments are assigned while maintaining balance between treatment groups. However, they are not always feasible, and we aim to estimate causal effects from observational data, where imbalance between treatment groups often arises. To correct this imbalance, propensity scores or balancing weights have been proposed.

In ATE estimation, the Riesz representer corresponds to inverse propensity weights, so propensity-score estimation is central. Maximum likelihood is a standard choice, but Riesz regression estimates the representer directly and applies to target functionals beyond the ATE \citep{Chernozhukov2021automaticdebiased,Chernozhukov2022riesznet,Lee2025rieszboost}. Covariate balancing provides another direct approach. Because the propensity score is the coarsest balancing score \citep{Rosenbaum1983centralrole}, many studies estimate either the propensity score or the weights by imposing covariate balance \citep{Li2017balancingcovariates,Imai2013covariatebalancing,Hainmueller2012entropybalancing,Zubizarreta2015stableweights,Tarr2025estimatingaverage,Chan2016globallyefficient,Wong2017kernelbased}. Moreover, Riesz regression and covariate balancing correspond to the same optimization problem under the stated duality conditions \citep{BrunsSmith2022outcomeassumptions,BrunsSmith2025augmentedbalancing,Benmichael2021balancingact,Zhao2019covariatebalancing}. Exact and approximate balance also underlie double-robustness and efficiency results for entropy and minimal-dispersion weights \citep{Zhao2016entropybalancing,Wang2019minimaldispersion}. Inverse probability tilting provides a likelihood-based balancing construction for missing-data moment models \citep{Graham2012inverseprobability}.

\paragraph{Covariate balancing.}
Covariate balancing is a popular approach for propensity score or balancing weight estimation. The propensity score is a balancing score, and based on this property, existing studies propose estimating the propensity score or the weights so that weighted covariate moments match between treated and control groups. \citet{Imai2013covariatebalancing} propose estimating the propensity score by matching first moments, and \citet{Hazlett2020kernelbalancing} extends this idea to higher-moment matching by mapping covariates into a high-dimensional space via basis functions. Other methods do not directly specify a propensity-score model. These empirical balancing methods include entropy balancing \citep{Hainmueller2012entropybalancing} and stable weights \citep{Zubizarreta2015stableweights}. These two approaches may appear different, but \citet{Zhao2019covariatebalancing} and \citet{BrunsSmith2025augmentedbalancing} show that they are essentially equivalent through a duality relationship.

\subsection{Density Ratio Estimation}
A parallel line of work is density ratio estimation, which has been extensively studied in machine learning. We refer to the ratio between two densities as the density ratio. The density ratio is a useful tool in semiparametric inference, as used in covariate shift adaptation \citep{Shimodaira2000improvingpredictive,Kato2024doubledebiasedcovariateshift}, and we show that the same density-ratio representation extends to other linear functionals, including the ATE.

While the density ratio can be estimated by separately estimating each density, such an approach may magnify estimation errors by compounding the errors from two separate estimators. To address this issue, direct density-ratio estimators have been studied, including moment matching \citep{Huang2007correctingsample,Gretton2009covariateshift}, probabilistic classification \citep{Qin1998inferencesfor,Cheng2004semiparametricdensity}, density matching \citep{Nguyen2010estimatingdivergence}, density ratio fitting \citep{Kanamori2009aleastsquares}, and PU learning \citep{Kato2019learningfrom}. It is also known that when complicated models such as neural networks are used for this task, the loss function can diverge in finite samples \citep{Kiryo2017positiveunlabeledlearning}. Therefore, density ratio estimation methods with neural networks have been investigated \citep{Kato2021nonnegativebregman,Rhodes2020telescopingdensityratio}.

As discussed in this study and in existing work such as \citet{Uehara2020offpolicy} and \citet{Lin2023estimationbased}, density ratio estimation is closely related to propensity score estimation. In particular, this study shows that the formulations of Riesz regression and LSIF in density ratio estimation are essentially the same. While Riesz regression applies to more general problems, the LSIF literature provides a range of theoretical and empirical results. One important extension is to generalize LSIF via Bregman divergence minimization \citep{Sugiyama2011densityratio}, and this study is strongly inspired by that work.

Beyond their role in semiparametric analysis, density ratios are also used in tasks such as learning with noisy labels \citep{Liu2014classificationwith,Fang2020rethinkingimportance}, anomaly detection \citep{Smola2009relativenovelty,Hido2008inlierbased,Abe2019anomalydetection,Nam2015directdensityratio}, two-sample testing \citep{Keziou2005testof,Kanamori2010fdivergence,Sugiyama2011leastsquarestwosample}, change-point detection \citep{Kawahara2009changepointdetection}, causal inference \citep{Uehara2020offpolicy}, and recommendation systems \citep{Togashi2021densityratiobased}. In causal inference, \citet{Uehara2020offpolicy} investigate efficient ATE estimation and policy learning under covariate shift. \citet{Kato2024activeadaptive} apply this approach to adaptive experimental design, and \citet{Kato2025puate} extend the framework to a PU learning setup. Density ratio estimation is discussed from the viewpoint of large language models (LLMs) by \citet{Higuchi2025directdensity}.

\section{Riesz Regression and Density Ratios}
\label{appdx:rieszdens}
As explained in the main text and in \citet{Kato2025rieszregression}, the Riesz representer is closely connected to density ratio estimation. In particular, for ATE, the Riesz representer can be expressed in terms of two density ratios relative to the marginal covariate distribution, which leads to a decomposition of the squared loss objective into two LSIF problems.

\paragraph{Riesz representer and density ratio.}
Let $p_Z$ denote the marginal density of $Z$ and $p_{Z\mid D=d}$ the conditional density of $Z$ given $D=d$. Let $\kappa_d\coloneqq P_0\p{D=d}$. By Bayes' rule,
\[
p_{Z\mid D=d}(z)
=
\frac{p_Z(z)P_0\p{D=d\mid Z=z}}{P_0\p{D=d}}
=
\frac{p_Z(z)e_0(z)^d\p{1-e_0(z)}^{1-d}}{\kappa_d},
\]
where $e_0(z)\coloneqq P_0\p{D=1\mid Z=z}$.

Define the density ratios with respect to the marginal distribution of $Z$ by
\[
r_1(z)\coloneqq \frac{p_Z(z)}{p_{Z\mid D=1}(z)},
\qquad
r_0(z)\coloneqq \frac{p_Z(z)}{p_{Z\mid D=0}(z)}.
\]
From the expression above,
\[
r_1(z)=\frac{\kappa_1}{e_0(z)},
\qquad
r_0(z)=\frac{\kappa_0}{1-e_0(z)}.
\]
Therefore, the ATE Riesz representer can be written as
\[
\alpha^{\text{ATE}}_0\p{D,Z}
=
\frac{\bbI\sqb{D=1}}{e_0\p{Z}}
-
\frac{\bbI\sqb{D=0}}{1-e_0\p{Z}}
=
\bbI\sqb{D=1}\frac{r_1\p{Z}}{\kappa_1}
-
\bbI\sqb{D=0}\frac{r_0\p{Z}}{\kappa_0}.
\]
Equivalently, estimating $\alpha^{\text{ATE}}_0$ reduces to estimating the two ratios $r_1$ and $r_0$, which compare the marginal covariate distribution with the treated and control covariate distributions, respectively.

\paragraph{Squared loss objective and decomposition into two LSIF problems.}
We next connect this representation to LSIF, a density ratio estimation method proposed in \citet{Kanamori2009aleastsquares}.
Let $g^{\text{SQ}}\p{u}\coloneqq \p{u-1}^2$ be the squared loss. The corresponding population squared loss Bregman objective can be written as
\[
\mathrm{BD}_{g^{\text{SQ}}}\p{\alpha}
=
\bbE\Bigsqb{-2\p{\alpha\p{1,Z}-\alpha\p{0,Z}}+\alpha\p{D,Z}^2},
\]
where $\alpha\p{d,Z}$ denotes the value of the representer evaluated at treatment status $d$ and covariates $Z$. Under the parameterization
\[
\alpha\p{D,Z}
=
\bbI\sqb{D=1}\frac{r_1\p{Z}}{\kappa_1}
-
\bbI\sqb{D=0}\frac{r_0\p{Z}}{\kappa_0},
\]
we have $\alpha\p{1,Z}=r_1\p{Z}/\kappa_1$ and $\alpha\p{0,Z}=-r_0\p{Z}/\kappa_0$, hence $\alpha\p{1,Z}-\alpha\p{0,Z}=r_1\p{Z}/\kappa_1+r_0\p{Z}/\kappa_0$. Substituting this into $\mathrm{BD}_{g^{\text{SQ}}}\p{\alpha}$ and using the law of total expectation yields
\begin{align}
\mathrm{BD}_{g^{\text{SQ}}}\p{\alpha}
&=
-2\bbE\sqb{\frac{r_1\p{Z}}{\kappa_1}+\frac{r_0\p{Z}}{\kappa_0}}
+\bbE\sqb{\alpha\p{D,Z}^2}.
\label{eq:riesz-nn-br-split-0}
\end{align}
Also,
\[
\bbE\sqb{\alpha\p{D,Z}^2}
=
\kappa_1\bbE\sqb{\p{\frac{r_1\p{Z}}{\kappa_1}}^2\mid D=1}
+
\kappa_0\bbE\sqb{\p{\frac{r_0\p{Z}}{\kappa_0}}^2\mid D=0}.
\]
Rewriting \eqref{eq:riesz-nn-br-split-0} in terms of expectations with respect to $p_Z$ and $p_{Z\mid D=d}$ gives, up to an additive constant,
\begin{align}
\mathrm{BD}_{g^{\text{SQ}}}\p{\alpha}
&\equiv
\frac{1}{\kappa_1}
\cb{-2\bbE_Z\sqb{r_1\p{Z}}+\bbE_{Z\mid D=1}\sqb{r_1\p{Z}^2}}
\notag\\
&\quad+
\frac{1}{\kappa_0}
\cb{-2\bbE_Z\sqb{r_0\p{Z}}+\bbE_{Z\mid D=0}\sqb{r_0\p{Z}^2}}.
\label{eq:riesz-nn-br-split}
\end{align}
If $r_1$ and $r_0$ are treated as independent functions, the two positive factors do not affect their separate minimizers. Hence, minimizing \eqref{eq:riesz-nn-br-split} over $\p{r_1,r_0}$ separates into two independent LSIF problems
\begin{align*}
r_1^*
&=
\argmin_{r_1}
\cb{-2\bbE_Z\sqb{r_1\p{Z}}+\bbE_{Z\mid D=1}\sqb{r_1\p{Z}^2}},\\
r_0^*
&=
\argmin_{r_0}
\cb{-2\bbE_Z\sqb{r_0\p{Z}}+\bbE_{Z\mid D=0}\sqb{r_0\p{Z}^2}},
\end{align*}
where $\bbE_Z$ and $\bbE_{Z\mid D=d}$ denote expectations under $P_0\p{Z}$ and $P_0\p{Z\mid D=d}$.

At the sample level, define
\[
n_1\coloneqq \sum^n_{i=1}\bbI\sqb{D_i=1},
\qquad
n_0\coloneqq \sum^n_{i=1}\bbI\sqb{D_i=0}.
\]
The empirical LSIF objectives can be written as
\begin{align*}
\widehat R_1\p{r_1}
&\coloneqq
-\frac{2}{n}\sum^n_{i=1} r_1\p{Z_i}
+
\frac{1}{\sum^n_{i=1}\bbI\sqb{D_i=1}}
\sum^n_{i=1}\bbI\sqb{D_i=1}r_1\p{Z_i}^2,\\
\widehat R_0\p{r_0}
&\coloneqq
-\frac{2}{n}\sum^n_{i=1} r_0\p{Z_i}
+
\frac{1}{\sum^n_{i=1}\bbI\sqb{D_i=0}}
\sum^n_{i=1}\bbI\sqb{D_i=0}r_0\p{Z_i}^2.
\end{align*}

\section{Detailed Relations Among Existing Methods}
\label{app:relations_dre_rre}
Table~\ref{tbl:dre_rre} relates density-ratio estimation, moment balance, and Riesz representer estimation. The generator $g$ determines the Bregman divergence, while the model class $\calH$ determines the admissible representer functions. Automatic regressor balancing requires the additional loss--link condition that $u=\partial g\circ\alpha$ be linear in the model coefficients. Entries in the same row of the table need not express the same type of relationship: some have identical objectives, some use the same objective over different model classes, and others share the dual moment equations. The target functional $m$ and the domain restrictions must therefore be stated separately.

\subsection{SQ-Riesz}
\label{app:sq_family}
With $g^{\mathrm{SQ}}\p{\alpha}=\p{\alpha-C}^2$, the empirical Bregman objective reduces to a quadratic ERM. In covariate shift, choosing $C=1$ yields the classical LSIF criterion up to constants \citep{Kanamori2009aleastsquares}, and its kernelized form KuLSIF when $\calH$ is an RKHS \citep{Kanamori2012statisticalanalysis}. In Riesz representer estimation, the same quadratic criterion is used in Riesz regression, and the same loss underlies a variety of implementations based on series, RKHS, forests, and boosting.

\subsubsection{LSIF}
\label{app:lsif_as_sqriesz}
LSIF is a density-ratio estimation method that trains a density ratio model $r$ by fitting it to the true density ratio $r_0\p{Z}=p_1\p{Z}/p_0\p{Z}$ under the squared loss \citep{Kanamori2009aleastsquares}. The population objective reduces to
\[
\mathrm{BD}^{\mathrm{CS}}_{g}\p{\alpha}
=
\bbE_{p_0}\sqb{\partial g\p{\alpha\p{Z}}\alpha\p{Z}-g\p{\alpha\p{Z}}}
-
\bbE_{p_1}\sqb{\partial g\p{\alpha\p{Z^{e}}}}.
\]
Its empirical counterpart replaces expectations by sample averages. For $g^{\mathrm{SQ}}\p{\alpha}=\p{\alpha-1}^2$, this yields, up to constants,
\[
\widehat{\mathrm{BD}}^{\mathrm{CS}}_{g^{\mathrm{SQ}}}\p{\alpha}
=
\frac{1}{n}\sum^n_{i=1}\p{\alpha\p{Z_i}^2}
-
\frac{2}{m}\sum^m_{j=1}\p{\alpha\p{Z^{e}_j}},
\]
which is exactly the LSIF objective. KuLSIF is obtained by taking $\calH$ to be an RKHS and adding RKHS-norm regularization.

\subsubsection{Riesz Regression}
\label{app:riesz_regression_sq}
The same quadratic empirical objective is used for direct estimation of the Riesz representer and for its subsequent algorithmic variants \citep{Chernozhukov2021automaticdebiased}. Table~\ref{tbl:dre_rre} includes RieszNet and RieszForest \citep{Chernozhukov2022riesznet}, RieszBoost \citep{Lee2025rieszboost}, and kernel ridge Riesz regression (KRRR) \citep{Singh2024kernelridge}, because they differ primarily in the choice of $\calH$ and optimization strategy.

\subsubsection{Duality: Moment Matching and Covariate Shift}
\label{app:dual_sq}
Under a linear link, so that $u=\partial g\circ\alpha$ is linear in parameters, quadratic ERM admits a quadratic-program dual whose constraints are approximate moment balance. SQ-Riesz with a linear link is therefore the same optimization problem as stable balancing weights under the stated convexity and duality conditions. Approximate residual balancing uses related moment constraints with a different criterion and regularization scheme. The dual row of Table~\ref{tbl:dre_rre} records this convex-dual and KKT relation.

\subsubsection{Model Specification}
The choice of $\calH$ yields different representer estimators.
\begin{itemize}
\item An RKHS gives a kernel ridge estimator of the representer. KuLSIF treats density-ratio estimation \citep{Kanamori2012statisticalanalysis}, while KRRR treats a general Riesz representer \citep{Singh2024kernelridge}.
\item Neural networks approximate composite smooth functions \citep{SchmidtHieber2020nonparametric} and piecewise nonsmooth functions \citep{Suzuki2019adaptivitydeep,Imaizumi2019deepneural}. \citet{Kato2021nonnegativebregman} and \citet{Zheng2022anerror} study neural-network density-ratio estimators.
\item Trees and forests represent the fitted function by indicator functions for their terminal nodes \citep{Chernozhukov2022riesznet}.
\item Boosting constructs the fitted function by stagewise updates \citep{Lee2025rieszboost}.
\end{itemize}

\paragraph{RKHS.}
For RKHS models, KuLSIF, KRRR, and kernel balancing use closely related quadratic objectives or moment equations:
\begin{align*}
\text{KuLSIF \citep{Kanamori2012statisticalanalysis}}
&\ \approx\ \text{KRRR \citep{Singh2024kernelridge}}\\
&\ \approx\ \text{kernel balancing \citep{Wong2017kernelbased}}.
\end{align*}
The approximation symbols indicate a common squared-loss Bregman divergence with different target functionals, or the primal and dual forms of the same moment equations.

KuLSIF is the RKHS version of LSIF, squared-loss density-ratio fitting with an RKHS hypothesis class and RKHS regularization. KRRR is kernel ridge estimation of the Riesz representer in an RKHS. Since density-ratio estimation under covariate shift is itself a Riesz representer estimation problem, KRRR reduces to the KuLSIF-style estimator when the functional $m$ corresponds to the covariate-shift parameter and the Riesz representer is the density ratio. Conversely, KuLSIF can be read as KRRR applied to the density-ratio representer.

Kernel balancing methods impose weighted equality of functional moments over an RKHS $\calH$:
\[
\frac{1}{n}\sum^n_{i=1}\p{\alpha_i\gamma\p{Z_i}}
=
\frac{1}{n_1}\sum^n_{i=1}\bbI\sqb{D_i=1}\gamma\p{Z_i}
\quad\text{for all }\gamma\in\calH.
\]
The reproducing property reduces these moment equations to finite-dimensional calculations with the kernel matrix.

\paragraph{Nearest neighbor matching and causal trees and forests.}
Table~\ref{tbl:dre_rre} also lists nearest neighbor matching and causal trees and forests in the squared-loss row. The point is not that these methods are quadratic programs in their usual presentation, but that they can be interpreted as Riesz representer estimation with a particular choice of hypothesis class. For more details, see Appendix~\ref{sec:nnmaching}.

A nearest neighbor matching estimator implicitly restricts attention to a particular class of weighting rules, namely weights that are piecewise constant on nearest-neighbor cells and matched sets. Equivalently, it specifies a low-complexity $\calH$, spanned by indicator-like basis functions, for the representer. In this sense, matching can be interpreted as estimating the Riesz representer within a constrained function class. In the density-ratio literature, matching-based estimators can be connected to LSIF, and hence back to Riesz regression.

Regression trees and random forests provide adaptive representations based on leaf indicators and forest weights. These representations are used for both regression functions and weighting functions. Once one views the representer as living in the span of a learned set of basis functions, a squared-loss Riesz objective becomes a natural ERM criterion over that set of basis functions. Trees, sieves, and RKHS models therefore differ in the functions used to represent and estimate the Riesz representer.

\subsection{UKL-Riesz}
\label{app:ukl_family}
Calibrated estimation and the tailored loss with parameters $\alpha = \beta = -1$ share the same objective and can be viewed as special cases of UKL-Riesz \citep{Tan2019regularizedcalbrated,Zhao2019covariatebalancing}. The dual form corresponds to entropy balancing weights \citep{Hainmueller2012entropybalancing}, while KLIEP in density-ratio estimation also belongs to this class. 

\subsection{BKL-Riesz}
\label{app:bkl_family}
Binary KL (BKL) losses are connected to logistic probability models. For a positive density ratio $r$, logistic classification corresponds to the convex generator $g_{\mathrm{log}}(r)=r\log r-(1+r)\log(1+r)$. The signed shifted BKL generator in Section~\ref{sec:generalizedrieszregression} is related to $g_{\mathrm{log}}$ by an affine change of coordinates on the positive component of the domain. Source--target classification is therefore related to the BKL row of Table~\ref{tbl:dre_rre}, but it is not direct BKL-Riesz fitting of the original density-ratio representer.

For causal parameters, the same row records two distinct relationships. Direct BKL-Riesz regression uses the BKL generator with a representer link that is compatible with the target functional. Logistic propensity-score estimation is a separate plug-in method related through binary classification. Under a sigmoid propensity specification, the Bernoulli likelihood is not the ATE BKL-Riesz objective and does not automatically give the ATE balance equations.

\paragraph{Relation to binary probability losses.}
\label{sec:classification_discussion}
Results on proper binary losses study probability calibration and transformations between a class probability and a density ratio \citep{Menon2016linkinglosses,Zellinger2025binarylosses}. Proposition~\ref{prop:compatible_characterization} asks when differentiation of a Riesz objective gives the empirical Riesz equations for the target functional. The questions coincide in some binary density-ratio models but not for general signed representers.

\subsection{BP-Riesz}
\label{app:bp_family}
The Basu power (BP) divergence gives a one-parameter family between the squared and UKL generators. The parameter $\omega$ changes the curvature of the loss. With a compatible power link, the dual coordinate remains linear in the coefficients, so BP-Riesz retains automatic regressor balancing.

\subsection{PU-Riesz}
\label{app:pu_row}

The PU row in Table~\ref{tbl:dre_rre} records that the unbiased case--control positive--unlabeled objective has a Bregman form on $\alpha\in(0,1)$. It applies when the target representer has the probability-ratio form implied by that sampling design.

\subsection{General Bregman Formulation}
\label{app:general_bregman}
The last row states the general identity shared by density-ratio matching under a Bregman divergence and generalized Riesz regression after the target functional is specified. Deep direct density-ratio estimation modifies the empirical density-ratio objective by a nonnegative correction \citep{Kato2021nonnegativebregman}.

\subsection{\texttt{genriesz}}
\label{app:genriesz}
The \texttt{genriesz} package implements the objectives studied here. Related software for causal and structural parameter estimation includes \texttt{EconML} and \texttt{DoubleML} \citep{Battocchi2019econmla,Bach2022doublemlpython,Bach2024doublemlr}. Users specify the functional, the regressor functions, the generator, and the representer model, while the built-in generators use exact derivatives and inverse derivatives. A fit is reported as invalid when the model leaves the domain or the optimization conditions are not met, and the implementation retains the specified loss. The package also implements ARW, RW, RA, and TMLE estimators for the applications in Section~\ref{sec:application}.\footnote{Code: \url{https://github.com/MasaKat0/genriesz}. Documentation: \url{https://genriesz.readthedocs.io/en/latest/}.}

\clearpage

\section{Sieve Riesz Representers}
\label{appdx:sieve_relation}
The finite-dimensional Riesz representer associated with a linear space $\calH_p$ is the element $\alpha_{0,p}\in\calH_p$ such that $\bbE[m(W,h)]=\bbE[\alpha_{0,p}(X)h(X)]$ for every $h\in\calH_p$. The sample analogue replaces both expectations by empirical averages. When the dual coordinate of generalized Riesz regression is linear in the basis functions spanning $\calH_p$, the first-order conditions are these sample equations, with equality in the unregularized case and a regularization-dependent discrepancy otherwise. This observation connects the objective in Section~\ref{sec:generalizedrieszregression} to the sieve constructions in \citet{Chen2014sieveinference}, \citet{Chen2015sievesemiparametric}, and \citet{Chen2015sievewald}.

\subsection{Regressor Balancing as Moment Matching Under Bregman Projection}
\label{sec:lambda0_guarantees}

This appendix connects three perspectives on representer fitting:
(i) the sieve Riesz equations in a Hilbert space,
(ii) Bregman projection geometry, and
(iii) the KKT balancing conditions in generalized Riesz regression.

One can estimate the Riesz representer by directly solving the equations implied by the Riesz representation theorem.
This approach is a variant of moment matching, often called the \emph{sieve Riesz representer}, and it has been used
in the semiparametric and sieve inference literature, see, e.g., \citet{Chen2015sievesemiparametric} and \citet{Chen2015sievewald}.
This viewpoint is closely connected to exact balancing. Understanding this approach clarifies how the choice of the loss function
$g$ in the Bregman divergence determines the geometry of the representer fit.

For the linear functional $\gamma\mapsto \bbE\sqb{m(W,\gamma)}$, the Riesz representer
$\alpha_0\in\calH$ satisfies $\bbE\sqb{m(W,\gamma)}=\langle \alpha_0,\gamma\rangle$ for all $\gamma\in\calH$.
On a finite-dimensional sieve space $H_p\coloneqq\mathrm{span}\{\phi_1,\ldots,\phi_p\}$, the sieve Riesz representer
$\alpha_p\in H_p$ is characterized by linear equations
$\langle \alpha_p,\phi_j\rangle=\bbE\sqb{m(W,\phi_j)}$.
Our KKT balancing equations are the empirical counterpart, and the Bregman geometry viewpoint is developed in
Appendix~\ref{appdx:kkt_riesz_linear_equation}.

\paragraph{Riesz representer as a linear equation in a Hilbert space.}
Let $\calH\coloneqq L_2(P_X)$ with inner product $\langle f,g\rangle \coloneqq \bbE\sqb{f(X)g(X)}$.
For the linear map $\gamma\mapsto \bbE\sqb{m(W,\gamma)}$ (Section~\ref{sec:setup}), the Riesz representation theorem yields $\alpha_0\in\calH$
such that
\begin{align}
\bbE\sqb{m(W,\gamma)} = \langle \alpha_0,\gamma\rangle
\qquad \forall \gamma\in\calH.
\label{eq:riesz_def_appdx}
\end{align}
Restricting to a finite-dimensional sieve space $\calH_p\coloneqq \mathrm{span}\{\phi_1,\ldots,\phi_p\}$, the sieve Riesz representer
$\alpha_p\in\calH_p$ is characterized by
\begin{align}
\langle \alpha_p,\phi_j\rangle = \bbE\sqb{m(W,\phi_j)}
\qquad j=1,\ldots,p.
\label{eq:riesz_sieve_equations}
\end{align}
Writing $\alpha_p(x)=\bmphi(x)^\top\bmbeta$ with $\bmphi\coloneqq (\phi_1,\ldots,\phi_p)^\top$,
\eqref{eq:riesz_sieve_equations} becomes the linear system
\begin{align}
\underbrace{\bbE\sqb{\bmphi(X)\bmphi(X)^\top}}_{=:G}\,\bmbeta
=
\underbrace{\bbE\sqb{m(W,\bmphi)}}_{=:b}.
\label{eq:normal_equation_riesz}
\end{align}

\paragraph{First-order conditions in the primal and dual coordinates.}
Recall the pointwise Bregman divergence
\[
\mathrm{BD}^\dagger_g\p{\alpha_0(x)\mid \alpha(x)}
\coloneqq g(\alpha_0(x)) - g(\alpha(x)) - \partial g(\alpha(x))\big(\alpha_0(x)-\alpha(x)\big),
\]
and let $\alpha^*$ minimize its expectation over a convex model class $\calH$. For any feasible direction $h$, differentiation with respect to the candidate representer gives
\[
\left.\frac{\rmd}{\rmd t}\bbE\sqb{\mathrm{BD}^\dagger_g\p{\alpha_0(X)\mid \alpha^*(X)+t h(X)}}\right|_{t=0}
=
\Big\langle g''(\alpha^*)\p{\alpha^*-\alpha_0},h\Big\rangle.
\]
Consequently, the first-order condition for the reverse Bregman divergence is curvature weighted. In particular,
\[
\Big\langle g''(\alpha^*)\p{\alpha^*-\alpha_0},\alpha-\alpha^*\Big\rangle\geq0
\qquad\text{for every }\alpha\in\calH,
\]
with equality along two-sided feasible directions.

The dual coordinate gives a simpler form. For each $x$, let $\calA_x$ be the open interval allowed by the representer model at $x$, and define
\[
g_x^*(v)\coloneqq\sup_{a\in\calA_x}\{av-g(a)\}.
\]
Set $u(x)=\partial g(\alpha(x))$ and $\calU=\{\partial g\circ\alpha:\alpha\in\calH\}$. The Fenchel--Young identity gives $g_x^*(u(x))=\alpha(x)u(x)-g(\alpha(x))$, and the population objective can therefore be written, up to an additive constant, as
\[
\bbE\sqb{g_X^*(u(X))-m(W,u)}
=
\bbE\sqb{g_X^*(u(X))-\alpha_0(X)u(X)}.
\]
If $\calU$ is convex and $u^*=\partial g\circ\alpha^*$ minimizes this objective, then it holds that
\begin{align}
\Big\langle \alpha^*-\alpha_0,u-u^*\Big\rangle\geq0
\qquad \text{for every }u\in\calU.
\label{eq:bregman_orthogonality}
\end{align}
Equality holds along two-sided feasible directions. Thus, the primal condition contains the curvature $g''$, whereas the dual-coordinate condition compares the representers directly.

\paragraph{Finite-dimensional dual models and KKT.}
Consider a model specified in dual coordinates by
\begin{align}
u_\bmbeta(x)=\bmphi(x)^\top\bmbeta,
\qquad
\alpha_\bmbeta(x)=\Bigp{\partial g\big|_{\calA_x}}^{-1}\p{u_\bmbeta(x)}.
\end{align}
For the penalized empirical objective
\[
\widehat\bmbeta
\in \arg\min_{\bmbeta\in\bbR^p}
\Biggcb{\frac1n\sum_{i=1}^n g_{X_i}^*\p{u_\bmbeta(X_i)}
-\frac1n\sum_{i=1}^n m(W_i,u_\bmbeta)
+\frac{\lambda}{q}\|\bmbeta\|_q^q},
\]
the KKT conditions yield
\begin{align}
\bbE_n\Big[\widehat\alpha(X)\phi_j(X)-m(W,\phi_j)\Big]
\in -\lambda\,\partial\Bigp{\frac1q|\beta_j|^q},
\qquad j=1,\ldots,p,
\label{eq:kkt_general_appdx}
\end{align}
where $\widehat\alpha=\alpha_{\widehat\bmbeta}$. When $\lambda=0$ and no parameter constraint is active, Equation~\eqref{eq:kkt_general_appdx} reduces to the empirical sieve Riesz equations, the finite-sample analogue of Equation~\eqref{eq:riesz_sieve_equations}.

\subsection{Special Case: Linear Riesz Models With Linear Regression Models}

A particularly instructive regime is when both the representer and regression are modeled in the same linear span.
This is the regime behind the augmented balancing equivalences emphasized by \citet{BrunsSmith2025augmentedbalancing}.

Let $\Phi\in\bbR^{n\times p}$ be the design matrix with $\Phi_{ij}=\phi_j(X_i)$, and let $b\in\bbR^p$ be the target moment vector with
\[
b_j=\frac{1}{n}\sum^n_{i=1}m\p{W_i,\phi_j}.
\]
Exact balance in the linear representer model $\widehat\alpha(X)=\bmphi(X)^\top\widehat\bmbeta$ means $\frac{1}{n}\Phi^\top\Phi\,\widehat\bmbeta=b$ (when solvable).
Let $\widehat\bmrho^{\mathrm{OLS}}=\p{\Phi^\top\Phi}^\dagger\Phi^\top Y$ and $\widehat\gamma^{\mathrm{OLS}}(x)=\bmphi(x)^\top\widehat\bmrho^{\mathrm{OLS}}$.

\begin{proposition}[Exact-balance RW equals an OLS plug-in functional]
\label{prop:rw_equals_ols}
If $\widehat\alpha(X)=\bmphi(X)^\top\widehat\bmbeta$ and $\frac{1}{n}\Phi^\top\Phi\,\widehat\bmbeta=b$, then it holds that
\[
\widehat\theta^{\mathrm{RW}}
=
\frac{1}{n}\sum^n_{i=1} \widehat\alpha(X_i)Y_i
=
\Bigp{\frac{1}{n}\sum^n_{i=1} m\p{W_i,\bmphi}}^\top \widehat\bmrho^{\mathrm{OLS}}.
\]
\end{proposition}

\paragraph{Consequences for the estimator.}
Proposition~\ref{prop:rw_equals_ols} shows that, in the linear--linear exact-balance regime, weighting and regression solve the same normal equations. Consequently, some balancing-weight estimators collapse to regression estimators, while particular regularization choices can yield a single effective regression estimator. \citet{BrunsSmith2025augmentedbalancing} give general augmented-balancing identities together with ridge and kernel-ridge special cases, and \citet{Robins2007commentperformance} give the related double-robustness argument for ordinary least squares. Once orthogonality holds on the entire working regression space, the RW estimator is already the orthogonal estimator for that space.

\paragraph{Augmented balancing weights collapse to regression estimators.}
\citet{BrunsSmith2025augmentedbalancing} show that when both models are linear, the augmented estimator is equivalent to a single linear regression whose coefficients are affine combinations of the original regression coefficients and unpenalized OLS coefficients.
Under certain regularization choices, the augmented estimator collapses to OLS itself.
This makes explicit that aggressive balancing can implicitly undo outcome regularization, which can reduce functional bias at the cost of higher variance.

\paragraph{Kernel ridge special case.}
In RKHS settings, this algebraic collapse yields an interpretable undersmoothing rule.
If both the outcome regression and the representer are fit by ridge in a common RKHS, the augmented estimator can be written as a single undersmoothed ridge regression with an effective regularization level, rather than as two separate nuisance estimators.

\paragraph{Special case: linear Riesz and linear regression under inexact balance.}
When both $\alpha$ and $\gamma$ are fitted in the same linear span with $\ell_2$ penalties, \citet{BrunsSmith2025augmentedbalancing} show that the resulting ARW estimator can be written as a single ridge-type regression estimator with an effective regularization parameter.
This makes explicit that augmentation can be interpreted as a data-dependent undersmoothing rule: regularization choices that are prediction-optimal for $\gamma$ can be too aggressive for inference on $\theta_0$, and the representer fit provides a principled correction.

\subsection{Neyman Error With Separate Samples}
When nuisance parameters are estimated on observations separate from those used to evaluate the score, the balancing identity holds on the estimation observations but need not hold on the evaluation observations. Consequently, exact balance does not transfer as a finite-sample identity, and the drift term in \eqref{eq:neyman_error_decomp} must instead be controlled by the stated convergence rate conditions. This issue arises under a single sample split as well as under cross-fitting.

\paragraph{Empirical-process conditions.}
If the nuisance parameters and score are evaluated on the same observations, an appropriate Donsker condition can control the empirical-process term. Alternatively, the nuisance parameters may be estimated via sample splitting or cross-fitting. Under the latter construction, the balance equations remain statements about the observations used to estimate the representer, whereas the score is evaluated elsewhere.

\paragraph{Basis functions depending only on $Z$.}
In ATE estimation with standard propensity-score modeling, it is common to use basis functions depending only on $Z$, that is, $\bmphi\colon \calZ \to \bbR^p$.
Under this choice, for ATE-type $m$ one typically has $m\p{W,\phi_j}=0$, so automatic covariate balancing controls only the weighted moments $\frac{1}{n}\sum^n_{i=1}\widehat\alpha(X_i)\phi_j(Z_i)$.
Automatic Neyman orthogonalization of RW without regression adjustment is then strongest only for outcome-model components that lie in the same $Z$-only working span.
If treatment effects are heterogeneous, the relevant regression components generally depend on $X=\p{D,Z}$, so one should use a richer set of basis functions, such as treatment-specific features $\p{D\phi(Z),(1-D)\phi(Z)}$ or other $X$-dependent bases, even though the Riesz representer corresponds to inverse propensity weights.

\subsection{Estimation Equation and TMLE Approaches}
\label{appdx:tmle_details}
TMLE is another approach in debiased machine learning \citep{vanderLaan2006targetedmaximum}.
TMLE adds a perturbation to an initial estimate of $\gamma_0$ and targets an empirical score equation.

To make the contrast concrete, consider the sample Neyman error written as
\[
\text{NeymanError}
=
\frac{1}{n}\sum^n_{i=1}\Bigp{
\annot{\widehat{\alpha}(X_i)\bigp{Y_i-\widehat{\gamma}(X_i)}}{= $(\star)$}
+
m\p{W_i,\widehat{\gamma}}
-
m\p{W_i,\gamma_0}
}.
\]
TMLE updates $\widehat{\gamma}$ so that the empirical mean of $(\star)$ becomes zero, that is, it enforces
\[
\frac{1}{n}\sum^n_{i=1}\widehat{\alpha}(X_i)\bigp{Y_i-\widehat{\gamma}^{(1)}(X_i)}=0.
\]

In contrast, generalized Riesz regression targets the representer, and it controls the empirical drift term induced by imbalance.
Equivalently, using $\widehat{\alpha}(X_i)Y_i=\widehat{\alpha}(X_i)\bigp{Y_i-\widehat{\gamma}(X_i)}+\widehat{\alpha}(X_i)\widehat{\gamma}(X_i)$, we can also rewrite NeymanError as
\[
\text{NeymanError}
=
\frac{1}{n}\sum^n_{i=1}\Bigp{
\widehat{\alpha}(X_i)Y_i
+
\annot{m\p{W_i,\widehat{\gamma}}-\widehat{\alpha}(X_i)\widehat{\gamma}(X_i)}{= $(\star\star)$}
-
m\p{W_i,\gamma_0}
}.
\]
Generalized Riesz regression uses approximate empirical Riesz equations on a prescribed working class to reduce the sample average of $(\star\star)$.

\paragraph{TMLE and generalized Riesz regression target different nuisances.}
\begin{itemize}
    \item TMLE puts the difficulty of efficient estimation of $\theta_0$ into the estimation of $\gamma_0$ by targeting the regression update so that the empirical score equation holds given $\widehat{\alpha}$.
    \item Generalized Riesz regression puts the difficulty of efficient estimation of $\theta_0$ into the estimation of $\alpha_0$ by targeting $\widehat{\alpha}$ so that the empirical Riesz equations hold, approximately, for a working regression class given $\widehat{\gamma}$.
\end{itemize}
Because the two procedures target different nuisance parameters, one may estimate $\widehat{\alpha}$ by generalized Riesz regression and then apply a TMLE-type fluctuation to $\widehat{\gamma}$.

\begin{remark}[TMLE]
Let $\widehat{\gamma}^{(0)}$ be an initial estimate of $\gamma_0$.
Given $\widehat{\gamma}^{(0)}$ and a representer estimate $\widehat{\alpha}$, a simple linear TMLE update sets
\[
\widehat{\gamma}^{(1)}(x)
\coloneqq
\widehat{\gamma}^{(0)}(x)
+
\widehat{\epsilon}\widehat{\alpha}(x),
\qquad
\widehat{\epsilon}
\coloneqq
\frac{\sum^n_{i=1}\widehat{\alpha}(X_i)\bigp{Y_i-\widehat{\gamma}^{(0)}(X_i)}}
     {\sum^n_{i=1}\widehat{\alpha}(X_i)^2}.
\]
This choice of $\widehat{\epsilon}$ solves
\[
\sum^n_{i=1}\widehat{\alpha}(X_i)\Bigp{Y_i-\p{\widehat{\gamma}^{(0)}(X_i)+\epsilon\widehat{\alpha}(X_i)}}=0,
\]
thereby eliminating the empirical mean of the $(\star)$ term.
The final TMLE estimator is
\[
\widehat{\theta}^{\text{TMLE}}
\coloneqq
\frac{1}{n}\sum^n_{i=1}m\p{W_i,\widehat{\gamma}^{(1)}}.
\]
\end{remark}

\paragraph{Direct estimation of the Riesz representer.}
Automatic debiasing estimates $\alpha_0$ without requiring a closed-form representer. The estimator minimizes an objective whose population value is minimized at $\alpha_0$, and the model for the representer may be nonparametric.
This includes Riesz regression and its extensions to generalized regressions \citep{Chernozhukov2022automaticdebiased}, as well as multitask and targeted-regularization architectures \citep{Chernozhukov2022riesznet}.
The first-order conditions extend the finite-dimensional moment equations to the selected model class. Empirical imbalance, the tails of the fitted weights, and the optimization residual describe different aspects of the fitted representer.

\paragraph{Relation to TMLE.}
TMLE updates $\gamma$ for a given $\alpha$, whereas generalized Riesz regression estimates $\alpha$ from the empirical Riesz equations. The two procedures therefore act on different nuisance parameters. A targeted update of $\gamma$ can also be applied after estimating $\alpha$, and Equation~\eqref{eq:inexact_balance_bound} describes the regression component that remains after the representer has been fitted.

\subsection{Regression Function and Riesz Representer}
\label{sec:outcomemodeling}

The regressor functions used for balance should approximate the components of the regression function that enter the target parameter.

\paragraph{Convergence of the nuisance parameter estimators.}
Under a Donsker condition or nuisance parameter estimation via cross-fitting, the leading remainder is bounded by
\[
\|\widehat{\gamma}-\gamma_0\|_2\|\widehat{\alpha}-\alpha_0\|_2.
\]
The required convergence rate can therefore be obtained through a sufficiently accurate regression estimator, a sufficiently accurate representer estimator, or a combination of the two.

\paragraph{Bias correction by the ARW estimator.}
In modern applications, $\widehat{\gamma}$ is typically regularized (lasso, ridge, kernels, neural nets), so it can be biased for $\gamma_0$, and this bias can propagate to $\theta_0$ through $m\p{W,\widehat{\gamma}}$.
The orthogonal ARW form corrects this bias to first order, leaving a second-order product remainder controlled by \eqref{eq:orth_score_bias_identity}.
The regression and representer estimators enter the remaining error through their product.

\paragraph{Misspecification of the regressor functions.}
If one balances only low-order moments of $Z$ but uses a rich estimator for $\gamma_0(D,Z)$ with interactions and nonlinearities, then the balanced span need not contain the functions needed to approximate the regression error.
Orthogonality defects can persist even if covariate balance diagnostics look excellent.
The regressor functions should therefore include functions of $X=\p{D,Z}$ that approximate both treatment-specific regression components.

\paragraph{Balance over an RKHS.}
Kernel-based methods impose balance over an RKHS through its mean embedding. The reproducing property and the representer theorem reduce the resulting equations to finite-dimensional calculations with the kernel matrix.

\paragraph{Misspecification of the regression and representer models.}
Complementing regression-centric undersmoothing, \citet{Singh2024kernelridge} studies kernel ridge estimation of the Riesz representer and analyzes its generalization error in population $L_2$.
For an orthogonal estimator based on approximations to the regression function and the representer, consistency can still follow when one nuisance parameter is estimated sufficiently accurately and the remaining stochastic terms satisfy the stated convergence rate conditions.

\begin{remark}[Minimax rate and definition of common support]
\citet{Mou2023kernelbased} provide a complementary minimax viewpoint for estimating weighted linear functionals from observational data, including regimes where strict overlap fails and semiparametric efficiency bounds may be infinite.
Two aspects are especially relevant for RR-based debiasing: the functional difficulty is governed by a modulus of continuity, and for RKHS classes the lower bound can be achieved up to constants by computationally simple outcome-regression estimators that do not require knowledge of a behavioral policy.
The attainable risk in that result depends on the function class and its Riesz representer.
\end{remark}

\paragraph{Covariate functions in ATE estimation.}
In ATE applications, it is common to set $\bmphi=\bmphi(Z)$ so that balancing is interpreted as covariate balance.
If, instead, the regression function $\gamma_0(D,Z)$ contains heterogeneous components that are poorly approximated by a $Z$-only linear span, then \eqref{eq:inexact_balance_bound} does not directly control the orthogonality defect for those components.
Treatment-specific functions, such as separate basis functions by treatment arm or $D\times f(Z)$ interactions, are then needed to control the corresponding regression components.

\section{Details for the Applications}
\label{appdx:application_details}

\subsection{ATE Estimation}
\label{supp:ate_details}
In ATE estimation, the linear functional is
\[
m^{\text{ATE}}(W,\gamma)\coloneqq \gamma(1,Z)-\gamma(0,Z),
\]
and the Riesz representer is
\[
\alpha^{\text{ATE}}_0(X)=\frac{D}{e_0(Z)}-\frac{1-D}{1-e_0(Z)},
\]
where $e_0(Z) = P(D = 1\mid Z)$ is the propensity score. Let $r_0(1,Z) \coloneqq \frac{1}{e_0(Z)}$ and $r_0(0,Z) \coloneqq \frac{1}{1 - e_0(Z)}$ denote the two inverse-propensity components. They are proportional to the density ratios defined in Appendix~\ref{appdx:rieszdens}.
We estimate $\alpha^{\text{ATE}}_0$ by minimizing the empirical Bregman divergence objective $\widehat{\text{BD}}_g(\alpha)$ introduced in Section~\ref{sec:generalizedrieszregression}, with $m=m^{\text{ATE}}$, an application-specific choice of $g$, and a model class for $\alpha$.

\paragraph{SQ-Riesz regression.}
We take the squared loss,
\[
g^{\text{SQ}}(\alpha)=\alpha^2,
\]
and minimize the corresponding empirical Bregman objective.
By substituting $g^{\text{SQ}}$ into \eqref{eq:empbregman} and using $m^{\text{ATE}}(W,\gamma)=\gamma(1,Z)-\gamma(0,Z)$, we obtain, up to an additive constant that does not depend on $\alpha$,
\[
\text{BD}_{g^{\text{SQ}}}(\alpha)
=
\bbE\sqb{
\alpha(D,Z)^2
-2\bigp{\alpha(1,Z)-\alpha(0,Z)}
}.
\]
Thus, SQ-Riesz regression estimates $\alpha^{\text{ATE}}_0$ by
\[
\widehat{\alpha}
\coloneqq
\argmin_{\alpha\in\calH}
\widehat{\text{BD}}_{g^{\text{SQ}}}(\alpha)+\lambda J(\alpha),
\]
where
\[
\widehat{\text{BD}}_{g^{\text{SQ}}}(\alpha)
\coloneqq
\frac{1}{n}\sum_{i=1}^n
\p{
\alpha(D_i,Z_i)^2
-2\bigp{\alpha(1,Z_i)-\alpha(0,Z_i)}
}.
\]
This coincides with Riesz regression in \citet{Chernozhukov2021automaticdebiased} and corresponds to LSIF in density ratio estimation \citep{Kanamori2009aleastsquares}.
With appropriate choices of $\calH$, it also recovers nearest neighbor matching-based constructions, as discussed in \citet{Kato2025nearestneighbor}.

\begin{remark}[Linear link function]
    A recommended Riesz representer model is
\[
\alpha_\bmbeta(X)=\bmphi(X)^\top \bmbeta,
\]
where $\bmphi\colon\calX\to\bbR^p$ is a vector of basis functions.
Under this model, minimizing the SQ-Riesz objective yields an estimator that satisfies an automatic covariate balancing property, as discussed in Section~\ref{sec:automaticcovariatebalancing} and \citet{Zhao2019covariatebalancing}.

Concretely, letting $\widehat{\alpha}=\alpha_{\widehat{\bmbeta}}$, we estimate $\bmbeta$ by
\begin{align*}
\widehat{\bmbeta}
&\coloneqq
\argmin_{\bmbeta}
\frac{1}{n}\sum_{i=1}^n
\Biggp{\Bigp{\bmphi(X_i)^\top \bmbeta}^2 - 2\Bigp{\bmphi(1, Z_i)^\top - \bmphi(0, Z_i)^\top} \bmbeta} + \frac{\lambda}{q} \|\bmbeta\|_q^q.
\end{align*}

By duality, if $q=1$, SQ-Riesz regression is equivalent to the following covariate balancing problem:
\begin{align*}
\min_{\alpha\in\bbR^n}\ &\frac{1}{n}\sum_{i=1}^n \alpha_i^2\\
\text{s.t.}\ &\left|\frac{1}{n}\sum_{i=1}^n\Bigp{\alpha_i\phi_j(X_i)-\phi_j(1,Z_i)+\phi_j(0,Z_i)}\right|\leq\frac{\lambda}{2},
\qquad j=1,\ldots,p.
\end{align*}
The factor $1/2$ in the tolerance follows from $\partial g^{\mathrm{SQ}}(\alpha)=2\alpha$ under the stated normalization.
Here, $\widehat{w}_i$ estimates $\alpha_0(X_i)$ for treated observations and $-\alpha_0(X_i)$ for control observations:
\[\widehat{w}_i=
    \begin{cases}
    \widehat{\alpha}(1,Z_i) & \text{if } D_i=1,\\
    - \widehat{\alpha}(0,Z_i) & \text{if } D_i=0.
    \end{cases}.\]
\end{remark}

\paragraph{UKL-Riesz regression.}
Consider Riesz representer models $\alpha$ such that $\alpha(1, x) > 1$ and $\alpha(0, x) < -1$ for all $x$.
We next use the UKL divergence loss with $C=1$,
\[
g^{\text{UKL}}(\alpha)=(|\alpha|-1)\log\p{|\alpha|-1}-|\alpha|.
\]
For a representer model $\alpha$, define
\[
f_\alpha(d,z)\coloneqq\sign(\alpha(d,z))\log(|\alpha(d,z)|-1).
\]
Substitution into Equation~\eqref{eq:population_bregman} gives
\begin{align*}
\mathrm{BD}_{g^{\mathrm{UKL}}}(\alpha)
=\bbE\Bigp{\log(|\alpha(X)|-1)+|\alpha(X)|-f_\alpha(1,Z)+f_\alpha(0,Z)}.
\end{align*}
Thus, UKL-Riesz regression estimates $\alpha^{\mathrm{ATE}}_0$ by
\[
\widehat\alpha\in\argmin_{\alpha\in\calH}
\widehat{\mathrm{BD}}_{g^{\mathrm{UKL}}}(\alpha)+\lambda J(\alpha),
\]
where
\begin{align*}
\widehat{\mathrm{BD}}_{g^{\mathrm{UKL}}}(\alpha)
=\frac1n\sum_{i=1}^n\Bigp{\log(|\alpha(X_i)|-1)+|\alpha(X_i)|-f_\alpha(1,Z_i)+f_\alpha(0,Z_i)}.
\end{align*}
This coincides with the tailored loss minimization with $\alpha = \beta = -1$ in \citet{Zhao2019covariatebalancing} and corresponds to KLIEP in density ratio estimation \citep{Sugiyama2008directimportance}.

\begin{remark}[Log link function]
    A recommended Riesz representer model is
\[
\alpha_\bmbeta(X)=\1[D=1]\Bigp{1+\exp\bigp{-\bmphi(X)^\top \bmbeta}}-\1[D=0]\Bigp{1+\exp\bigp{\bmphi(X)^\top \bmbeta}},
\]
where $\bmphi\colon\calX\to\bbR^p$ is a vector of basis functions.
Under this model, minimizing the UKL-Riesz objective yields an estimator that satisfies an automatic covariate balancing property, as discussed in Section~\ref{sec:automaticcovariatebalancing} and \citet{Zhao2019covariatebalancing}.

Concretely, letting $\widehat{\alpha}=\alpha_{\widehat{\bmbeta}}$, we estimate $\bmbeta$ by
\begin{align*}
\widehat{\bmbeta}
&\coloneqq
\argmin_{\bmbeta}
\frac{1}{n}\sum_{i=1}^n
\Biggp{
\1[D_i=1]\p{ - \bmphi(1, Z_i)^\top \bmbeta + 1 + \exp\bigp{-\bmphi(1, Z_i)^\top \bmbeta}}
\\
&\ \ \ \ \ \ \ \ \ \ \ \ \ \ \ \ \ \ \ \ \ \ \ \ \ \ \ \ +
\1[D_i=0]\p{ \bmphi(0, Z_i)^\top \bmbeta + 1 + \exp\bigp{\bmphi(0, Z_i)^\top \bmbeta}}\\
&\ \ \ \ \ \ \ \ \ \ \ \ \ \ \ \ \ \ \ \ \ \ \ \ \ \ \ \ + \Bigp{\bmphi(1, Z_i)^\top - \bmphi(0, Z_i)^\top} \bmbeta} + \frac{\lambda}{q} \|\bmbeta\|_q^q.
\end{align*}

For $j=1,\ldots,p$, define
\[
\Delta_j(\bmw)\coloneqq
\frac{1}{n}\sum_{i=1}^n\Bigp{\1[D_i=1]w_i\phi_j(X_i)-\1[D_i=0]w_i\phi_j(X_i)
-\phi_j(1,Z_i)+\phi_j(0,Z_i)}.
\]
By duality, if $q=1$, UKL-Riesz regression is equivalent to
\begin{align*}
\min_{\bmw\in(1,\infty)^n}\quad &\frac{1}{n}\sum_{i=1}^n\Bigp{(w_i-1)\log(w_i-1)-w_i}\\
\text{subject to}\quad &|\Delta_j(\bmw)|\leq\lambda,
\qquad j=1,\ldots,p.
\end{align*}
Here, $\widehat{w}_i$ estimates $\alpha_0(X_i)$ for treated observations and $-\alpha_0(X_i)$ for control observations:
\[\widehat{w}_i=
    \begin{cases}
    \widehat{\alpha}(1,Z_i) & \text{if } D_i=1,\\
    - \widehat{\alpha}(0,Z_i) & \text{if } D_i=0.
    \end{cases}.\]
This model enforces the correct signs of the Riesz representer and the nonnegativity of its magnitude.
\end{remark}

\begin{remark}[Propensity score modeling]
The Riesz representer model can be written in terms of a propensity score model:
\[
\alpha_\bmbeta(X)=\1[D=1]r_\bmbeta(1,Z)-\1[D=0]r_\bmbeta(0,Z),
\]
where
\begin{align*}
r_\bmbeta(1,Z)&=\frac{1}{e_\bmbeta(Z)},
\qquad
r_\bmbeta(0,Z)=\frac{1}{1-e_\bmbeta(Z)},\\
e_\bmbeta(Z)&\coloneqq \frac{1}{1+\exp\bigp{-\bmphi(Z)^\top \bmbeta}},
\end{align*}
and $\bmphi\colon\calZ\to\bbR^p$ is a vector of basis functions.
Under this model, minimizing the UKL-type empirical Bregman objective yields an estimator that satisfies an automatic covariate balancing property, as discussed in Section~\ref{sec:automaticcovariatebalancing} and \citet{Zhao2019covariatebalancing}.

Concretely, letting $\widehat\alpha=\alpha_{\widehat\bmbeta}$, we estimate $\bmbeta$ by
\begin{align*}
\widehat\bmbeta\in\argmin_{\bmbeta}\quad
&\frac1n\sum_{i:D_i=1}\Bigp{-\log\left(\frac1{r_\bmbeta(1,Z_i)-1}\right)+r_\bmbeta(1,Z_i)}\\
&+\frac1n\sum_{i:D_i=0}\Bigp{-\log\left(\frac1{r_\bmbeta(0,Z_i)-1}\right)+r_\bmbeta(0,Z_i)}
+\frac{\lambda}{q}\|\bmbeta\|_q^q.
\end{align*}

By duality, if $q=1$, this KL divergence objective is equivalent to a covariate balancing program:
\begin{align*}
\min_{\bmw\in(1,\infty)^n}\ &\frac{1}{n}\sum_{i=1}^n\Bigp{(w_i-1)\log(w_i-1)-w_i}\\
\text{s.t.}\ &\left|\frac{1}{n}\sum_{i=1}^n\Bigp{\1[D_i=1]w_i\phi_j(Z_i)-\1[D_i=0]w_i\phi_j(Z_i)}\right|\leq\lambda,
\qquad j=1,\ldots,p.
\end{align*}
The solution $\widehat{w}_i$ estimates $\alpha_0(X_i)$ for treated observations and $-\alpha_0(X_i)$ for control observations:
\[\widehat{w}_i=
    \begin{cases}
    \widehat{r}(1,Z_i) & \text{if } D_i=1,\\
    \widehat{r}(0,Z_i) & \text{if } D_i=0.
    \end{cases},\]
and $\widehat{r}$ is an estimator of the density ratio $r_0$.
This constrained optimization has the entropy form underlying entropy balancing \citep{Hainmueller2012entropybalancing}.
In particular, when $\lambda=0$ we obtain exact balance,
\begin{align*}
    &\sum_{i=1}^n\Bigp{\1[D_i=1]\widehat{w}_i\phi_j(Z_i)-\1[D_i=0]\widehat{w}_i\phi_j(Z_i)}=0\quad \text{for}\ j=1,2,\dots,p.
\end{align*}
This specification has the practical advantage that $\phi_j(Z)$ can be chosen independently of $D$, which reduces the dimension of the model.
\end{remark}

\paragraph{BP-Riesz regression.}
BP-Riesz regression uses Basu's power divergence with $C=1$ and $\omega\in(0,\infty)$:
\[
g^{\text{BP}}(\alpha)
\coloneqq
\frac{\bigp{|\alpha|-1}^{1+\omega}-\bigp{|\alpha|-1}}{\omega}-|\alpha|.
\]
For the signed BP generator with a general shift $C$, let $\upsilon\coloneqq 1+1/\omega$ and define
\begin{align}
t_{C,\alpha}(x)
&\coloneqq |\alpha(x)|-C,\nonumber\\
u_{\omega,C,\alpha}(x)
&\coloneqq \upsilon\sign\p{\alpha(x)}\Bigp{t_{C,\alpha}(x)^\omega-1},\nonumber\\
A_{\omega,C,\alpha}(x)
&\coloneqq |\alpha(x)|t_{C,\alpha}(x)^\omega
+\frac{C}{\omega}\Bigp{t_{C,\alpha}(x)^\omega-1}.
\label{eq:signed_bp_terms}
\end{align}
Then, we have $u_{\omega,C,\alpha}=\partial g^{\text{BP}}\circ\alpha$ and
$A_{\omega,C,\alpha}=\alpha\p{\partial g^{\text{BP}}\circ\alpha}-g^{\text{BP}}\circ\alpha$.
Setting $C=1$ and using $m^{\text{ATE}}$ yields
\begin{align*}
\widehat{\mathrm{BD}}_{g^{\text{BP}}}(\alpha)
\coloneqq
\frac{1}{n}\sum_{i=1}^n
\Bigp{
A_{\omega,1,\alpha}(D_i,Z_i)
-u_{\omega,1,\alpha}(1,Z_i)
+u_{\omega,1,\alpha}(0,Z_i)
}.
\end{align*}
We then estimate $\alpha^{\text{ATE}}_0$ by
\[
\widehat{\alpha}
\coloneqq
\argmin_{\alpha\in\calH}\widehat{\mathrm{BD}}_{g^{\text{BP}}}(\alpha)+\lambda J(\alpha).
\]

\begin{remark}[Power link function]
    A convenient parametric specification that is consistent with Section~\ref{sec:automaticcovariatebalancing} models $\alpha$ through a power link function for the inverse propensity components:
\[
\alpha_\bmbeta(X)=\1[D=1]r_\bmbeta(1,Z)-\1[D=0]r_\bmbeta(0,Z),
\]
with
\[
r_\bmbeta(1,Z)
\coloneqq
1+\p{1+\frac{\bmphi(1, Z)^\top\bmbeta}{\upsilon }}^{1/\omega},
\qquad
r_\bmbeta(0,Z)
\coloneqq
1+\p{1-\frac{\bmphi(0, Z)^\top\bmbeta}{\upsilon }}^{1/\omega},
\]
on the domain where the above powers are well defined.
At $\omega=1$, the generator agrees with a sign-specific squared generator up to affine terms, whereas the limit $\omega\to0$ gives the UKL generator, as discussed in Section~\ref{sec:generalizedrieszregression}. The compatible power links retain the KKT balancing equations while changing the curvature across the family.
\end{remark}

\paragraph{BKL-Riesz regression.}
Consider Riesz representer models $\alpha$ such that $\alpha(1, x) > 1$ and $\alpha(0, x) < -1$ for all $x$.
BKL-Riesz regression uses BKL divergence with $C=1$:
\[g^{\text{BKL}}(\alpha) \coloneqq (|\alpha| - 1)\log \bigp{|\alpha| - 1} - (|\alpha| + 1)\log(|\alpha| + 1).\]
Substituting $g^{\text{BKL}}$ into \eqref{eq:empbregman} and using $m^{\text{ATE}}$ gives the empirical objective
\begin{align*}
\widehat{\text{BD}}_{g^{\text{BKL}}}(\alpha)
&\coloneqq
\frac{1}{n}\sum^n_{i=1}\Bigp{\log\p{|\alpha(X_i)|^2-1}\\
&\ \ \ \ \ \ - \Bigp{\log\p{\frac{\alpha(1, Z_i) - 1}{\alpha(1, Z_i) + 1}} + \log\p{\frac{-\alpha(0, Z_i) - 1}{-\alpha(0, Z_i) + 1}}}}.
\end{align*}
We then estimate $\alpha^{\text{ATE}}_0$ by
\[
\widehat{\alpha}
\coloneqq
\argmin_{\alpha\in\calH}\widehat{\text{BD}}_{g^{\text{BKL}}}(\alpha)+\lambda J(\alpha).
\]

\begin{remark}[Logistic propensity-score plug-in baseline]
Let
\[
e_\bmbeta(Z)\coloneqq \frac{1}{1+\exp\bigp{-\bmphi(Z)^\top \bmbeta}},
\qquad
\alpha_\bmbeta(X)\coloneqq \frac{D}{e_\bmbeta(Z)}-\frac{1-D}{1-e_\bmbeta(Z)}.
\]
The direct BKL-Riesz objective does not reduce to the Bernoulli likelihood under this sigmoid propensity model. Writing $e=e_\bmbeta(Z)$, its observationwise loss is
\[
\ell^{\mathrm{BKL}}(D,e)
=
\begin{cases}
2\log(1+e)+\log(2-e)-3\log e, & D=1,\\
\log(1+e)+2\log(2-e)-3\log(1-e), & D=0.
\end{cases}
\]
This expression is not a positive affine transformation of the Bernoulli negative log-likelihood.

As a separate comparison method, the logistic propensity-score plug-in estimates
\[
\widehat{\bmbeta}
\coloneqq
\argmin_{\bmbeta}
-\frac{1}{n}\sum_{i=1}^n
\Bigp{D_i\log e_\bmbeta(Z_i)+(1-D_i)\log\bigp{1-e_\bmbeta(Z_i)}}
+\lambda\|\bmbeta\|_2^2,
\]
and sets
\[
\widehat{\alpha}(X)\coloneqq \frac{D}{e_{\widehat{\bmbeta}}(Z)}-\frac{1-D}{1-e_{\widehat{\bmbeta}}(Z)}.
\]
This plug-in method is related to probabilistic classification for density-ratio estimation \citep{Qin1998inferencesfor,Cheng2004semiparametricdensity}, but it is not the direct ATE BKL-Riesz objective.
\end{remark}

\subsection{AME Estimation}
\label{appdx:ame_estimation}
We consider the AME setup described in Section~\ref{sec:setup}. Let $X=(D,Z)$, where $D$ is a scalar continuous regressor and $Z$ is a vector of covariates. The target parameter is
\[
\theta^{\text{AME}}_0 \coloneqq \BigExp{\partial_d \gamma_0(D,Z)},
\qquad 
m^{\text{AME}}(W,\gamma)\coloneqq \partial_d \gamma(D,Z).
\]
Assume that $X$ admits a density $f_0$ that is continuously differentiable and that an integration by parts argument is valid, for example, $\gamma(x)f_0(x)$ vanishes on the boundary of the support in the $d$ direction. Then, integration by parts gives
\[
\BigExp{\partial_d \gamma(X)}
=
\BigExp{\alpha^{\text{AME}}_0(X)\gamma(X)},
\qquad
\alpha^{\text{AME}}_0(X)=-\partial_d \log f_0(X),
\]
so the AME Riesz representer is the negative score of the marginal density of $X$ with respect to $d$. Since $\partial_d \log f_0(D,Z)=\partial_d \log f_0(D\mid Z)$, we can equivalently view $\alpha^{\text{AME}}_0$ as the negative score of the conditional density of $D$ given $Z$.

To estimate $\alpha^{\text{AME}}_0$, we apply generalized Riesz regression with $m=m^{\text{AME}}$. The population objective in Section~\ref{sec:generalizedrieszregression} becomes
\[
\text{BD}^{\text{AME}}_g(\alpha)
\coloneqq
\BigExp{
- g\bigp{\alpha(X)}
+ \partial g\bigp{\alpha(X)}\alpha(X)
- \partial_d\Bigp{\partial g\bigp{\alpha(X)}}},
\]
and we minimize its empirical analogue over a differentiable model class $\calH$ (so that $\partial_d \alpha(X)$ and $\partial_d\{\partial g(\alpha(X))\}$ are well defined), possibly with regularization.

\paragraph{SQ-Riesz regression.}
Let $g^{\text{SQ}}(\alpha)=(\alpha-C)^2$ for an arbitrary constant $C\in\bbR$, so that $\partial g^{\text{SQ}}(\alpha)=2(\alpha-C)$. Substituting into $\text{BD}^{\text{AME}}_g$ yields
\begin{align*}
\text{BD}^{\text{AME}}_{g^{\text{SQ}}}(\alpha)
&=
\BigExp{
-\bigp{\alpha(X)-C}^2
+2\bigp{\alpha(X)-C}\alpha(X)
-\partial_d\Bigp{2\bigp{\alpha(X)-C}}}\\
&=
\BigExp{\alpha(X)^2-2\partial_d \alpha(X)}+\text{const},
\end{align*}
where the constant does not depend on $\alpha$. Hence, SQ-Riesz regression targets $\alpha^{\text{AME}}_0$ in $L_2$. This objective is also a score matching style criterion for estimating the score, written here in terms of the negative score $\alpha^{\text{AME}}_0$. The empirical estimator is
\begin{align*}
\widehat\alpha\in\argmin_{\alpha\in\calH}
\left\{
\begin{gathered}
\frac1n\sum_{i=1}^n\Bigp{\alpha(X_i)^2-2\partial_d\alpha(X_i)}\\
+\lambda J(\alpha)
\end{gathered}
\right\}.
\end{align*}
\citet{Chernozhukov2021automaticdebiased} use the same objective for Riesz regression in AME estimation.

\paragraph{UKL-Riesz regression.}
To obtain a KL motivated loss that allows signed $\alpha$, we use the signed KL type convex function
\[
g^{\text{UKL}}(\alpha)=|\alpha|\log|\alpha|-|\alpha|,
\qquad
\partial g^{\text{UKL}}(\alpha)=\sign(\alpha)\log|\alpha|,
\]
on a domain that excludes $\alpha=0$. Substitution into $\mathrm{BD}^{\mathrm{AME}}_g$ gives
\begin{align*}
\mathrm{BD}^{\mathrm{AME}}_{g^{\mathrm{UKL}}}(\alpha)
=\bbE\Bigp{|\alpha(X)|-\partial_d\{\sign(\alpha(X))\log|\alpha(X)|\}}+\mathrm{const}.
\end{align*}
Accordingly, we estimate $\alpha^{\mathrm{AME}}_0$ by minimizing
\[
\frac1n\sum_{i=1}^n\Bigp{|\alpha(X_i)|-\partial_d\{\sign(\alpha(X_i))\log|\alpha(X_i)|\}}+\lambda J(\alpha)
\]
over $\calH$. The shifted UKL loss in Section~\ref{sec:generalizedrieszregression} avoids the singularity at zero. Its use still requires a model that remains on a sign-specific part of the domain.

\paragraph{BP-Riesz regression.}
BP-Riesz regression interpolates between squared distance and UKL divergence. We present the unshifted form with $C=0$:
\begin{align*}
g^{\mathrm{BP}}(\alpha)
&\coloneqq\frac{|\alpha|^{1+\gamma}-|\alpha|}{\gamma}-|\alpha|,\\
\partial g^{\mathrm{BP}}(\alpha)
&=\left(1+\frac1\gamma\right)\sign(\alpha)(|\alpha|^\gamma-1),
\qquad \gamma\in(0,\infty).
\end{align*}
Let $k\coloneqq 1+1/\gamma$. Then, $\text{BD}^{\text{AME}}_g$ simplifies to
\[
\text{BD}^{\text{AME}}_{g^{\text{BP}}}(\alpha)
=
\BigExp{
|\alpha(X)|^{1+\gamma}
-
\partial_d\Bigp{k\sign\bigp{\alpha(X)}\Bigp{|\alpha(X)|^\gamma-1}}}
+\text{const}.
\]
We estimate $\alpha^{\text{AME}}_0$ by minimizing the empirical objective:
\[
\widehat{\alpha}
\coloneqq
\argmin_{\alpha\in\calH}
\frac{1}{n}\sum_{i=1}^n
\Bigp{
|\alpha(X_i)|^{1+\gamma}
-
\partial_d\Bigp{k\sign\bigp{\alpha(X_i)}\Bigp{|\alpha(X_i)|^\gamma-1}}}
+\lambda J(\alpha).
\]
As in Section~\ref{sec:generalizedrieszregression}, $\gamma=1$ recovers the squared loss behavior (up to scaling), while $\gamma\to 0$ approaches a KL type criterion through the identity $\lim_{\gamma\to 0}(|\alpha|^\gamma-1)/\gamma=\log|\alpha|$.

\paragraph{BKL-Riesz regression.}
Finally, we can use the BKL loss from Section~\ref{sec:generalizedrieszregression} to obtain a logistic motivated criterion. Let $C>0$ and define
\begin{align*}
g^{\mathrm{BKL}}(\alpha)
&\coloneqq(|\alpha|-C)\log(|\alpha|-C)
-(|\alpha|+C)\log(|\alpha|+C),\\
\partial g^{\mathrm{BKL}}(\alpha)
&=\sign(\alpha)\log\left(\frac{|\alpha|-C}{|\alpha|+C}\right).
\end{align*}
Then, the AME objective is
\begin{align*}
\mathrm{BD}^{\mathrm{AME}}_{g^{\mathrm{BKL}}}(\alpha)
={}&\bbE\Bigp{C\log(|\alpha(X)|^2-C^2)}\\
&-\bbE\left[\partial_d\left\{\sign(\alpha(X))
\log\left(\frac{|\alpha(X)|-C}{|\alpha(X)|+C}\right)\right\}\right].
\end{align*}
The estimator minimizes its empirical counterpart over $\calH$ with regularization. As in the ATE case, this loss is naturally paired with a logistic style link for the magnitude of $\alpha$, while sign changes can be handled by fitting the construction separately on the positive and negative parts of the domain as described in Section~\ref{sec:automaticcovariatebalancing}.

Once we obtain $\widehat{\alpha}^{\text{AME}}$ and an outcome regression estimator $\widehat{\gamma}$, we plug them into the Neyman orthogonal score in Section~\ref{sec:setup} to form an estimator of $\theta_0^{\text{AME}}$.

\subsection{Covariate Shift Adaptation (Density Ratio Estimation)}
\label{supp:covariate_shift_details}
We consider the covariate shift setting in Section~\ref{sec:setup}. Let $X$ denote a source covariate and $\widetilde{X}$ a target covariate. The labeled sample $\{(X_i,Y_i)\}^n_{i=1}$ is drawn from the source law, and the independent unlabeled sample $\{\widetilde{X}_j\}^m_{j=1}$ is drawn from the target law. Let $p_0(x)$ and $p_1(x)$ denote the corresponding densities. We assume that $p_0(x), p_1(x) > 0$ for all $x\in\calX$. 
The Riesz representer for covariate shift adaptation is the density ratio
\[
\alpha^{\text{CS}}_0(X)=r_0(X)\coloneqq \frac{p_1(X)}{p_0(X)}.
\]
We estimate $r_0$ directly by density ratio fitting under a Bregman divergence, avoiding separate density estimation for $p_0(X)$ and $p_1(X)$.

Let $g\colon \bbR_+\to\bbR$ be differentiable and strictly convex.
The Bregman divergence between $r_0$ and a candidate ratio model $\alpha$ is
\[
\text{BD}^\dagger_g\bigp{r_0\mid \alpha}
\coloneqq
\bbE_X\Bigp{
g\bigp{r_0(X)}-g\bigp{\alpha(X)}-\partial g\bigp{\alpha(X)}\bigp{r_0(X)-\alpha(X)}
}.
\]
Dropping the constant $\bbE_X\bigsqb{g(r_0(X))}$ and using the identity $\bbE_X\bigsqb{r_0(X)h(X)}=\bbE_{\widetilde{X}}\bigsqb{h(X)}$, we obtain the equivalent population objective
\begin{align*}
\text{BD}^{\text{CS}}_g(\alpha)
&\coloneqq
\bbE_X\bigsqb{\partial g(\alpha(X))\alpha(X)-g(\alpha(X))}
-
\bbE_{\widetilde{X}}\bigsqb{\partial g(\alpha(X))}.
\end{align*}
Given samples $\{X_i\}_{i\in\calI_S}$ and $\{\widetilde{X}_j\}_{j\in\calI_T}$, the empirical objective is
\begin{align}
\label{eq:cs_bd_empirical}
\widehat{\text{BD}}^{\text{CS}}_g(\alpha)
&\coloneqq
\frac{1}{n}\sum^n_{i=1}\Bigp{\partial g\bigp{\alpha(X_i)}\alpha(X_i)-g\bigp{\alpha(X_i)}}
-
\frac{1}{m}\sum^m_{j=1}\partial g\bigp{\alpha(\widetilde{X}_j)}.
\end{align}
We estimate the density ratio by
\[
\widehat{\alpha}
\coloneqq
\argmin_{\alpha\in\calH}
\widehat{\text{BD}}^{\text{CS}}_g(\alpha)+\lambda J(\alpha),
\]
where $\calH$ is a model class and $J$ is a regularizer.
A convenient way to enforce $\alpha(x)\ge 0$ is to use a link specification such as $\alpha(x)=\exp\bigp{f(x)}$ with a flexible regression model $f$.

\paragraph{SQ-Riesz regression.}
For the squared loss, take
\[
g^{\text{SQ}}(\alpha)=(\alpha-1)^2,
\qquad
\partial g^{\text{SQ}}(\alpha)=2(\alpha-1).
\]
Substituting into \eqref{eq:cs_bd_empirical} and dropping constants that do not depend on $\alpha$, we obtain
\[
\widehat{\text{BD}}^{\text{CS}}_{g^{\text{SQ}}}(\alpha)
=
\frac{1}{n}\sum^n_{i=1}\alpha(X_i)^2
-
\frac{2}{m}\sum^m_{j=1}\alpha(\widetilde{X}_j).
\]
This is the classical least-squares importance fitting (LSIF) objective in density ratio estimation \citep{Kanamori2009aleastsquares}.
In the debiased machine learning literature, the same squared loss criterion is also used as Riesz regression for covariate shift adaptation \citep{Chernozhukov2025automaticdebiased}.
Related extensions include doubly robust covariate shift adaptation schemes that combine density ratio estimation with regression adjustment \citep{Kato2024doubledebiasedcovariateshift}.

\paragraph{UKL-Riesz regression.}
For a KL motivated objective on $\bbR_+$, take
\[
g^{\text{UKL}}(\alpha)=\alpha\log\alpha-\alpha,
\qquad
\partial g^{\text{UKL}}(\alpha)=\log\alpha.
\]
Then, Equation~\eqref{eq:cs_bd_empirical} becomes, up to an additive constant that does not depend on $\alpha$,
\[
\widehat{\text{BD}}^{\text{CS}}_{g^{\text{UKL}}}(\alpha)
=
\frac{1}{n}\sum^n_{i=1}\alpha(X_i)
-
\frac{1}{m}\sum^m_{j=1}\log \alpha(\widetilde{X}_j).
\]
A standard implementation imposes the normalization constraint
$\frac{1}{n}\sum^n_{i=1}\alpha(X_i)=1$,
in which case minimizing $\widehat{\text{BD}}^{\text{CS}}_{g^{\text{UKL}}}$ is equivalent to maximizing the target log-likelihood
$\frac{1}{m}\sum^m_{j=1}\log \alpha(\widetilde{X}_j)$ subject to normalization and nonnegativity, which yields KLIEP style procedures \citep{Sugiyama2008directimportance}.
This constrained view is also useful for understanding the dual characterization and the associated moment matching property.

\paragraph{BP-Riesz regression.}
BP-Riesz regression interpolates between squared loss and KL type objectives.
For $\gamma\in(0,\infty)$, consider the BP choice on $\bbR_+$ with $C=0$,
\[
g^{\text{BP}}(\alpha)
\coloneqq
\frac{\alpha^{1+\gamma}-\alpha}{\gamma}-\alpha,
\qquad
\partial g^{\text{BP}}(\alpha)
=
\p{1+\frac{1}{\gamma}}\Bigp{\alpha^\gamma-1}.
\]
A useful simplification is that
$\partial g^{\text{BP}}(\alpha)\alpha-g^{\text{BP}}(\alpha)=\alpha^{1+\gamma}$,
so \eqref{eq:cs_bd_empirical} reduces, up to constants, to
\[
\widehat{\text{BD}}^{\text{CS}}_{g^{\text{BP}}}(\alpha)
=
\frac{1}{n}\sum^n_{i=1}\alpha(X_i)^{1+\gamma}
-
\p{1+\frac{1}{\gamma}}\frac{1}{m}\sum^m_{j=1}\alpha(\widetilde{X}_j)^{\gamma}.
\]
When $\gamma=1$, this objective coincides with the SQ-Riesz objective, up to scaling and constants.
As $\gamma\to 0$, it approaches a KL-type criterion via the expansion $\alpha^\gamma=1+\gamma\log \alpha+o(\gamma)$, providing a continuous bridge between LSIF and KLIEP, and offering a robustness device against extreme ratios \citep{Basu1998robustandefficient,Sugiyama2012densityratio}.

\paragraph{BKL-Riesz regression.}
BKL-Riesz regression corresponds to probabilistic classification based density ratio estimation, which estimates the log density ratio by fitting a classifier that discriminates target covariates from source covariates \citep{Qin1998inferencesfor,Cheng2004semiparametricdensity}.
Let $S\in\{0,1\}$ denote a domain indicator, where $S=1$ for target and $S=0$ for source, and let $\pi\coloneqq P(S=1)$ denote the mixture class prior.
Under Bayes' rule, the density ratio satisfies
\[
r_0(x)=\frac{p_1(x)}{p_0(x)}
=
\frac{1-\pi}{\pi}\frac{P(S=1\mid X=x)}{P(S=0\mid X=x)}.
\]
We model $P(S=1\mid X=x)$ by a logistic specification
\[
p_\bmbeta(S=1\mid X=x)\coloneqq \frac{1}{1+\exp\bigp{-\bmphi(x)^\top\bmbeta}},
\]
and estimate $\bmbeta$ by regularized Bernoulli likelihood on the pooled sample:
\begin{align*}
\widehat\bmbeta\in\argmin_{\bmbeta}\quad
&-\frac1{n+m}\sum_{i=1}^n\log\{1-p_\bmbeta(S=1\mid X_i)\}\\
&-\frac1{n+m}\sum_{j=1}^m\log p_\bmbeta(S=1\mid\widetilde X_j)
+\lambda\|\bmbeta\|_2^2.
\end{align*}
With $\widehat{\pi}\coloneqq \frac{m}{n+m}$, we then set
\[
\widehat{\alpha}(x)
\coloneqq
\frac{1-\widehat{\pi}}{\widehat{\pi}}
\frac{p_{\widehat{\bmbeta}}(S=1\mid X=x)}{1-p_{\widehat{\bmbeta}}(S=1\mid X=x)}
=
\frac{1-\widehat{\pi}}{\widehat{\pi}}\exp\bigp{\bmphi(x)^\top\widehat{\bmbeta}}.
\]
This construction enforces nonnegativity by design and connects density ratio estimation to standard classification tools.

\begin{remark}[From density ratio estimation to covariate shift adaptation]
    Once we obtain $\widehat{\alpha}$ and an outcome regression estimator $\widehat{\gamma}$, we plug them into the covariate shift Neyman score in Section~\ref{sec:setup}.
In particular, a doubly robust estimator that accommodates separate source and target samples is
\[
\widehat{\theta}^{\text{CS}}
\coloneqq
\frac{1}{m}\sum^m_{j=1}\widehat{\gamma}(\widetilde{X}_j)
+
\frac{1}{n}\sum^n_{i=1}\widehat{\alpha}(X_i)\Bigp{Y_i-\widehat{\gamma}(X_i)}.
\]
The corresponding IPW estimator is obtained by dropping the regression adjustment term and using
$\widehat{\theta}^{\text{CS}}_{\text{IPW}}=\frac{1}{n}\sum^n_{i=1}\widehat{\alpha}(X_i)Y_i$.
When cross-fitting is used, $\widehat{\alpha}$ and $\widehat{\gamma}$ for an observation are estimated without using the fold that contains that observation.
\end{remark}

\begin{table}[!t]
\caption{Synthetic-data results. MSE is mean squared error, and CR is the empirical coverage of nominal $95\%$ confidence intervals over 100 replications.}
\vspace{5mm}
    \centering
    \resizebox{\linewidth}{!}{
\begin{tabular}{l|rrr|rrr|rrr|rrr|rrr|rrr}
\hline
      & \multicolumn{3}{|c|}{True} & \multicolumn{3}{|c|}{SQ-Riesz (Linear)} & \multicolumn{3}{|c|}{SQ-Riesz (Logit)} &  \multicolumn{3}{|c|}{UKL-Riesz ($\bmphi(Z)$)}  &  \multicolumn{3}{|c|}{UKL-Riesz ($\bmphi(X)$)}  &  \multicolumn{3}{|c}{Logistic plug-in} \\
    & RA & RW & ARW & RA & RW & ARW & RA & RW & ARW & RA & RW & ARW & RA & RW & ARW & RA & RW & ARW\\
\hline
MSE & 0.00 & 1.44 & 0.01 & 0.39 & 0.49 & 0.19 & 0.39 & 1.38 & 0.08 & 0.38 & 1.50 & 0.10 & 0.40 & 1.52 & 0.10 & 0.39 & 3.79 & 0.23 \\
CR & 1.00 & 0.84 & 0.98 & 0.06 & 0.98 & 0.87 & 0.12 & 0.80 & 0.89 & 0.08 & 0.73 & 0.77 & 0.06 & 0.68 & 0.81 & 0.06 & 0.32 & 0.60 \\
\hline
\end{tabular}
}
    \label{tab:simulation2}
\end{table}
The ``True'' columns use the true nuisance functions. SQ-Riesz is fitted with linear and logarithmic representer models. The two UKL specifications use functions of $Z$ and functions of $X=(D,Z)$, respectively. The logistic plug-in uses the sigmoid propensity-score model. Each fitted representer is reported with RA, RW, and ARW.

\section{Additional Experimental Results}
\label{appdx:additional_experiments}

\subsection{Loss--Link Compatibility in an ATE Design}
\label{appdx:compatibility_figure}
\begin{figure}[!t]
    \centering
    \subfigure[Compatible loss--link pairs.]{\includegraphics[width=0.94\linewidth]{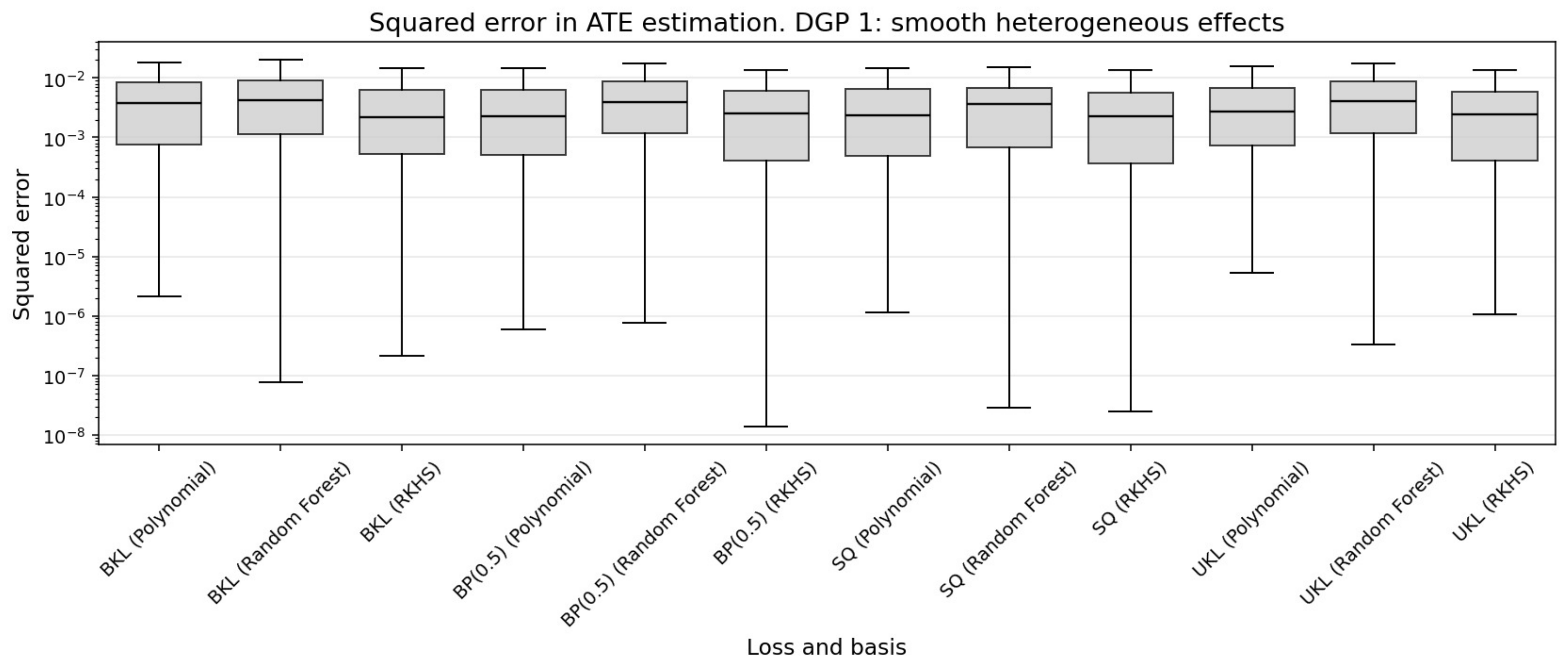}}\\[0.6em]
    \subfigure[Incompatible loss--link pairs.]{\includegraphics[width=0.94\linewidth]{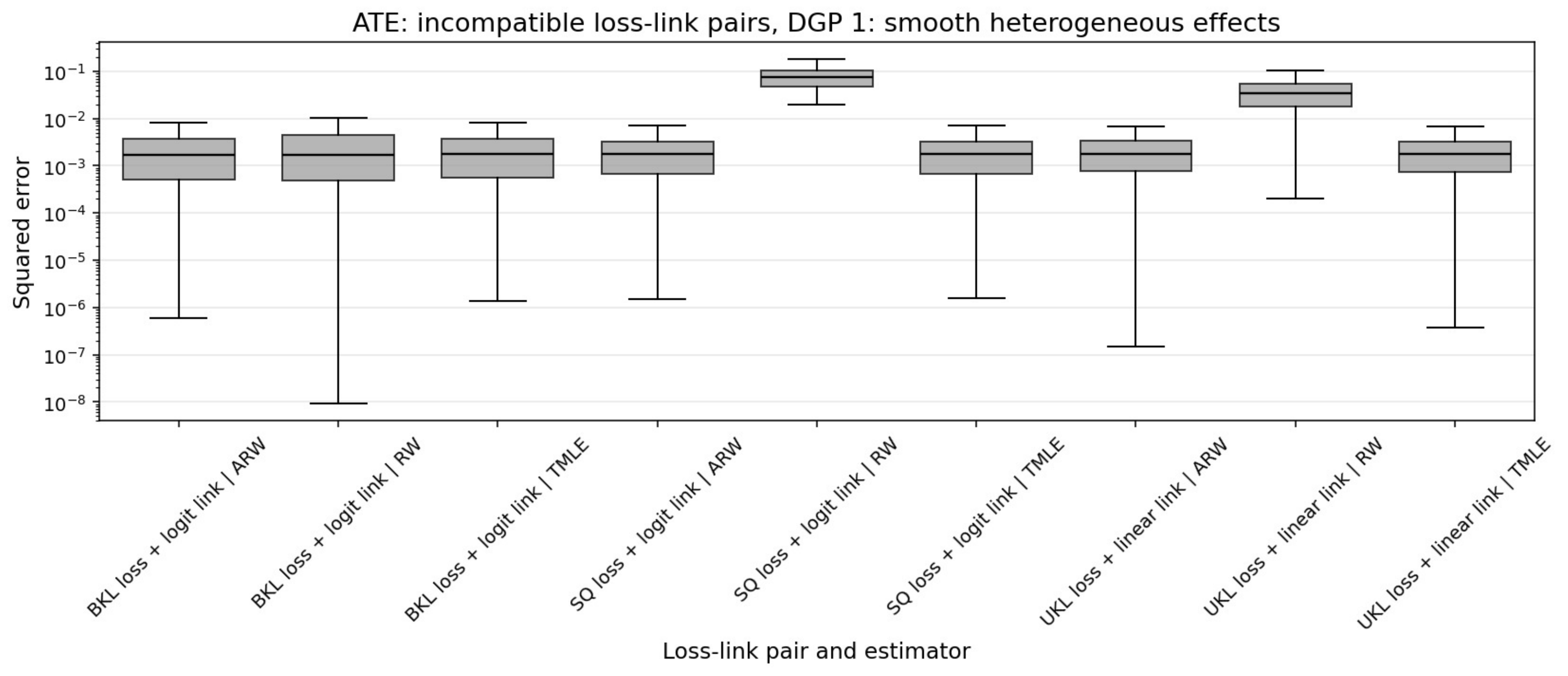}}
    \caption{Loss--link compatibility in the first ATE simulation design. Panel~(a) reports ARW squared errors across compatible specifications and three representer dictionaries. Panel~(b) reports RW, ARW, and TMLE under deliberately incompatible pairs. The vertical axis is logarithmic in both panels.}
    \label{fig:compatibility_ate_dgp1}
\end{figure}

The publication notebook \texttt{01\_main\_simulation\_study.ipynb} compares compatible specifications with deliberately incompatible loss--link pairs. Figure~\ref{fig:compatibility_ate_dgp1} reports the first ATE design, which has smooth heterogeneous effects. The simulation uses $n=1200$, 100 replications, and five-fold cross-fitting. Panel~(a) reports ARW squared errors when each generator is paired with its compatible link and the representer is estimated with polynomial, random-forest leaf, or RKHS dictionaries. Panel~(b) reports RW, ARW, and TMLE under three incompatible pairs.

The two panels serve different purposes and do not rank the losses on a common estimator and dictionary. They illustrate Proposition~\ref{prop:arbitrary_pair_foc}. An incompatible pair can satisfy its own first-order conditions while balancing model-dependent tangent directions rather than the regressors selected in advance. In this design, the largest deterioration appears for RW, while outcome augmentation reduces the effect. The corresponding ATE and ATT panels for the other two data-generating processes remain in the replication files and are omitted here to avoid repeating the same set of comparisons.

\clearpage

\subsection{Synthetic Data}
\paragraph{Design.}
The covariates are three-dimensional, $K = 3$, and we fix the sample size at $n = 3000$. In each Monte Carlo replication, we generate covariates $Z_i \in \bbR^3$ from a multivariate normal distribution $\mathcal{N}(0, I_3)$ and construct a nonlinear propensity score model with polynomial and interaction terms as
\[e_0(Z_i) = \frac{1}{1 + \exp\bigl(-h(Z_i)\bigr)},\]
where
\[h(Z_i) = \sum_{j=1}^3 a_j Z_{i,j} + \sum_{j=1}^3 b_j Z_{i,j}^2 + c_{1} Z_{i,1} Z_{i,2} + c_{2} Z_{i,2} Z_{i,3} + c_{3} Z_{i,1} Z_{i,3}.\]
The coefficients \(a_j\), \(b_j\), and \(c_j\) are independently drawn from \(\mathcal{N}(0,0.5)\). Given these propensity scores, the treatment assignment \(D_i\) is sampled accordingly. We then generate the outcome as
\[Y_i = 1.0 + \p{\sum^3_{j=1}Z_{i,j} \widetilde{a}_j}^2 + 1/\p{1 + \exp\p{-\p{\sum^3_{j=1}Z^2_{i,j} \widetilde{b}_j} }} + \tau_0 D_i + \varepsilon_i,\]
where \(\varepsilon_i \sim \mathcal{N}(0,1)\) and \(\tau_0 = 5.0\).

\paragraph{Estimators and implementation.}
We estimate the Riesz representer using the following variants, matched to Table~\ref{tab:simulation2}.
\begin{itemize}
\item \textbf{SQ-Riesz (Linear)} and \textbf{SQ-Riesz (Logit)}: squared-loss generalized Riesz regression with two different link specifications for the Riesz-representer model.
\item \textbf{UKL-Riesz ($\bmphi(Z)$)} and \textbf{UKL-Riesz ($\bmphi(X)$)}: UKL generalized Riesz regression with a log-type link, comparing two feature sets. Here $X=(D,Z)$ and $\bmphi(Z)$ uses only $Z$, while $\bmphi(X)$ uses the full regressor (allowing treatment-dependent features).
\item \textbf{Logistic plug-in}: Bernoulli-likelihood estimation of a sigmoid propensity score followed by substitution of $\widehat e(Z)$ into the ATE Riesz representer.
\end{itemize}

For the representer and regression models, we use a neural network with one hidden layer of $100$ nodes. To avoid relying on a Donsker condition, we estimate the nuisance parameters via two-fold cross-fitting: each fold is used once to estimate the nuisance parameters and once to evaluate the score contributions. The final estimator averages the contributions from the two folds.

This sample-splitting scheme does not impose exact balance on the observations used to evaluate the score, because the representer is estimated on the other fold and the regression estimator need not belong to the same linear space as the regressor functions used for balance. Cross-fitting separates nuisance estimation from score evaluation. It does not by itself establish automatic Neyman orthogonalization.

We repeat the simulation 100 times. The ``True'' columns in Table~\ref{tab:simulation2} report infeasible oracle performance using the true nuisance functions.

\paragraph{Results.}
The oracle columns in Table~\ref{tab:simulation2} separate estimation error from the variance of the estimators. The oracle ARW estimator is close to the efficiency benchmark (MSE $=0.01$) and has near-nominal coverage (CR $=0.98$). However, oracle RW remains noisy even with the true propensity score (MSE $=1.44$) and undercovers (CR $=0.84$).

The RA estimator has moderate MSE (about $0.38$–$0.40$) across the fitted specifications, but its coverage ranges only from $0.06$ to $0.12$. Thus, in this design, regression adjustment alone does not provide reliable confidence intervals.

The method used to estimate the representer matters more for RW. SQ-Riesz with the linear link has MSE $0.49$ and coverage $0.98$, whereas the two UKL specifications have MSE near $1.50$ and coverage between $0.68$ and $0.73$. The logistic propensity-score plug-in is still more variable (MSE $=3.79$, CR $=0.32$).

ARW reduces the differences in MSE, although its coverage still depends on the representer estimate. SQ-Riesz with the logarithmic link has the smallest ARW MSE (0.08) and coverage of $0.89$, while the two UKL specifications have MSE of $0.10$ and coverage between $0.77$ and $0.81$. The logistic plug-in has ARW MSE of $0.23$ and coverage of $0.60$. These results show that the loss and link affect pure weighting most strongly, but they also affect the finite-sample calibration of ARW.

\begin{table}[t!]
    \centering
        \caption{IHDP results under neural-network and RKHS nuisance estimators.}
        \vspace{5mm}
\scalebox{0.8}{
% Neural network
\begin{tabular}{l|rrr|rrr|rrr}
\hline
 & \multicolumn{9}{|c}{Neural network} \\
       & \multicolumn{3}{|c|}{SQ-Riesz} &  \multicolumn{3}{|c|}{UKL-Riesz} &  \multicolumn{3}{|c}{Logistic plug-in} \\
    & RA & RW & ARW & RA & RW & ARW & RA & RW & ARW \\
\hline
MSE & 1.52 & 6.82 & 0.31 & 1.55 & 2.84 & 0.32 & 1.58 & 3.00 & 0.43 \\
CR & 0.03 & 0.41 & 1.00 & 0.03 & 0.73 & 0.94 & 0.01 & 0.61 & 0.90 \\
\hline
\end{tabular}}
\scalebox{0.8}{
% RKHS
\begin{tabular}{l|rrr|rrr|rrr}
\hline
 & \multicolumn{9}{|c}{RKHS} \\
       & \multicolumn{3}{|c|}{SQ-Riesz} &  \multicolumn{3}{|c|}{UKL-Riesz} &  \multicolumn{3}{|c}{Logistic plug-in} \\
    & RA & RW & ARW & RA & RW & ARW & RA & RW & ARW \\
\hline
MSE & 19.98 & 3.56 & 19.97 & 2.59 & 1.78 & 4.45 & 2.48 & 1.22 & 2.32 \\
CR & 0.00 & 0.00 & 0.00 & 0.48 & 0.93 & 0.88 & 0.39 & 0.81 & 0.84 \\
\hline
\end{tabular}
}
    \label{tab:ihdp}
\end{table}

\subsection{An Analytic Example for the Summary-Space Diagnostic}
\label{appdx:summary_space_diagnostic}
Let $Z\sim\mathcal N(0,I_3)$, $U=\bmbeta^\top Z$, and $q(Z)=Z_2^2-1$. Suppose that treatment assignment depends on $Z$ only through $U$, so $D\perp Z\mid U$. If the summary contains $(D,U)$, then
\[
\bbE[q(Z)\mid D,U]=\bbE[q(Z)\mid U].
\]
Write $b=\beta_2/\|\bmbeta\|$. The conditional distribution of $Z_2$ given $U$ is normal with conditional variance $1-b^2$ and conditional mean $bU/\|\bmbeta\|$. Direct calculation gives
\begin{align*}
\bbE\sqb{\Var(q(Z)\mid U)}=2(1-b^4).
\end{align*}
Because $\Var(q(Z))=2$, the unexplained-variance ratio in Section~\ref{subsec:specification_reporting} is
\[
R_q(D,U)=1-b^4.
\]
Thus, the summary represents the quadratic direction well only when the propensity index is strongly aligned with $Z_2$.

Table~\ref{tab:summary_space_diagnostic} verifies the formula by Monte Carlo simulation. We use $1.5\times10^6$ draws and approximate conditioning on $U$ with 300 equal-frequency bins. The discrepancy in the last row is the binning error at the degenerate limit $b=1$.

\begin{table}[!t]
\centering
\caption{Analytic and Monte Carlo values of $\bbE[\Var(Z_2^2-1\mid U)]$.}
\label{tab:summary_space_diagnostic}
\begin{tabular}{lrrr}
\toprule
$\bmbeta$ & $b$ & Analytic value & Monte Carlo value \\
\midrule
$(1,0,0)$ & 0.0000 & 2.0000 & 1.9994 \\
$(1,0.5,1)$ & 0.3333 & 1.9753 & 1.9732 \\
$(1,1,1)$ & 0.5774 & 1.7778 & 1.7859 \\
$(0.5,1,0.5)$ & 0.8165 & 1.1111 & 1.1213 \\
$(0,1,0)$ & 1.0000 & 0.0000 & 0.0250 \\
\bottomrule
\end{tabular}
\end{table}

\subsection{IHDP Semi-Synthetic Data}
\label{appdx:semisynthetic}
Following \citet{Chernozhukov2022riesznet}, we next evaluate the same family of estimators on the semi-synthetic IHDP benchmark using the standard setting ``A'' in the \texttt{npci} package. We generate 1000 replications, each with $n=747$ observations, a binary treatment, an outcome, and $p=25$ covariates. The estimand is the ATE.

We report RA, RW, and ARW for SQ-Riesz, UKL-Riesz, and the logistic propensity-score plug-in. We consider two nuisance-estimator families:
\begin{itemize}
\item a feedforward neural network with one hidden layer of 100 units,
\item an RKHS estimator with 100 Gaussian basis functions (with tuning by cross validation).
\end{itemize}
For each configuration, we compute the MSE of the ATE estimate and the coverage rate (CR) of nominal $95$\% confidence intervals based on the estimated influence-function variance. Results appear in Table~\ref{tab:ihdp}.

With neural-network nuisance estimators, ARW has MSE between $0.31$ and $0.43$. Its coverage is $1.00$ for SQ-Riesz, $0.94$ for UKL-Riesz, and $0.90$ for the logistic plug-in. By contrast, RA has coverage close to zero, while RW is substantially more variable, especially under SQ-Riesz (MSE $=6.82$).

With RKHS nuisance estimators, SQ-Riesz has MSE near $20$ and zero coverage for both RA and ARW. UKL-Riesz is less variable for RW (MSE $=1.78$, CR $=0.93$), while the logistic plug-in has RW MSE of $1.22$ and coverage of $0.81$. Thus, the ranking changes with the nuisance estimator used in the experiment.

\section{Extensions}
\label{appdx:extensions}

\subsection{Nearest Neighbor Matching}
\label{sec:nnmaching}
Following this study, \citet{Kato2025nearestneighbor} shows that nearest neighbor matching for ATE estimation can be interpreted as a special case of SQ-Riesz regression, that is, Riesz regression or LSIF. The key step is to express the ATE Riesz representer $\alpha^{\text{ATE}}_0\p{D,Z}$ in terms of density ratios with respect to the marginal covariate distribution, and to approximate these density ratios via nearest neighbor cells, following the density ratio interpretation in \citet{Lin2023estimationbased}. \citet{Kato2026vectorsearch} applies the result for policy learning methods based on Retrieval-Augmented Generation (RAG).  

\paragraph{Nearest-neighbor matching ATE estimator.}
Let
\[
J_M\p{i}\subset\cb{1,\ldots,n}
\]
be the index set of the $M$ nearest neighbors of $Z_i$ among the units with $D_j=1-D_i$. We define $\widehat Y_i\p{d}$ as
\begin{align*}
\widehat Y_i\p{0}
&\coloneqq
\begin{cases}
Y_i, & \text{if } D_i=0\\
\frac{1}{M}\sum_{j\in J_M\p{i}} Y_j, & \text{if } D_i=1,
\end{cases}
\\
\widehat Y_i\p{1}
&\coloneqq
\begin{cases}
\frac{1}{M}\sum_{j\in J_M\p{i}} Y_j, & \text{if } D_i=0\\
Y_i, & \text{if } D_i=1.
\end{cases}
\end{align*}
Then, the NN matching ATE estimator is given by
\[
\widehat\theta_M\coloneqq \frac{1}{n}\sum^n_{i=1}\p{\widehat Y_i\p{1}-\widehat Y_i\p{0}}.
\]
Introduce the \emph{matched-times count}, the number of times unit $i$ is used as a match by units in the opposite group, as
\[
K_M\p{i}
\coloneqq
\sum^n_{j=1}\bbI\sqb{D_j=1-D_i}\bbI\sqb{i\in J_M\p{j}}.
\]
Then, $\widehat\theta_M$ can be written as
\begin{align*}
\widehat\theta_M
&=\frac1n\left\{\sum_{i:D_i=1}\left(1+\frac{K_M(i)}{M}\right)Y_i
-\sum_{i:D_i=0}\left(1+\frac{K_M(i)}{M}\right)Y_i\right\}\\
&=\frac1n\sum_{i=1}^n(2D_i-1)\left(1+\frac{K_M(i)}{M}\right)Y_i.
\end{align*}

\paragraph{Nearest neighbor matching as density ratio estimation.}
\citet{Lin2023estimationbased} show that nearest neighbor matching can be interpreted as a method for density ratio estimation. Let $Z^{(0)},Z^{(1)}\in\calZ$ be independent with densities $p_0$ and $p_1$. Assume that $p_0(z)>0$ and $p_1(z)>0$ for all $z\in\calZ$. We observe i.i.d.\ samples $\cb{Z^{(0)}_i}^{N_0}_{i=1}$ and $\cb{Z^{(1)}_j}^{N_1}_{j=1}$ and aim to estimate the density ratio
\[
r^\dagger_0\p{z}\coloneqq \frac{p_1\p{z}}{p_0\p{z}}.
\]

For $M\in\cb{1,\dots,N_0}$ and $z\in\calZ$, let $\calZ_{(M)}\p{z}$ be the $M$th nearest neighbor of $z$ in $\cb{Z^{(0)}_i}^{N_0}_{i=1}$ under a given metric $\norm{\cdot}$. Define the \emph{catchment area} of $c\in\calZ$ as
\[
A_M\p{c}
\coloneqq
\cb{z\in\calZ:\norm{c-z}\le \norm{\calZ_{(M)}\p{z}-z}},
\]
and the \emph{matched-times count} as
\[
K_M\p{c}\coloneqq \sum^{N_1}_{j=1}\bbI\sqb{Z^{(1)}_j\in A_M\p{c}}.
\]
\citet{Lin2023estimationbased} propose the one-step estimator
\[
\widehat r^\dagger_M\p{c}
=
\frac{N_0}{N_1}\frac{K_M\p{c}}{M}.
\]
They also show that this method is computationally efficient and rate-optimal for Lipschitz densities.

\paragraph{Nearest neighbor matching as LSIF.}
We next show that the one-step estimator of \citet{Lin2023estimationbased} can be obtained as a squared-loss density ratio estimator of LSIF type. Since SQ-Riesz regression for density ratios coincides with LSIF, this yields a direct connection between nearest neighbor matching and SQ-Riesz regression.

Consider the empirical LSIF objective for estimating $r^\dagger_0\p{z}=p_1\p{z}/p_0\p{z}$:
\[
\widehat R^{\text{LSIF}}\p{r}
\coloneqq
\frac{1}{2N_0}\sum^{N_0}_{i=1} r\p{Z^{(0)}_i}^2
-
\frac{1}{N_1}\sum^{N_1}_{j=1} r\p{Z^{(1)}_j}.
\]
Following the nearest neighbor construction, let $J_M\p{j}\subset\cb{1,\dots,N_0}$ be the indices of the $M$ nearest neighbors of $Z^{(1)}_j$ in $\cb{Z^{(0)}_i}^{N_0}_{i=1}$. Consider the kNN linear model indexed by coefficients $\bmbeta=\p{\beta_1,\dots,\beta_{N_0}}^\top$:
\[
r_{\bmbeta}\p{Z^{(0)}_i}\coloneqq \beta_i,
\qquad
r_{\bmbeta}\p{Z^{(1)}_j}\coloneqq \frac{1}{M}\sum_{i\in J_M\p{j}}\beta_i.
\]
Plugging this model into the LSIF objective yields
\begin{align*}
\widehat R^{\text{LSIF}}\p{r_{\bmbeta}}
&=
\frac{1}{2N_0}\sum^{N_0}_{i=1}\beta_i^2
-
\frac{1}{N_1}\sum^{N_1}_{j=1}\frac{1}{M}\sum_{i\in J_M\p{j}}\beta_i\\
&=
\frac{1}{2N_0}\sum^{N_0}_{i=1}\beta_i^2
-
\frac{1}{N_1}\sum^{N_0}_{i=1}\frac{K_M\p{i}}{M}\beta_i,
\end{align*}
where $K_M\p{i}\coloneqq \sum^{N_1}_{j=1}\bbI\sqb{i\in J_M\p{j}}$ counts how many times $Z^{(0)}_i$ appears among the $M$ nearest neighbors of the $p_1$ sample points. The objective is separable in the coordinates $\beta_i$, and the unique minimizer satisfies
\[
\frac{1}{N_0}\widehat\beta_i-\frac{1}{N_1}\frac{K_M\p{i}}{M}=0,
\qquad i=1,\dots,N_0,
\]
and hence we have
\[
\widehat\beta_i
=
\frac{N_0}{N_1}\frac{K_M\p{i}}{M}.
\]
Therefore, the fitted density ratio evaluated at the denominator sample points is
\[
\widehat r^\dagger_M\p{Z^{(0)}_i}
=
r_{\widehat\bmbeta}\p{Z^{(0)}_i}
=
\frac{N_0}{N_1}\frac{K_M\p{i}}{M},
\]
which matches the one-step estimator of \citet{Lin2023estimationbased}.

This density ratio viewpoint can be transferred to ATE estimation by applying the same kNN idea to the density ratios that define the ATE Riesz representer, as discussed in \citet{Kato2025nearestneighbor}.

\subsection{Causal Tree and Causal Forest}
Causal trees and causal forests estimate the conditional average treatment effect (CATE) by constructing a partition of the covariate space and estimating a local ATE within each cell, as in \citet{Wager2018estimationinference}. We emphasize that this procedure implicitly constructs an estimator of the corresponding Riesz representer. In particular, once a partition is fixed, the leafwise CATE estimator can be rewritten as an inverse probability weighting type estimator, with weights that coincide with a leafwise Riesz representer estimator.

\paragraph{Leafwise CATE as a Riesz representer plug-in.}
Let $\Pi=\cb{\ell}$ be a partition of the covariate space $\calZ$ produced by a causal tree, and let $\ell\p{z}\in\Pi$ denote the leaf containing $z\in\calZ$. For a leaf $\ell$, define
\begin{align*}
n_\ell&\coloneqq\sum_{i=1}^n\bbI[Z_i\in\ell],\\
n_{1,\ell}&\coloneqq\sum_{i=1}^n\bbI[D_i=1,Z_i\in\ell],
\qquad
n_{0,\ell}\coloneqq\sum_{i=1}^n\bbI[D_i=0,Z_i\in\ell].
\end{align*}
The CATE estimator obtained by a causal tree is the leafwise difference in means
\[
\widehat\theta\p{z}
\coloneqq
\frac{1}{n_{1,\ell(z)}}\sum^n_{i=1}\bbI\sqb{D_i=1,Z_i\in\ell\p{z}}Y_i
-
\frac{1}{n_{0,\ell(z)}}\sum^n_{i=1}\bbI\sqb{D_i=0,Z_i\in\ell\p{z}}Y_i.
\]
This estimator admits the weighted representation
\[
\widehat\theta\p{z}
=
\frac{1}{n}\sum^n_{i=1}\p{\widehat{\alpha}\p{D_i,Z_i;z}Y_i},
\]
where
\[
\widehat{\alpha}\p{D,Z;z}
\coloneqq
\bbI\sqb{Z\in\ell\p{z}}
\p{\frac{D}{\widehat{\pi}_{1,\ell(z)}}-\frac{1-D}{\widehat{\pi}_{0,\ell(z)}}}
\frac{1}{\widehat p_{\ell(z)}},
\qquad
\widehat{\pi}_{d,\ell}\coloneqq \frac{n_{d,\ell}}{n_\ell},
\qquad
\widehat p_\ell\coloneqq \frac{n_\ell}{n}.
\]
Hence, $\widehat\theta\p{z}$ is an inverse probability weighting type estimator with a weight function $\widehat{\alpha}\p{\cdot,\cdot;z}$. This weight function is a plug-in estimator of the leafwise Riesz representer for the local ATE
\[
\theta\p{\ell}\coloneqq \bbE\sqb{Y\p{1}-Y\p{0}\mid Z\in\ell},
\]
because the corresponding population representer takes the same form, with $\p{\widehat{\pi}_{d,\ell},\widehat p_\ell}$ replaced by their population counterparts. Therefore, conditional on the partition, causal trees estimate the CATE by implicitly estimating a Riesz representer that is constant on each leaf.

\paragraph{Connection to SQ-Riesz regression and adaptive nearest neighbors.}
The expression above shows that a causal tree is a histogram-type estimator of the Riesz representer, with leaf indicators $\cb{\bbI\sqb{Z\in\ell}}_{\ell\in\Pi}$ as its basis functions. This is directly analogous to the nearest neighbor histogram model in the previous subsection, except that the partition is learned from the data rather than fixed a priori. From this viewpoint, the splitting criterion in a causal tree can be interpreted as choosing an adaptive partition that reduces the error of the induced leafwise Riesz representer approximation, and hence reduces the error of the resulting local CATE estimator.

A causal forest averages many such trees, built on subsamples and random feature choices, and therefore produces weights that average the leafwise Riesz representer estimators across trees. Equivalently, causal forests produce an adaptive nearest neighbor type representation for CATE, where the neighborhood structure is learned via the random partitions. This clarifies why causal trees and causal forests fit naturally into the same squared loss Bregman divergence, namely SQ-Riesz, perspective as nearest neighbor matching, with the main difference being that causal forests learn the partition adaptively to target CATE estimation accuracy.

\subsection{AME Estimation by Score Matching}
Subsequent work by \citet{Kato2025scorematchingriesz} shows that, for derivative-type linear functionals, the Riesz representer can be estimated via score matching. This principle also underlies score-based diffusion models \citep{Song2020generativemodeling,Song2021denoisingdiffusion}. This viewpoint is useful for AME and APE estimation, and for mitigating overfitting in flexible Riesz representer models, because score matching objectives introduce smoothing through derivatives or noise perturbations.

\paragraph{Score matching identity for AME.}
Recall the AME example in Section~\ref{sec:setup}, where
\[
m^{\text{AME}}\p{W,\gamma}=\partial_d\gamma\p{D,Z},
\qquad
\alpha^{\text{AME}}_0\p{D,Z}=-\partial_d\log f_0\p{D,Z},
\]
with $f_0$ denoting the joint density of $X=\p{D,Z}$. Let $s_{0,d}\p{x}\coloneqq \partial_d\log f_0\p{x}$ be the $d$th component of the score. Consider a sufficiently smooth candidate function $\alpha\p{x}$ such that integration by parts is valid and boundary terms vanish. Then, integration by parts gives
\[
\bbE\sqb{\partial_d\alpha\p{X}}
=
\int \partial_d\alpha\p{x}f_0\p{x}\rmd x
=
-\int \alpha\p{x}\partial_d f_0\p{x}\rmd x
=
-\bbE\sqb{\alpha\p{X}s_{0,d}\p{X}}.
\]
Therefore, the squared loss Bregman objective for AME can be rewritten as
\begin{align*}
\bbE\sqb{\alpha\p{X}^2-2\partial_d\alpha\p{X}}
&=
\bbE\sqb{\alpha\p{X}^2+2\alpha\p{X}s_{0,d}\p{X}}\\
&=
\bbE\sqb{\p{\alpha\p{X}+s_{0,d}\p{X}}^2}-\bbE\sqb{s_{0,d}\p{X}^2}.
\end{align*}
The last term does not depend on $\alpha$. Hence, minimizing $\bbE\sqb{\alpha\p{X}^2-2\partial_d\alpha\p{X}}$ is equivalent to minimizing $\bbE\sqb{\p{\alpha\p{X}-\alpha^{\text{AME}}_0\p{X}}^2}$, and the population minimizer is $\alpha^{\text{AME}}_0=-s_{0,d}$. This is a coordinatewise form of the classical score matching principle and shows that, for derivative-type $m$, our squared loss Bregman risk coincides with an $L_2$ score matching risk for the Riesz representer.

\paragraph{Denoising score matching via diffusion.}
In high dimensions, directly learning the score $x\mapsto \nabla_x\log f_0\p{x}$ can be unstable. Score-based diffusion models address this issue by estimating scores of noise-perturbed distributions via denoising score matching \citep{Song2020improvedtechniques,Song2021denoisingdiffusion}. Let $T$ be a noise index, continuous or discrete, and generate noisy covariates by
\[
X_T\coloneqq X+\sigma\p{T}Z,
\qquad
Z\sim\mathcal{N}\p{0,I},
\]
independent of $X\sim f_0$. Let $p_T$ denote the density of $X_T$. A time-dependent score model $s_\theta\p{\cdot,T}$ is trained by minimizing the denoising objective
\[
\bbE\sqb{\norm{\sigma\p{T}s_\theta\p{X_T,T}+Z}^2},
\]
which is equivalent, up to an additive constant, to matching $s_\theta\p{\cdot,T}$ to the true score $\nabla_x\log p_T\p{x}$ under an $L_2$ risk. Once $s_\theta$ is trained, we can recover an estimator of the original score $\nabla_x\log f_0\p{x}$ by evaluating at small noise levels and then extracting the relevant component to estimate
\[
\alpha_0^{\text{AME}}\p{x}=-\partial_d\log f_0\p{x}.
\]
Operationally, this replaces the derivative term $\partial_d \alpha\p{X}$ in the score matching objective with a denoising criterion that learns a smoothed score field. This smoothing can mitigate overfitting in high-capacity models and can be combined with flexible neural architectures through automatic differentiation.

\subsection{Riesz Representer Estimation Via Infinitesimal Classification}
\label{sec:rr_infclass}
Following \citet{Kato2025scorematchingriesz}, we introduce Riesz representer estimation via infinitesimal classification, which also reduces to score matching. This approach applies beyond AME estimation.

\paragraph{Density ratio estimation via infinitesimal classification.}
We first review density ratio estimation via infinitesimal classification, proposed in \citet{Choi2022densityratio}. Let $p_0\p{x}$ and $p_1\p{x}$ be two probability density functions such that $p_0\p{x}>0$ holds for all $x\in\calX$. For $x\in\calX$, define the density ratio
\[
r_0\p{x}\coloneqq \frac{p_0\p{x}}{p_1\p{x}}.
\]
We aim to estimate $r_0$.

We define a continuum of bridge densities $\cb{p_t}_{t\in\sqb{0,1}}$ through a simple sampling procedure. Let $p_t\p{x}$ be the density of the random variable
\[
X_t=\beta^{(1)}\p{t}X_0+\beta^{(2)}\p{t}X_1,
\]
where $\beta^{(1)}\p{\cdot},\beta^{(2)}\p{\cdot}\colon\sqb{0,1}\to\sqb{0,1}$ are $C^2$ and monotonic, and satisfy the boundary conditions $\beta^{(1)}\p{0}=1$, $\beta^{(2)}\p{0}=0$, $\beta^{(1)}\p{1}=0$, and $\beta^{(2)}\p{1}=1$.
Using intermediate density ratios, we decompose $r_0$ as
\[
r_0\p{x}
=
\prod^T_{t=1}\frac{p_{(t-1)/T}\p{x}}{p_{t/T}\p{x}}.
\]
We can choose $\beta^{(1)}$ and $\beta^{(2)}$ to improve numerical stability. For example, \citet{Choi2022densityratio} use $\beta^{(1)}\p{t}=1-t$ and $\beta^{(2)}\p{t}=t$ in some applications.

In practice, when optimizing objectives that integrate over $t$, we sample $t$ jointly with $\p{X_0,X_1}$. Specifically, for each stochastic gradient step we draw a minibatch $\cb{\p{X_{0,i},X_{1,i}}}^{B}_{i=1}$ with $X_{0,i}\sim p_0$ and $X_{1,i}\sim p_1$ independently, and we draw times $\cb{t_i}^B_{i=1}$ i.i.d.\ from a reference density $q\p{t}$ on $\sqb{0,1}$. We then form $X_{t_i,i}=\beta^{(1)}\p{t_i}X_{0,i}+\beta^{(2)}\p{t_i}X_{1,i}$ and approximate time integrals using importance weights. For example, an integral term of the form $\int_0^1 \bbE_{X_t\sim p_t}\sqb{h\p{X_t,t}}\rmd t$ is estimated by
\[
\int_0^1 \bbE_{X_t\sim p_t}\sqb{h\p{X_t,t}}\rmd t
\approx
\frac{1}{B}\sum^B_{i=1}\frac{h\p{X_{t_i,i},t_i}}{q\p{t_i}}.
\]
Endpoint expectations, such as $\bbE_{X_0\sim p_0}\sqb{\cdot}$ and $\bbE_{X_1\sim p_1}\sqb{\cdot}$, are approximated by sample averages over $\cb{X_{0,i}}$ and $\cb{X_{1,i}}$, respectively. All derivatives with respect to $t$ that appear in the objective, such as $\partial_t\p{\lambda\p{t}s_\beta\p{X_t,t}}$, can be computed by automatic differentiation through the explicit dependence of $X_t$ on $t$ via $\beta^{(1)}\p{t}$ and $\beta^{(2)}\p{t}$.

By taking logarithms,
\[
\log r_0\p{x}
=
\sum^T_{t=1}\log\p{\frac{p_{(t-1)/T}\p{x}}{p_{t/T}\p{x}}}.
\]
As $T\to\infty$, it holds that \citep{Choi2022densityratio,Chen2025dequantifieddiffusion}
\[
\log r_0\p{x}
=
\log\p{\frac{p_0\p{x}}{p_1\p{x}}}
=
\int_1^0 \partial_t\log p_t\p{x}\rmd t
=
-\int_0^1 \partial_t\log p_t\p{x}\rmd t.
\]

Let $s^{\text{time}}_{\beta}\p{x,t}$ be a time score model that approximates the time score $\partial_t\log p_t\p{x}$. We train $s^{\text{time}}_{\beta}$ by minimizing the time score matching loss \citep{Choi2022densityratio,Chen2025dequantifieddiffusion}
\[
\calR^\dagger\p{s^{\text{time}}_{\beta}}
\coloneqq
\int_0^1 \bbE_{X_t\sim p_t}\sqb{\lambda\p{t}\Bigp{\partial_t\log p_t\p{X_t}-s^{\text{time}}_{\beta}\p{X_t,t}}^2}\rmd t,
\]
where $\lambda\colon\sqb{0,1}\to\bbR_+$ is a positive weighting function. Because $\log p_t$ is unknown in practice, we use the following objective, which is equivalent to $\calR^\dagger\p{s^{\text{time}}_{\beta}}$ up to an additive constant:
\begin{align*}
\calR\p{s^{\text{time}}_{\beta}}
&\coloneqq
\bbE_{X_0\sim p_0}\sqb{\lambda\p{0}s^{\text{time}}_{\beta}\p{X_0,0}}
-
\bbE_{X_1\sim p_1}\sqb{\lambda\p{1}s^{\text{time}}_{\beta}\p{X_1,1}}\\
&\quad
+
\int_0^1 \bbE_{X_t\sim p_t}\sqb{
\partial_t\p{\lambda\p{t}s^{\text{time}}_{\beta}\p{X_t,t}}
+
\frac{1}{2}\lambda\p{t}s^{\text{time}}_{\beta}\p{X_t,t}^2
}\rmd t.
\end{align*}

To generate a sample from $p_t$, first draw independent endpoint samples $X_0\sim p_0$ and $X_1\sim p_1$. Second, for a given $t\in\sqb{0,1}$, construct the bridge sample by the deterministic map
\[
X_t\coloneqq \beta^{(1)}\p{t}X_0+\beta^{(2)}\p{t}X_1.
\]
We define $p_t$ as the probability law of $X_t$ induced by this procedure, that is, $p_t$ is the pushforward of the product measure $p_0\otimes p_1$ through the map $\p{x_0,x_1}\mapsto \beta^{(1)}\p{t}x_0+\beta^{(2)}\p{t}x_1$. With this definition, expectations under $p_t$ can be evaluated by Monte Carlo as
\[
\bbE_{X_t\sim p_t}\sqb{f\p{X_t,t}}
=
\bbE\sqb{f\p{\beta^{(1)}\p{t}X_0+\beta^{(2)}\p{t}X_1,t}},
\]
where the outer expectation is taken over $\p{X_0,X_1}\sim p_0\otimes p_1$.

\paragraph{Riesz representer estimation via infinitesimal classification.}
\citet{Kato2025scorematchingriesz} extends density ratio estimation via infinitesimal classification to Riesz representer estimation. The following example applies the method to APE estimation. Implementations for the other applications appear in \citet{Kato2025scorematchingriesz}.

In APE estimation, the Riesz representer is given by
\[
\alpha^{\text{APE}}\p{X}\coloneqq \frac{p_1\p{X}-p_{-1}\p{X}}{p_0\p{X}}.
\]
Using intermediate density ratios,
\begin{align*}
\frac{p_1\p{x}}{p_0\p{x}}
&=
\prod^T_{t=1}\frac{p_{t/T}\p{x}}{p_{(t-1)/T}\p{x}},
&
\frac{p_{-1}\p{x}}{p_0\p{x}}
&=
\prod^T_{t=1}\frac{p_{-t/T}\p{x}}{p_{-(t-1)/T}\p{x}}.
\end{align*}
Then, we have
\begin{align*}
\log\frac{p_1\p{x}}{p_0\p{x}}
&=
\sum^T_{t=1}\log\p{\frac{p_{t/T}\p{x}}{p_{(t-1)/T}\p{x}}}
\to
\int_0^1 \partial_t\log p_t\p{x}\rmd t
\qquad \p{T\to\infty},
\\
\log\frac{p_{-1}\p{x}}{p_0\p{x}}
&=
\sum^T_{t=1}\log\p{\frac{p_{-t/T}\p{x}}{p_{-(t-1)/T}\p{x}}}
\to
\int_0^{-1} \partial_t\log p_t\p{x}\rmd t
\qquad \p{T\to\infty}.
\end{align*}

Define the bridge random variable $X_t$ on $t\in\sqb{-1,1}$ by
\[
X_t
\coloneqq
\begin{cases}
\beta^{(1)}\p{t}X_1+\beta^{(2)}\p{t}X_0, & \text{if } t\ge 0,\\
\beta^{(1)}\p{t}X_{-1}+\beta^{(2)}\p{t}X_0, & \text{if } t<0,
\end{cases}
\]
where $\beta^{(1)}\p{\cdot},\beta^{(2)}\p{\cdot}\colon\sqb{-1,1}\to\sqb{0,1}$ are $C^2$ and monotonic, with $\beta^{(1)}$ increasing and $\beta^{(2)}$ decreasing for $t\ge 0$, $\beta^{(1)}$ decreasing and $\beta^{(2)}$ increasing for $t<0$, and satisfying the boundary conditions $\beta^{(1)}\p{0}=0$, $\beta^{(2)}\p{0}=1$, $\beta^{(1)}\p{-1}=1$, $\beta^{(2)}\p{-1}=0$, $\beta^{(1)}\p{1}=1$, and $\beta^{(2)}\p{1}=0$.

Let $p_t$ denote the density of $X_t$. Let $s^{\text{time}}_{\beta}\p{x,t}$ be a time score model that approximates $\partial_t\log p_t\p{x}$. We train the score model by minimizing
\[
\calR^{\text{APE}\dagger}\p{s^{\text{time}}_{\beta}}
\coloneqq
\int_{-1}^1 \bbE_{X_t\sim p_t}\sqb{\lambda\p{t}\Bigp{\partial_t\log p_t\p{X_t}-s^{\text{time}}_{\beta}\p{X_t,t}}^2}\rmd t,
\]
where $\lambda\colon\sqb{-1,1}\to\bbR_+$ is a positive weighting function. Since $\partial_t\log p_t$ is unknown, we minimize the following risk:
\begin{align*}
\calR^{\text{APE}}\p{s^{\text{time}}_{\beta}}
&\coloneqq
\bbE_{X_{-1}\sim p_{-1}}\sqb{\lambda\p{-1}s^{\text{time}}_{\beta}\p{X_{-1},-1}}
-
\bbE_{X_1\sim p_1}\sqb{\lambda\p{1}s^{\text{time}}_{\beta}\p{X_1,1}}\\
&\quad
+
\int_{-1}^1 \bbE_{X_t\sim p_t}\sqb{
\partial_t\p{\lambda\p{t}s^{\text{time}}_{\beta}\p{X_t,t}}
+
\frac{1}{2}\lambda\p{t}s^{\text{time}}_{\beta}\p{X_t,t}^2
}\rmd t.
\end{align*}

\section{From Automatic Regressor Balancing to Classical Covariate Balancing}
\label{appdx:automatic_covariate_balancing_supp}

Section~\ref{sec:automaticcovariatebalancing} derives automatic regressor balancing from the KKT conditions of generalized Riesz regression when the dual coordinate $u=\partial g\circ\alpha$ is linear in the model coefficients. This appendix specializes the generic equations to ATE and density-ratio estimation and records the corresponding loss--link formulas. It also gives a short KKT derivation, which makes clear that the equations apply to the observations used to estimate the representer and need not remain exact on other observations.

\subsection{Generalized Linear Models in Dual Coordinates}
\label{appdx:glm_dual_coordinates}

Let $\bmphi=\p{\phi_1,\dots,\phi_p}^\top$ be basis functions $\bmphi\colon\calX\to\bbR^p$. As in Section~\ref{subsec:lin_score_models}, we consider a generalized linear representer model
\begin{align}
\alpha_{\bmbeta}\p{x}
=
\zeta^{-1}\p{x,\bmphi\p{x}^\top\bmbeta},
\qquad
\bmbeta\in\bbR^p,
\label{eq:appdx_alpha_glm}
\end{align}
but the balancing phenomenon is most transparent when we view \eqref{eq:appdx_alpha_glm} through the dual coordinate $u_{\bmbeta}\p{x}\coloneqq \partial g\p{\alpha_{\bmbeta}\p{x}}$. Automatic balancing arises when $u_{\bmbeta}$ is linear in $\bmbeta$. To state this conveniently, define feature functions
\[
\widetilde\phi_j\p{x}\coloneqq \widetilde g\p{x,\phi_j\p{x}},
\qquad j=1,\dots,p,
\]
and suppose
\begin{align}
u_{\bmbeta}\p{x}
=
\partial g\p{\alpha_{\bmbeta}\p{x}}
=
\sum^p_{j=1}\beta_j\widetilde\phi_j\p{x}.
\label{eq:appdx_dual_linearity}
\end{align}
When $g$ is strictly convex and differentiable, a direct way to enforce \eqref{eq:appdx_dual_linearity} is to choose the link so that, possibly separately on the positive and negative parts of the domain,
\[
\alpha_{\bmbeta}\p{x}
=
\p{\partial g}^{-1}\Bigp{\sum^p_{j=1}\beta_j\widetilde\phi_j\p{x}}.
\]
Section~\ref{subsec:loss_link_pairs} lists the main loss--link pairs used in this study. Below we add a compact summary in Appendix~\ref{appdx:loss_link_recipes}.

\subsection{KKT Derivation: Balancing Equations as First-Order Optimality}
\label{appdx:kkt_derivation_balance}

This subsection provides a short derivation of the balancing identities and inequalities in Theorem~\ref{thm:autocovariance}, emphasizing the role of the dual linearity condition \eqref{eq:appdx_dual_linearity}.

Recall the penalized ERM in Section~\ref{subsec:acb_kkt}:
\[
\widehat{\bmbeta}
\in
\arg\min_{\bmbeta\in\bbR^p}
\Biggcb{
\widehat{\mathrm{BD}}_g\p{\alpha_{\bmbeta}}
+\frac{\lambda}{q}\norm{\bmbeta}_q^q
},
\qquad q\ge 1.
\]
A convenient way to see the gradient structure is to re-express the empirical Bregman objective in dual form. For each observation $i$, let $\calA_i$ be the interval allowed by the representer model and define
\[
g_i^*(u)\coloneqq\sup_{\alpha\in\calA_i}\{\alpha u-g(\alpha)\}.
\]
Let $u_{\bmbeta}=\partial g\circ\alpha_{\bmbeta}$. The Fenchel--Young identity gives $g_i^*(u)=\alpha u-g(\alpha)$ at $u=\partial g(\alpha)$ for $\alpha$ in the interior of $\calA_i$. Hence, up to terms not depending on $\bmbeta$, the empirical objective can be written as
\begin{align}
\widehat{\mathrm{BD}}_g\p{\alpha_{\bmbeta}}
=
\frac{1}{n}\sum^n_{i=1}g_i^*\p{u_{\bmbeta}\p{X_i}}
-
\frac{1}{n}\sum^n_{i=1}m\p{W_i,u_{\bmbeta}}.
\label{eq:appdx_bd_dual_form}
\end{align}
Under \eqref{eq:appdx_dual_linearity}, we have $u_{\bmbeta}\p{x}=\sum^p_{j=1}\beta_j\widetilde\phi_j\p{x}$ and therefore $\partial_{\beta_j}u_{\bmbeta}\p{x}=\widetilde\phi_j\p{x}$. On the range of $\partial g$ over $\calA_i$, it holds that $\partial g_i^*(u)=\p{\partial g}^{-1}(u)=\alpha$. Consequently, we have
\[
\partial_{\beta_j}\Bigp{\frac{1}{n}\sum^n_{i=1}g_i^*\p{u_{\bmbeta}\p{X_i}}}
=
\frac{1}{n}\sum^n_{i=1}\alpha_{\bmbeta}\p{X_i}\widetilde\phi_j\p{X_i},
\]
and by linearity of $m$,
\[
\partial_{\beta_j}\Bigp{\frac{1}{n}\sum^n_{i=1}\p{m\p{W_i,u_{\bmbeta}}}}
=
\frac{1}{n}\sum^n_{i=1}\p{m\p{W_i,\widetilde\phi_j}}.
\]
Therefore, the gradient of the empirical Bregman objective is exactly the imbalance:
\begin{align}
\partial_{\beta_j}\widehat{\mathrm{BD}}_g\p{\alpha_{\bmbeta}}
=
\widehat\Delta_j\p{\alpha_{\bmbeta}}
\coloneqq
\frac{1}{n}\sum^n_{i=1}\Bigp{\alpha_{\bmbeta}\p{X_i}\widetilde\phi_j\p{X_i}-m\p{W_i,\widetilde\phi_j}}.
\label{eq:appdx_grad_equals_imbalance}
\end{align}
Applying the KKT conditions to the penalized problem yields
\[
\widehat\Delta_j\p{\widehat\alpha}+\lambda s_j=0,
\qquad
s_j\in\partial\Bigp{\frac{1}{q}\abs{\beta_j}^q}\Big|_{\beta_j=\widehat\beta_j},
\]
which is Theorem~\ref{thm:autocovariance}.

\paragraph{Balance on other observations.}
The identity \eqref{eq:appdx_grad_equals_imbalance} holds for the observations used to estimate $\widehat\bmbeta$. If $\widehat\alpha$ is evaluated on other observations, the corresponding imbalance need not be zero, regardless of whether the separation arises from a single sample split or cross-fitting. Under cross-fitting, each score summand uses nuisance parameter estimates obtained without the fold containing that summand, so the KKT equations for the estimation folds do not automatically hold on that fold.

\subsection{Specialization to ATE: Covariate Balance as a Special Case}
\label{appdx:ate_covariate_balance_specialization}

We now show how the generic regressor-balancing equations reduce to classical covariate balancing conditions in the ATE setup.

In the ATE model with $X=\p{D,Z}$ and
\[
m^{\text{ATE}}\p{W,\gamma}=\gamma\p{1,Z}-\gamma\p{0,Z},
\qquad
\alpha_0^{\text{ATE}}\p{D,Z}=\frac{D}{e_0\p{Z}}-\frac{1-D}{1-e_0\p{Z}},
\]
consider a feature set of basis functions $\cb{\widetilde\phi_j}_{j=1}^p$ and the imbalance
\[
\widehat\Delta_j\p{\alpha}
=
\frac{1}{n}\sum^n_{i=1}\Bigp{\alpha\p{D_i,Z_i}\widetilde\phi_j\p{D_i,Z_i}-m^{\text{ATE}}\p{W_i,\widetilde\phi_j}}.
\]

\paragraph{The standard $Z$-only covariate balancing regime.}
A classical choice sets $\widetilde\phi_j\p{D,Z}=\phi_j\p{Z}$. Then, it holds that
\[
m^{\text{ATE}}\p{W,\phi_j\p{Z}}=\phi_j\p{Z}-\phi_j\p{Z}=0,
\]
so the exact-balance equations $\widehat\Delta_j\p{\widehat\alpha}=0$ become
\begin{align}
\frac{1}{n}\sum^n_{i=1}\p{\widehat\alpha\p{D_i,Z_i}\phi_j\p{Z_i}}=0
\qquad \p{j=1,\dots,p}.
\label{eq:appdx_z_only_balance}
\end{align}
Plugging in the ATE representer form $\widehat\alpha\p{D,Z}=D/\widehat e\p{Z}-\p{1-D}/\p{1-\widehat e\p{Z}}$, \eqref{eq:appdx_z_only_balance} is equivalent to
\begin{align}
\frac{1}{n}\sum^n_{i=1}\p{\frac{D_i}{\widehat e\p{Z_i}}\phi_j\p{Z_i}}
=
\frac{1}{n}\sum^n_{i=1}\p{\frac{1-D_i}{1-\widehat e\p{Z_i}}\phi_j\p{Z_i}},
\qquad j=1,\dots,p,
\label{eq:appdx_covariate_balance_treated_control}
\end{align}
which is the usual covariate balancing condition.

\paragraph{Regressor balancing for heterogeneous effects.}
If the goal is automatic orthogonalization against a richer working regression class, then $Z$-only balancing can be too restrictive under treatment-effect heterogeneity. A common remedy is to use treatment-specific features, for example,
\[
\widetilde\phi_j\p{D,Z}=D f_j\p{Z},
\qquad
\widetilde\phi_{j+p}\p{D,Z}=\p{1-D} f_j\p{Z},
\]
which makes the balanced span large enough to approximate $\gamma_0\p{D,Z}$ with arm-specific structure. In that case the balancing equations enforce moment conditions for both arms simultaneously, and the orthogonalization implications in Section~\ref{sec:automaticneyman} become stronger.

\paragraph{Approximate balance.}
Under $\ell_1$ regularization, Theorem~\ref{thm:autocovariance} yields $\abs{\widehat\Delta_j\p{\widehat\alpha}}\le \lambda$. In the $Z$-only regime, this means the treated-versus-control covariate-moment mismatch in \eqref{eq:appdx_covariate_balance_treated_control} is controlled by $\lambda$, providing an interpretable balance tolerance knob.

\subsection{Specialization to Density Ratio Estimation: Moment Matching}
\label{appdx:dr_moment_matching_specialization}

For covariate shift and density ratio estimation with source sample $\cb{Z_i}^{n_0}_{i=1}\sim p_0$ and target sample $\cb{Z^e_j}^{n_1}_{j=1}\sim p_1$, the density ratio $r_0\p{z}=p_1\p{z}/p_0\p{z}$ satisfies the moment identity
\[
\bbE_{p_0}\sqb{r_0\p{Z}f\p{Z}}=\bbE_{p_1}\sqb{f\p{Z}}
\qquad \text{for all integrable } f.
\]
For a finite set of basis functions $\cb{f_j}^p_{j=1}$, the empirical analogue is
\begin{align}
\frac{1}{n_0}\sum^{n_0}_{i=1}\p{\widehat r\p{Z_i}f_j\p{Z_i}}
\approx
\frac{1}{n_1}\sum^{n_1}_{j=1}\p{f_j\p{Z^e_j}},
\qquad j=1,\dots,p,
\label{eq:appdx_dr_moment_matching}
\end{align}
which is the moment matching interpretation in the density-ratio literature \citep{Sugiyama2007covariateshift,Sugiyama2011densityratio} and is closely related to kernel mean matching \citep{Gretton2009covariateshift}.

Generalized Riesz regression for covariate shift (Section~\ref{sec:application}) is an ERM for $\widehat r$ under a Bregman objective. When the dual coordinate is modeled linearly, the same KKT logic implies approximate versions of \eqref{eq:appdx_dr_moment_matching}, with the tolerance controlled by $\lambda$, and, in nonconvex implementations, also by optimization residuals.

\subsection{Loss--Link Recipes That Guarantee Dual Linearity}
\label{appdx:loss_link_recipes}

This subsection complements Section~\ref{subsec:loss_link_pairs} by collecting the derivative and inverse maps used in the compatible loss--link pairs. The link is chosen so that $\partial g\p{\alpha_\bmbeta(x)}$ is a linear index. In signed-representer problems, a known map $\xi(x)\in\cb{0,1}$ determines which component of the domain is used at $x$.

\paragraph{SQ-Riesz.}
Let $g^{\mathrm{SQ}}\p{\alpha}=\p{\alpha-C}^2$, so $\partial g\p{\alpha}=2\p{\alpha-C}$ and $\p{\partial g}^{-1}\p{v}=C+v/2$. Hence, the affine link
\[
\alpha_\bmbeta\p{x}=C+\frac{1}{2}\bmphi\p{x}^\top\bmbeta
\]
implies $\partial g\circ\alpha_\bmbeta=\bmphi^\top\bmbeta$ and therefore $\widetilde\phi_j=\phi_j$ in Theorem~\ref{thm:autocovariance}.

\paragraph{UKL-Riesz.}
Let $g^{\mathrm{UKL}}\p{\alpha}=\p{\abs{\alpha}-C}\log\p{\abs{\alpha}-C}-\abs{\alpha}$ on $\cb{\abs{\alpha}>C}$. Its derivative is
\[
\partial g\p{\alpha}=\sign\p{\alpha}\log\p{\abs{\alpha}-C},
\]
and its inverse on the positive and negative components is
\[
\p{\partial g_+}^{-1}\p{v}=C+\exp\p{v},
\qquad
\p{\partial g_-}^{-1}\p{v}=-C-\exp\p{-v}.
\]
It follows that the sign-specific log link
\[
\alpha_\bmbeta\p{x}
=
\xi\p{x}\Bigp{C+\exp\p{\bmphi\p{x}^\top\bmbeta}}
-
\p{1-\xi\p{x}}\Bigp{C+\exp\p{-\bmphi\p{x}^\top\bmbeta}}
\]
satisfies $\partial g\circ\alpha_\bmbeta=\bmphi^\top\bmbeta$ on both components of the domain. A sigmoid propensity model also induces this UKL-compatible link for the ATE. Appendix~\ref{appdx:sigmoid_implies_ukl} gives the calculation, following the estimand-driven argument of \citet{Zhao2019covariatebalancing}.

\paragraph{BP-Riesz.}
Let $\omega\in\p{0,\infty}$ and $k\coloneqq1+1/\omega$. The inverse derivative on the positive and negative components is
\[
\p{\partial g_+}^{-1}\p{v}=C+\Bigp{1+\frac{v}{k}}^{1/\omega},
\qquad
\p{\partial g_-}^{-1}\p{v}=-C-\Bigp{1-\frac{v}{k}}^{1/\omega},
\]
on the respective domains. Therefore, the power link
\[
\alpha_\bmbeta\p{x}
=
\xi\p{x}\Biggcb{C+\Bigp{1+\frac{\bmphi\p{x}^\top\bmbeta}{k}}^{1/\omega}}
-
\p{1-\xi\p{x}}\Biggcb{C+\Bigp{1-\frac{\bmphi\p{x}^\top\bmbeta}{k}}^{1/\omega}}
\]
also makes the dual coordinate linear and yields the KKT balancing equations.

\paragraph{BKL-Riesz and the logistic plug-in.}
A BKL generator can be paired with a compatible representer link, in which case the same KKT argument gives balancing. See Remark~\ref{rem:automaticbklriesz}. However, the Bernoulli likelihood with a sigmoid propensity link is a separate plug-in method. The induced ATE representer is given by Equation~\eqref{eq:alpha_sigmoid_loglink}, and its BKL dual coordinate is not linear in the logistic index. The direct compatible BKL specification and the logistic propensity-score plug-in are therefore distinct. Appendix~\ref{appdx:sigmoid_implies_ukl} gives the distinction.

\subsection{Practical Summary}
\label{appdx:practical_summary_balance}

The central message of Section~\ref{sec:automaticcovariatebalancing} and this appendix is:
\begin{itemize}
\item Automatic balancing is a KKT phenomenon that appears when we parameterize the dual coordinate $u=\partial g\circ\alpha$ linearly.
\item Classical covariate balance in ATE and moment matching in density ratio estimation are special cases obtained by choosing application-specific sets of basis functions.
\item Exact balance is a property of the observations used to solve an unregularized representer problem. Regularization, active parameter constraints, numerical error, or evaluation on other observations can make the empirical imbalance nonzero. The quantity $\max_j\abs{\widehat\Delta_j\p{\widehat\alpha}}$ records that discrepancy.
\item Loss choice matters because it determines which dual coordinate is linear and therefore which link preserves the KKT balancing mechanism, for example, sigmoid propensity and ATE naturally pair with UKL.
\end{itemize}

\section{Connections to \texorpdfstring{\citet{Wong2017kernelbased}}{wong} and \texorpdfstring{\citet{Hirano2003efficientestimation}}{hirano}}
\label{appdx:wong-hir}
This appendix collects formal statements related to the discussion in Section~\ref{sec:convanalysis} and focuses on the ATE specialization to keep notation concrete.

\subsection{Setup and Efficient Influence Function for ATE}
Let $W\coloneqq(Y,D,Z)$, where $D\in\cb{0,1}$ and $Z\in\calZ$, and let $Y(1),Y(0)$ be potential outcomes. Under unconfoundedness and overlap, we define the ATE by
\[
\theta_0 \coloneqq \bbE\sqb{Y(1)-Y(0)}.
\]
Let $\mu_d(z)\coloneqq \bbE\sqb{Y\mid D=d,Z=z}$ and $e_0(z)\coloneqq P(D=1\mid Z=z)$.
The efficient influence function is
\[
\psi^{\text{ATE}}(W;\eta_0,\theta_0)
\coloneqq
\mu_1(Z) - \mu_0(Z)
+
\frac{D}{e_0(Z)}\bigp{Y-\mu_1(Z)}
-
\frac{1-D}{1-e_0(Z)}\bigp{Y-\mu_0(Z)}
-
\theta_0,
\]
and the semiparametric efficiency bound equals $V^*\coloneqq \bbE\sqb{\psi^{\text{ATE}}(W;\eta_0,\theta_0)^2}$.

\subsection{A Balancing-to-Orthogonalization Identity}
Let $\bmphi\colon\calX\to\bbR^p$ and suppose $\gamma_0$ lies in its linear span.
If an estimated weight function $\widehat{\alpha}$ satisfies exact sample balancing
\[
\frac{1}{n}\sum^n_{i=1}\widehat{\alpha}(X_i)\phi_j(X_i)
=
\frac{1}{n}\sum^n_{i=1} m\bigp{W_i,\phi_j},
\qquad
j=1,\dots,p,
\]
then by linearity of $m(W,\gamma)$ in $\gamma$,
\[
\frac{1}{n}\sum^n_{i=1}m(W_i,\gamma_0)
=
\frac{1}{n}\sum^n_{i=1}\widehat{\alpha}(X_i)\gamma_0(X_i),
\]
which implies the exact decomposition
\[
\frac{1}{n}\sum^n_{i=1}\widehat{\alpha}(X_i)Y_i
=
\frac{1}{n}\sum^n_{i=1}\Bigp{m(W_i,\gamma_0)+\widehat{\alpha}(X_i)\bigp{Y_i-\gamma_0(X_i)}}.
\]
This is the formal content behind Theorem~\ref{thm:automaticneymanorthogonalization} in the ATE specialization.

\subsection{A Representative Efficiency Theorem of the IPW Estimator}
Theorem statements in \citet{Hirano2003efficientestimation} are given under sieve estimation of $e_0$ with regularity conditions on the sieve dimension and smoothness of $e_0$.
A representative formulation is as follows.

\begin{theorem}[Representative formulation of the efficiency result in \citet{Hirano2003efficientestimation}]
\label{thm:hir_rep}
Assume overlap and regularity conditions for a sieve estimator $\widehat{e}$ of $e_0$ such that $\|\widehat{e}-e_0\|_{L_2(P_0)}=o_p(1)$ and the sieve dimension grows slowly enough.
Define the IPW estimator
\[
\widehat{\theta}^{\text{IPW}}
\coloneqq
\frac{1}{n}\sum^n_{i=1}\Bigp{\frac{D_iY_i}{\widehat{e}(Z_i)}-\frac{(1-D_i)Y_i}{1-\widehat{e}(Z_i)}}.
\]
Then, we have
\[
\sqrt{n}\Bigp{\widehat{\theta}^{\text{IPW}}-\theta_0}
\xrightarrow{\rmd}
\calN\p{0,V^*},
\]
where $V^*$ is the semiparametric efficiency bound for ATE.
\end{theorem}

The main message is that, under appropriate first-step estimation, IPW based on an estimated propensity score can achieve semiparametric efficiency.

\subsection{Relation to Kernel Balancing Approaches}
Kernel balancing approaches can be interpreted as constructing weights that approximately satisfy balancing restrictions over an RKHS feature space.
When the balancing space is rich enough to approximate the outcome regression, the induced estimator can be shown to be asymptotically equivalent to an estimator with an orthogonal-score representation.
This is consistent with Theorem~\ref{thm:automaticneymanorthogonalization} and the general identity above, with exact balancing replaced by an approximation argument plus a convergence-rate control of the imbalance.

\section{KKT Conditions as Bregman Projections}
\label{appdx:kkt_riesz_linear_equation}
In this section, we show how the KKT first-order conditions in generalized Riesz regression coincide with the characterization of a sieve Riesz representer as the solution to a linear equation in a Hilbert space discussed in \citet{Chen2015sievesemiparametric} and \citet{Chen2015sievewald}. These works show that the Riesz representer can be formulated via a linear equation in semiparametric generalized method of moments (GMM) and efficiency analysis.

\subsection{Riesz Representer as a Linear Equation in a Hilbert Space}
Let $\calH\coloneqq L_2\p{P_X}$ with inner product $\langle f,g\rangle \coloneqq \bbE\sqb{f\p{X}g\p{X}}$. For the linear map $\gamma\mapsto \bbE\sqb{m\p{W,\gamma}}$ (Section~\ref{sec:setup}), the Riesz representation theorem yields $\alpha_0\in\calH$ such that
\begin{align}
\bbE\sqb{m\p{W,\gamma}}=\langle \alpha_0,\gamma\rangle
\qquad \forall \gamma\in\calH.
\end{align}
If we restrict to a finite-dimensional sieve space $\calH_p\coloneqq \mathrm{span}\cb{\phi_1,\ldots,\phi_p}$, the sieve Riesz representer $\alpha_p\in\calH_p$ is the unique element satisfying
\begin{align}
\langle \alpha_p,\phi_j\rangle = \bbE\sqb{m\p{W,\phi_j}}
\qquad j=1,\ldots,p.
\label{eq:riesz_sieve_equations2}
\end{align}
Writing $\alpha_p\p{x}=\bmphi\p{x}^\top\bmbeta$ with $\bmphi\coloneqq \p{\phi_1,\ldots,\phi_p}^\top$, \eqref{eq:riesz_sieve_equations2} becomes the linear system
\begin{align}
\underbrace{\bbE\sqb{\bmphi\p{X}\bmphi\p{X}^\top}}_{=:G}\bmbeta
=
\underbrace{\bbE\sqb{m\p{W,\bmphi}}}_{=:b}.
\label{eq:normal_equation_riesz2}
\end{align}

\subsection{Bregman Objectives and Dual Coordinates}
Recall the pointwise Bregman divergence
\[
\mathrm{BD}^\dagger_g\p{\alpha_0\p{x}\mid \alpha\p{x}}
\coloneqq
g\p{\alpha_0\p{x}}-g\p{\alpha\p{x}}-\partial g\p{\alpha\p{x}}\p{\alpha_0\p{x}-\alpha\p{x}},
\]
and let $\alpha^*$ minimize its expectation over a convex set $\calH$. For a feasible direction $h$, differentiation with respect to the candidate representer gives
\[
\left.\frac{\rmd}{\rmd t}\bbE\sqb{\mathrm{BD}^\dagger_g\p{\alpha_0\p{X}\mid\alpha^*\p{X}+t h\p{X}}}\right|_{t=0}
=
\Big\langle g''\p{\alpha^*}\p{\alpha^*-\alpha_0},h\Big\rangle.
\]
Consequently, the first-order condition for the reverse Bregman divergence is
\[
\Big\langle g''\p{\alpha^*}\p{\alpha^*-\alpha_0},\alpha-\alpha^*\Big\rangle\geq0
\qquad\text{for every }\alpha\in\calH.
\]
It becomes an equality along two-sided feasible directions.

For each $x$, let $\calA_x$ be the open interval allowed by the representer model and define
\[
g_x^*(v)\coloneqq\sup_{a\in\calA_x}\{av-g(a)\}.
\]
Set $u\p{x}=\partial g\p{\alpha\p{x}}$ and $\calU=\cb{\partial g\circ\alpha:\alpha\in\calH}$. The Fenchel--Young identity gives $g_x^*(u)=\alpha u-g\p{\alpha}$ at $u=\partial g\p{\alpha}$, so the population objective is, up to an additive constant,
\[
Q^*\p{u}
=
\bbE\sqb{g_X^*\p{u\p{X}}-m\p{W,u}}
=
\bbE\sqb{g_X^*\p{u\p{X}}-\alpha_0\p{X}u\p{X}}.
\]
If $\calU$ is convex and $u^*=\partial g\circ\alpha^*$ minimizes $Q^*$, then it holds that
\begin{align}
\Big\langle \alpha^*-\alpha_0,u-u^*\Big\rangle\geq0
\qquad\text{for every }u\in\calU.
\label{eq:bregman_orthogonality2}
\end{align}
Equality holds along two-sided feasible directions. The primal condition is therefore curvature weighted, whereas Equation~\eqref{eq:bregman_orthogonality2} is expressed in the dual coordinate.

\paragraph{Finite-dimensional models and KKT.}
Consider
\[
u_\bmbeta\p{x}=\bmphi\p{x}^\top\bmbeta,
\qquad
\alpha_\bmbeta\p{x}=\Bigp{\partial g\big|_{\calA_x}}^{-1}\p{u_\bmbeta\p{x}}.
\]
The penalized empirical objective is
\[
\widehat\bmbeta
\in
\arg\min_{\bmbeta\in\bbR^p}
\Biggcb{
\frac1n\sum_{i=1}^n g_{X_i}^*\p{u_\bmbeta\p{X_i}}
-
\frac1n\sum_{i=1}^n m\p{W_i,u_\bmbeta}
+
\frac{\lambda}{q}\norm{\bmbeta}_q^q
}.
\]
Because $\partial g_x^*\p{u}=\alpha$ on the allowed interval, the KKT conditions are
\begin{align}
\frac1n\sum_{i=1}^n
\Bigp{\widehat\alpha\p{X_i}\phi_j\p{X_i}-m\p{W_i,\phi_j}}
\in
-\lambda\partial\Bigp{\frac1q|\beta_j|^q},
\qquad j=1,\ldots,p,
\label{eq:kkt_general}
\end{align}
where $\widehat\alpha=\alpha_{\widehat\bmbeta}$. Hence, when $\lambda=0$ and no parameter constraint is active, the KKT conditions reduce to the empirical sieve Riesz equations
\begin{align}
\frac{1}{n}\sum_{i=1}^n\widehat\alpha\p{X_i}\phi_j\p{X_i}
=
\frac{1}{n}\sum_{i=1}^n m\p{W_i,\phi_j},
\qquad j=1,\ldots,p.
\label{eq:kkt_equals_riesz}
\end{align}
With regularization or an active parameter constraint, the same equations hold with the corresponding KKT terms, as stated in Equation~\eqref{eq:kkt_normal_cone}. Thus, the empirical Riesz equations can be common across compatible loss--link pairs, although the objective and regularization determine which representer is selected when those equations do not identify a unique solution.

\begin{remark}[Derivation of Equation~\eqref{eq:bregman_orthogonality2}]
\label{rem:derivationbregman}
For the reverse Bregman divergence, the directional derivative along $\alpha+t h$ is
\[
\left.\frac{\rmd}{\rmd t}\bbE\sqb{\mathrm{BD}^\dagger_g\p{\alpha_0\p{X}\mid\alpha\p{X}+t h\p{X}}}\right|_{t=0}
=
\Big\langle g''\p{\alpha}\p{\alpha-\alpha_0},h\Big\rangle.
\]
If $\alpha^*$ minimizes the objective over a convex set $\calH$, then the one-sided derivative along $\alpha-\alpha^*$ is nonnegative. Equivalently, with
\[
N_{\calH}\p{\alpha^*}
\coloneqq
\cb{v:\langle v,\alpha-\alpha^*\rangle\leq0\text{ for every }\alpha\in\calH},
\]
we have $g''\p{\alpha^*}\p{\alpha_0-\alpha^*}\in N_{\calH}\p{\alpha^*}$.

Now set $u=\partial g\circ\alpha$. Since $\partial g_x^*\p{u}=\Bigp{\partial g\big|_{\calA_x}}^{-1}\p{u}=\alpha$, the directional derivative of
\[
Q^*\p{u}=\bbE\sqb{g_X^*\p{u\p{X}}-\alpha_0\p{X}u\p{X}}
\]
along $h$ is $\langle\alpha-\alpha_0,h\rangle$. Applying the one-sided first-order condition to the path $u_t=u^*+t\p{u-u^*}$ gives Equation~\eqref{eq:bregman_orthogonality2}. If the same direction is feasible for positive and negative $t$, the inner product equals zero.
\end{remark}

\subsection{SQ-Riesz With Linear Link and an \texorpdfstring{$L_2$}{L2} Projection}
Take $g^{\mathrm{SQ}}\p{\alpha}=\p{\alpha-C}^2$, so $\partial g\p{\alpha}=2\p{\alpha-C}$ and $\p{\partial g}^{-1}\p{u}=C+u/2$. Under the dual linear specification $u_\bmbeta\p{X}=\bmphi\p{X}^\top\bmbeta$, the primal model is
\begin{align}
\alpha_\bmbeta\p{X}=C+\frac12\bmphi\p{X}^\top\bmbeta.
\label{eq:sq_linear_link}
\end{align}
When $\lambda=0$, the KKT equations give
\begin{align}
\bbE\sqb{\bmphi\p{X}\bmphi\p{X}^\top}\bmbeta
=
2\bbE\sqb{m\p{W,\bmphi}}-2C\bbE\sqb{\bmphi\p{X}}.
\label{eq:sq_normal_equations}
\end{align}
Moreover, $u-u^*=2\p{\alpha-\alpha^*}$, so Equation~\eqref{eq:bregman_orthogonality2} reduces to the usual $L_2\p{P_X}$ projection condition. SQ-Riesz with a linear link is therefore an $L_2$ projection of $\alpha_0$ onto the affine sieve model.

\subsection{UKL-Riesz With Exponential and Logit Links}
For KL-type losses, the same projection geometry holds, but in the dual coordinate $u=\partial g\p{\alpha}$.

Consider the UKL generator separately on the positive and negative parts of the domain, shifted to avoid singularities and defined for $\abs{\alpha}>C$:
\[
g^{\mathrm{UKL}}\p{\alpha}
=
\p{\abs{\alpha}-C}\log\p{\abs{\alpha}-C}-\abs{\alpha},
\qquad
\partial g\p{\alpha}=\sign\p{\alpha}\log\p{\abs{\alpha}-C}.
\]
Let $\xi\p{X}\in\cb{0,1}$ be a known map that determines the sign of $\alpha_{\bmbeta}\p{X}$, for example, $\xi\p{X}=D$ in ATE, and impose the dual linear model
\begin{align}
u_{\bmbeta}\p{X}
=
\partial g\p{\alpha_{\bmbeta}\p{X}}
=
\bmphi\p{X}^\top\bmbeta.
\label{eq:ukl_dual_linear}
\end{align}
Inverting $\partial g$ on the interval determined by $\xi(X)$ yields the exponential form
\begin{align}
\alpha_{\bmbeta}\p{X}
=
\xi\p{X}\Bigp{C+\exp\p{\bmphi\p{X}^\top\bmbeta}}
-
\p{1-\xi\p{X}}\Bigp{C+\exp\p{-\bmphi\p{X}^\top\bmbeta}}.
\label{eq:ukl_exp_link}
\end{align}
Despite the nonlinearity in $\bmbeta$, the KKT conditions remain linear in the test functions. For $\lambda=0$, they are exactly the sieve Riesz equations \eqref{eq:kkt_equals_riesz}. Hence, UKL-Riesz returns the Bregman projection solution subject to the same Riesz linear equations that define the representer on the sieve.

\subsection{BKL-Riesz Regression With Logit Link}
For the BKL generator, again on $\abs{\alpha}>C$,
\begin{align*}
g^{\mathrm{BKL}}(\alpha)
&=(|\alpha|-C)\log(|\alpha|-C)-(|\alpha|+C)\log(|\alpha|+C),\\
\partial g(\alpha)
&=\sign(\alpha)\log\left(\frac{|\alpha|-C}{|\alpha|+C}\right).
\end{align*}
We impose the dual linear model $u_\bmbeta(X)=\bmphi(X)^\top\bmbeta$ on the interval determined by $\xi$. Inverting $\partial g$ gives a logit-type link for $|\alpha_\bmbeta|$, while $\xi$ determines the sign. The KKT conditions are again Equations~\eqref{eq:kkt_general}--\eqref{eq:kkt_equals_riesz}. For ATE, this is a direct BKL-Riesz specification only when the representer link makes the BKL dual coordinate linear. It is distinct from the Bernoulli likelihood used by the logistic propensity-score plug-in.

\section{Compatibility Between Sigmoid Propensity Model and UKL-Riesz}
\label{appdx:sigmoid_implies_ukl}

This section applies the automatic regressor balancing result in Section~\ref{sec:automaticcovariatebalancing} to a sigmoid propensity-score model. The induced ATE representer has the log link associated with UKL-Riesz, as in the estimand-driven loss construction of \citet{Zhao2019covariatebalancing}. However, replacing the loss while retaining the same link generally removes the linear dual coordinate required for the ATE balance equations.

\subsection{Compatibility Between Loss Choice and Covariate Balancing for the Target Estimand}
\citet{Zhao2019covariatebalancing} emphasizes that the loss used to estimate weights or a propensity score should be tied to the balance equations for the target estimand. In ATE estimation, different losses paired with the same logistic propensity model can correspond to different target weightings. Consequently, only particular loss--link pairs deliver the ATE balance equations under that propensity specification.

Generalized Riesz regression makes this condition explicit: automatic covariate balancing arises only when the loss generator $g$ and the link function are paired so that $\partial g\p{\alpha_{\bmbeta}\p{X}}$ is linear in the features used in the index (Theorem~\ref{thm:autocovariance} and Corollary~\ref{cor:automaticcovariate}).

\subsection{Sigmoid Propensity Model and a Log Link Function}
Consider the logistic propensity score model
\[
e_{\bmbeta}\p{Z}
\coloneqq
\Lambda\p{\eta_{\bmbeta}\p{Z}},
\qquad
\eta_{\bmbeta}\p{Z}\coloneqq \bmphi\p{Z}^\top\bmbeta,
\qquad
\Lambda\p{t}\coloneqq \frac{1}{1+\exp\p{-t}}.
\]
Then, the inverse-propensity components satisfy
\begin{align*}
r_{\bmbeta}\p{1,Z}
\coloneqq
\frac{1}{e_{\bmbeta}\p{Z}}
&=
1+\exp\p{-\eta_{\bmbeta}\p{Z}},
\\
r_{\bmbeta}\p{0,Z}
\coloneqq
\frac{1}{1-e_{\bmbeta}\p{Z}}
&=
1+\exp\p{\eta_{\bmbeta}\p{Z}}.
\end{align*}
Therefore, the induced ATE Riesz representer model
\[
\alpha_{\bmbeta}^{\mathrm{ATE}}\p{D,Z}
\coloneqq
\frac{D}{e_{\bmbeta}\p{Z}}-\frac{1-D}{1-e_{\bmbeta}\p{Z}}
\]
can be written in the sign-specific exponential form
\begin{align}
\alpha_{\bmbeta}^{\mathrm{ATE}}\p{D,Z}
&=
D\Bigp{1+\exp\p{-\eta_{\bmbeta}\p{Z}}}
-
\p{1-D}\Bigp{1+\exp\p{\eta_{\bmbeta}\p{Z}}}.
\label{eq:alpha_sigmoid_loglink}
\end{align}
This is the log-link Riesz representer specification described in Section~\ref{sec:automaticcovariatebalancing} with $\p{\xi,C}=\p{D,1}$. Moreover, Equation~\eqref{eq:alpha_sigmoid_loglink} implies the sign and domain restrictions
\[
\alpha_{\bmbeta}^{\mathrm{ATE}}\p{1,z}>1,
\qquad
\alpha_{\bmbeta}^{\mathrm{ATE}}\p{0,z}<-1,
\]
so the shifted domain $\abs{\alpha}>1$ is compatible with the shifted UKL/BKL generators used in Section~\ref{sec:generalizedrieszregression}.

\subsection{Automatic Covariate Balancing Under UKL-Riesz Regression}
The automatic covariate balancing theorem (Theorem~\ref{thm:autocovariance}) requires that $\partial g\p{\alpha_{\bmbeta}\p{X}}$ is linear in $\bmphi\p{X}^\top\bmbeta$, in the sense that it can be written as a linear combination of fixed feature transforms independent of $\bmbeta$.

For ATE with the sigmoid-induced model \eqref{eq:alpha_sigmoid_loglink}, consider the shifted UKL generator with $C=1$,
\[
g^{\mathrm{UKL}}\p{\alpha}
\coloneqq
\p{\abs{\alpha}-1}\log\p{\abs{\alpha}-1}-\abs{\alpha},
\qquad
\partial g\p{\alpha}
=
\sign\p{\alpha}\log\p{\abs{\alpha}-1}.
\]
Evaluate $\partial g$ at $\alpha_{\bmbeta}^{\mathrm{ATE}}\p{D,Z}$. Let $\eta=\eta_{\bmbeta}\p{Z}$.

\paragraph{Treated observations $(D=1)$.}
Then, it holds that $\alpha_{\bmbeta}^{\mathrm{ATE}}\p{1,Z}=1+\exp\p{-\eta}$, so $\abs{\alpha}-1=\exp\p{-\eta}$ and $\sign\p{\alpha}=1$. Hence, we have
\[
\partial g\Bigp{\alpha_{\bmbeta}^{\mathrm{ATE}}\p{1,Z}}
=
\log\p{\exp\p{-\eta}}
=
-\eta.
\]

\paragraph{Control observations $(D=0)$.}
Then, it holds that $\alpha_{\bmbeta}^{\mathrm{ATE}}\p{0,Z}=-\p{1+\exp\p{\eta}}$, so $\abs{\alpha}-1=\exp\p{\eta}$ and $\sign\p{\alpha}=-1$. Hence, we have
\[
\partial g\Bigp{\alpha_{\bmbeta}^{\mathrm{ATE}}\p{0,Z}}
=
-\log\p{\exp\p{\eta}}
=
-\eta.
\]

\paragraph{Key identity.}
The two cases yield the same linear index:
\begin{align}
\partial g\Bigp{\alpha_{\bmbeta}^{\mathrm{ATE}}\p{D,Z}}
=
-\eta_{\bmbeta}\p{Z}
=
-\bmphi\p{Z}^\top\bmbeta.
\label{eq:ukl_linear_index}
\end{align}
Thus, $\partial g\p{\alpha_{\bmbeta}^{\mathrm{ATE}}\p{X}}$ is linear in $\bmphi\p{Z}$. Therefore, Theorem~\ref{thm:autocovariance} and Corollary~\ref{cor:automaticcovariate} apply to the regressor functions used in the propensity index.

\subsection{Resulting Automatic Covariate Balancing}
Take $\ell_1$-penalized generalized Riesz regression for $\bmbeta$ as in Theorem~\ref{thm:autocovariance} and let $\widehat{\alpha}=\alpha_{\widehat{\bmbeta}}$. Because \eqref{eq:ukl_linear_index} makes $\partial g\p{\alpha_{\bmbeta}}$ linear in $\bmphi\p{Z}^\top\bmbeta$, the KKT conditions imply approximate balancing of the corresponding moments.

To see the standard ATE interpretation, suppose $\phi_j$ depends only on $Z$, as in standard propensity modeling, so that $m^{\mathrm{ATE}}\p{W,\phi_j}=0$. Then, Corollary~\ref{cor:automaticcovariate} yields, up to the penalty tolerance,
\begin{align}
\left|
\frac{1}{n}\sum^n_{i=1}\p{\widehat{\alpha}\p{D_i,Z_i}\phi_j\p{Z_i}}
\right|
\le
\lambda
\qquad \p{j=1,\dots,p},
\label{eq:ate_balance_sigmoid_ukl}
\end{align}
which is equivalent to the familiar treated versus control balancing condition
\[
\frac{1}{n}\sum^n_{i=1}\p{\frac{D_i}{\widehat e\p{Z_i}}\phi_j\p{Z_i}}
\approx
\frac{1}{n}\sum^n_{i=1}\p{\frac{1-D_i}{1-\widehat e\p{Z_i}}\phi_j\p{Z_i}},
\]
because $\widehat{\alpha}\p{D,Z}=D/\widehat e\p{Z}-\p{1-D}/\p{1-\widehat e\p{Z}}$. This is the covariate balancing condition that motivates the ATE-targeted loss in \citet{Zhao2019covariatebalancing}. It also agrees with the dual characterization of entropy balancing weights in Table~\ref{tbl:dre_rre}.

\subsection{Failure of Automatic Covariate Balancing Under Other Losses}
The key requirement behind automatic covariate balancing is loss--link compatibility: on each domain component, the link must equal the inverse map of $\partial g$ up to affine constants. When the propensity is parameterized by a sigmoid, the induced Riesz representer \eqref{eq:alpha_sigmoid_loglink} is of log-link form, which matches the inverse map of the UKL derivative (Section~\ref{sec:automaticcovariatebalancing}). If we keep the sigmoid model but replace the loss, this compatibility is broken and $\partial g\p{\alpha_{\bmbeta}}$ is no longer linear in the index.

The following two cases show the failure of dual linearity.

\paragraph{Squared loss (SQ-Riesz) + sigmoid propensity.}
With $g^{\mathrm{SQ}}\p{\alpha}=\p{\alpha-1}^2$ we have $\partial g\p{\alpha}=2\p{\alpha-1}$. Under \eqref{eq:alpha_sigmoid_loglink},
\[
\partial g\Bigp{\alpha_{\bmbeta}^{\mathrm{ATE}}\p{1,Z}}
=
2\exp\p{-\eta_{\bmbeta}\p{Z}},
\qquad
\partial g\Bigp{\alpha_{\bmbeta}^{\mathrm{ATE}}\p{0,Z}}
=
-2\Bigp{2+\exp\p{\eta_{\bmbeta}\p{Z}}}.
\]
These expressions are not linear in $\eta_{\bmbeta}\p{Z}=\bmphi\p{Z}^\top\bmbeta$, so the linearity condition in Theorem~\ref{thm:autocovariance} fails. Hence, the SQ-Riesz objective does not yield the ATE-style balancing equations \eqref{eq:ate_balance_sigmoid_ukl} when we insist on a sigmoid propensity model. Equivalently, to obtain balancing with squared loss, we must change the link to the linear link discussed in Section~\ref{sec:automaticcovariatebalancing}.

\paragraph{BKL generator with a sigmoid propensity plug-in.}
The BKL generator satisfies
\[
\partial g^{\mathrm{BKL}}\p{\alpha}
=
\sign\p{\alpha}\log\p{\frac{\abs{\alpha}-1}{\abs{\alpha}+1}}.
\]
Under \eqref{eq:alpha_sigmoid_loglink}, on the treated component $\abs{\alpha}-1=\exp\p{-\eta}$ but $\abs{\alpha}+1=2+\exp\p{-\eta}$, so
\begin{align*}
\partial g^{\mathrm{BKL}}\Bigp{\alpha_{\bmbeta}^{\mathrm{ATE}}(1,Z)}
&=\log\left(\frac{\exp\{-\eta_\bmbeta(Z)\}}{2+\exp\{-\eta_\bmbeta(Z)\}}\right)\\
&=-\eta_\bmbeta(Z)-\log\{2+\exp(-\eta_\bmbeta(Z))\}.
\end{align*}
This is not linear in $\eta_{\bmbeta}\p{Z}$. Therefore, evaluating the BKL generator at the sigmoid propensity plug-in does not satisfy the automatic covariate balancing conditions for the ATE. This conclusion agrees with the discussion in \citet{Zhao2019covariatebalancing}: within that tailored-loss family, the logistic likelihood corresponds to a different weighting and hence to a different estimand than the ATE (see also Section~\ref{sec:loss_model_bias_correction} in the main text).

\paragraph{BP-Riesz and other divergences.}
The same point applies more broadly: if we keep the sigmoid propensity link \eqref{eq:alpha_sigmoid_loglink}, then for $\omega\neq 0$ the BP derivative involves powers $\p{\abs{\alpha}-1}^\omega=\exp\p{\pm\omega\eta}$ and is not linear in $\eta$. Thus, BP-Riesz does not yield automatic balancing under the sigmoid link unless one also changes the link to the compatible power link in Section~\ref{sec:automaticcovariatebalancing}.

\section{Automatic Orthogonalization Under \texorpdfstring{$\ell_0$ and $\ell_1$}{ell0ell1} Penalties}
\label{app:auto_orth_l0_l1}

\subsection{\texorpdfstring{$\ell_0$}{ell0}-Penalized Generalized Riesz Regression}
\label{app:l0_penalty}

This subsection records the cleanest deterministic implication one can retain when replacing convex $\ell_a$ penalties, $a\ge 1$, by an $\ell_0$ penalty. The key point is that $\ell_0$ is nonconvex and discontinuous at $0$, so the global KKT-style imbalance bounds available under $\ell_1$, or $\ell_a$, regularization do not directly carry over.

Let the differentiable empirical objective be
\[
\mathcal L_n\p{\beta}\coloneqq \widehat{\mathrm{BD}}_g\p{\alpha_\beta},
\]
and consider the $\ell_0$-penalized estimator
\begin{align}
\widehat\beta^{\p{\ell_0}}
\in
\arg\min_{\beta\in\bbR^p}
\cb{\mathcal L_n\p{\beta}+\lambda\norm{\beta}_0},
\qquad
\norm{\beta}_0\coloneqq \abs{\cb{j:\beta_j\neq 0}}.
\label{eq:l0_penalized_problem}
\end{align}
Write the selected active set as $\widehat S\coloneqq \cb{j:\widehat\beta^{\p{\ell_0}}_j\neq 0}$.

\begin{proposition}[Active-set exact balance under an $\ell_0$ penalty]
\label{prop:l0_active_balance}
Suppose the dual linearity condition holds, that is,
\[
\p{\partial g}\p{\alpha_\beta\p{x}}
=
\sum^p_{j=1}\beta_j\widetilde\phi_j\p{x}
\qquad \text{for all relevant } x,
\]
and $\mathcal L_n\p{\beta}$ is differentiable in a neighborhood of $\widehat\beta^{\p{\ell_0}}$ restricted to the coordinates in $\widehat S$. Assume also that $\widehat\beta^{\p{\ell_0}}$ is a coordinatewise local minimizer of \eqref{eq:l0_penalized_problem} in the sense that for each $j\in\widehat S$ there exists $\varepsilon_j>0$ such that $\widehat\beta^{\p{\ell_0}}_j+t\neq 0$ for all $\abs{t}<\varepsilon_j$, and the map $t\mapsto \mathcal L_n\p{\widehat\beta^{\p{\ell_0}}+t e_j}$ has a local minimum at $t=0$.

Then, the fitted representer $\widehat\alpha^{\p{\ell_0}}\coloneqq \alpha_{\widehat\beta^{\p{\ell_0}}}$ satisfies exact training-sample balance on the active set:
\begin{align}
\widehat\Delta_j\p{\widehat\alpha^{\p{\ell_0}}}=0
\qquad \text{for all } j\in\widehat S,
\label{eq:l0_active_balance}
\end{align}
where $\widehat\Delta_j\p{\cdot}$ is the imbalance functional defined in \eqref{eq:imbalance_def_rewrite}. Consequently, for any $\gamma$ in the reduced span $\mathrm{span}\cb{\widetilde\phi_j:j\in\widehat S}$,
\[
\frac{1}{n}\sum^n_{i=1}\p{\widehat\alpha^{\p{\ell_0}}\p{X_i}\gamma\p{X_i}}
=
\frac{1}{n}\sum^n_{i=1}\p{m\p{W_i,\gamma}}.
\]
\end{proposition}

\begin{proof}
Fix $j\in\widehat S$. By the assumption that $\widehat\beta^{\p{\ell_0}}_j+t$ remains nonzero for small $t$, the penalty term $\lambda\norm{\widehat\beta^{\p{\ell_0}}+t e_j}_0$ is constant for $\abs{t}<\varepsilon_j$. Therefore, for such $t$ the local optimality of $\widehat\beta^{\p{\ell_0}}$ along coordinate $j$ reduces to local optimality of $t\mapsto \mathcal L_n\p{\widehat\beta^{\p{\ell_0}}+t e_j}$ at $t=0$. Differentiability then implies the first-order condition
\[
\left.\frac{\partial}{\partial \beta_j}\mathcal L_n\p{\beta}\right|_{\beta=\widehat\beta^{\p{\ell_0}}}=0.
\]

Under the dual linearity specification, the same calculation as in the $\ell_a$-penalized KKT derivation shows that $\partial\mathcal L_n\p{\beta}/\partial \beta_j=\widehat\Delta_j\p{\alpha_\beta}$. Therefore, $\widehat\Delta_j\p{\widehat\alpha^{\p{\ell_0}}}=0$ holds for every $j\in\widehat S$, which proves Equation~\eqref{eq:l0_active_balance}.

The final display follows by the same linear-span argument as in Theorem~\ref{thm:automaticneymanorthogonalization}, restricted to the active set of basis functions $\cb{\widetilde\phi_j:j\in\widehat S}$.
\end{proof}

\begin{remark}[On the $\ell_0$ penalty]
Proposition~\ref{prop:l0_active_balance} only yields balance on the selected coordinates. There is generally no uniform imbalance bound of the form $\max_{1\le j\le p}\abs{\widehat\Delta_j\p{\widehat\alpha^{\p{\ell_0}}}}\le \lambda$, because such bounds rely on convex subgradient and KKT structure. Thus, compared to $\ell_1$ regularization, $\ell_0$ trades a global, all-features, balancing control for an active-set, selected-features, balancing statement.
\end{remark}

\subsection{\texorpdfstring{$\ell_1$}{ell1}-Penalized Generalized Riesz Regression}
\label{app:l1_not_sparse}
In many regression contexts $\ell_1$ is introduced primarily to encourage sparsity. In balancing-weight and generalized Riesz regression contexts, an equally important interpretation is:
\begin{itemize}
\item $\ell_1$ regularization provides a direct, featurewise, handle on the worst-case imbalance through the KKT system, yielding bounds of the form $\abs{\widehat\Delta_j\p{\widehat\alpha}}\le \lambda$ simultaneously for all $j$ when the score is linear in $\beta$.
\item This connects $\lambda$ to a bias--variance, balance--stability, tradeoff, and to feasibility relaxation when exact balance is not attainable or leads to extreme weights.
\end{itemize}
In short, in this paper, $\ell_1$ can be used even if one does not believe $\beta$ is sparse. Sparsity may occur as a side-effect, but the primary role is to control imbalance in a convex and computationally stable way.

\section{Additional Applications}
\label{sec:additionalapplications}

\subsection{ATT Estimation}
\label{appdx:att_estimation}
\paragraph{Setup.}
We observe $W\coloneqq (Y,D,Z)$, where $D\in\cb{0,1}$ is a binary treatment and $Z$ is a vector of covariates.
Let $Y(1)$ and $Y(0)$ be the potential outcomes.
We target the average treatment effect on the treated
\[
\theta^{\text{ATT}}_0 \coloneqq \bbE\sqb{Y(1)-Y(0)\mid D=1}.
\]
Assume unconfoundedness and overlap, $(Y(1),Y(0))$ is independent of $D$ given $Z$, and $0<e_0(Z)<1$ almost surely, where $e_0(z)\coloneqq \bbP\p{D=1\mid Z=z}$.
The efficient influence-function representation for ATT is standard, see \citet{Hahn1998ontherole}.
Let $\pi_1\coloneqq \bbP\p{D=1}$.

\paragraph{Neyman orthogonal score and Riesz representer.}
Let $X\coloneqq (D,Z)$ and $\gamma_0(d,z)\coloneqq \bbE\sqb{Y\mid D=d,Z=z}$.
Then, we have
\[
\theta^{\text{ATT}}_0
=
\bbE\sqb{\frac{D}{\pi_1}\p{\gamma_0(1,Z)-\gamma_0(0,Z)}}.
\]
Define the linear functional
\[
m^{\text{ATT}}\p{W,\gamma}
\coloneqq
\frac{D}{\pi_1}\p{\gamma(1,Z)-\gamma(0,Z)}.
\]
The corresponding Riesz representer is
\[
\alpha^{\text{ATT}}_0(D,Z)
\coloneqq
\frac{D}{\pi_1}
-
\frac{1-D}{\pi_1}\frac{e_0(Z)}{1-e_0(Z)}.
\]
The Neyman-orthogonal score is
\[
\psi^{\text{ATT}}\p{W;\eta,\theta}
\coloneqq
m^{\text{ATT}}\p{W,\gamma}
+
\alpha^{\text{ATT}}(X)\p{Y-\gamma(X)}
-
\theta,
\]
with $\eta^{\text{ATT}}\coloneqq \p{\gamma,\alpha^{\text{ATT}}}$ and $\eta^{\text{ATT}}_0\coloneqq \p{\gamma_0,\alpha^{\text{ATT}}_0}$.
In implementation, we replace $\pi_1$ by $\widehat{\pi}_1\coloneqq \frac{1}{n}\sum^n_{i=1}D_i$ inside $m^{\text{ATT}}$.

\paragraph{SQ-Riesz regression.}
With $g^{\text{SQ}}\p{a}\coloneqq a^2$, the empirical Bregman objective specializes to
\[
\widehat{\text{BD}}_{g^{\text{SQ}}}\p{\alpha}
=
\frac{1}{n}\sum^n_{i=1}
\p{
\alpha(D_i,Z_i)^2
-
\frac{2D_i}{\widehat{\pi}_1}\p{\alpha(1,Z_i)-\alpha(0,Z_i)}
}.
\]
Minimizing $\widehat{\text{BD}}_{g^{\text{SQ}}}\p{\alpha}+\lambda J\p{\alpha}$ over a function class $\calH$ yields $\widehat{\alpha}^{\text{ATT}}$.

\paragraph{UKL-Riesz regression.}
With $g^{\mathrm{UKL}}$ in Section~\ref{sec:empbalancing}, define
\[
f_\alpha(x)\coloneqq\sign(\alpha(x))\log(|\alpha(x)|-C).
\]
The empirical objective is
\begin{align*}
\widehat{\text{BD}}_{g^{\text{UKL}}}\p{\alpha}
={}&\frac{1}{n}\sum^n_{i=1}
\Bigp{C\log\p{|\alpha(D_i,Z_i)|-C}+|\alpha(D_i,Z_i)|}\\
&-\frac{1}{n}\sum^n_{i=1}\frac{D_i}{\widehat{\pi}_1}
\p{f_\alpha(1,Z_i)-f_\alpha(0,Z_i)}.
\end{align*}

\paragraph{BKL-Riesz regression.}
With $g^{\mathrm{BKL}}$ defined in the BKL-Riesz regression subsection, define
\[
h_\alpha(x)\coloneqq\sign(\alpha(x))\log\left(\frac{|\alpha(x)|-C}{|\alpha(x)|+C}\right).
\]
Then, we have
\begin{align*}
\widehat{\text{BD}}_{g^{\text{BKL}}}\p{\alpha}
={}&\frac{C}{n}\sum^n_{i=1}\log\p{|\alpha(D_i,Z_i)|^2-C^2}\\
&-\frac{1}{n}\sum^n_{i=1}\frac{D_i}{\widehat{\pi}_1}
\p{h_\alpha(1,Z_i)-h_\alpha(0,Z_i)}.
\end{align*}

\subsection{ATC Estimation}
\paragraph{Setup.}
We use the same setup as in the previous subsection, but now target the average treatment effect on the controls
\[
\theta^{\text{ATC}}_0 \coloneqq \bbE\sqb{Y(1)-Y(0)\mid D=0}.
\]
Let $\pi_0\coloneqq \bbP\p{D=0}$.

\paragraph{Neyman orthogonal score and Riesz representer.}
Let $X\coloneqq (D,Z)$ and $\gamma_0(d,z)\coloneqq \bbE\sqb{Y\mid D=d,Z=z}$.
Write
\[
\theta^{\text{ATC}}_0
=
\bbE\sqb{\frac{1-D}{\pi_0}\p{\gamma_0(1,Z)-\gamma_0(0,Z)}}.
\]
Define
\[
m^{\text{ATC}}\p{W,\gamma}
\coloneqq
\frac{1-D}{\pi_0}\p{\gamma(1,Z)-\gamma(0,Z)}.
\]
The Riesz representer is
\[
\alpha^{\text{ATC}}_0(D,Z)
\coloneqq
\frac{D}{\pi_0}\frac{1-e_0(Z)}{e_0(Z)}
-
\frac{1-D}{\pi_0}.
\]
The orthogonal score is
\[
\psi^{\text{ATC}}\p{W;\eta,\theta}
\coloneqq
m^{\text{ATC}}\p{W,\gamma}
+
\alpha^{\text{ATC}}(X)\p{Y-\gamma(X)}
-
\theta.
\]

\paragraph{SQ-Riesz regression.}
Let $\widehat{\pi}_0\coloneqq \frac{1}{n}\sum^n_{i=1}\p{1-D_i}$ and define $m^{\text{ATC}}$ with $\widehat{\pi}_0$.
Then, we have
\[
\widehat{\text{BD}}_{g^{\text{SQ}}}\p{\alpha}
=
\frac{1}{n}\sum^n_{i=1}
\p{
\alpha(D_i,Z_i)^2
-
\frac{2(1-D_i)}{\widehat{\pi}_0}\p{\alpha(1,Z_i)-\alpha(0,Z_i)}
}.
\]

\paragraph{UKL-Riesz regression.}
With $f_\alpha(x)\coloneqq \sign\p{\alpha(x)}\log\p{|\alpha(x)|-C}$,
\begin{align*}
\widehat{\text{BD}}_{g^{\text{UKL}}}\p{\alpha}
={}&\frac{1}{n}\sum^n_{i=1}
\Bigp{C\log\p{|\alpha(D_i,Z_i)|-C}+|\alpha(D_i,Z_i)|}\\
&-\frac{1}{n}\sum^n_{i=1}\frac{1-D_i}{\widehat{\pi}_0}
\p{f_\alpha(1,Z_i)-f_\alpha(0,Z_i)}.
\end{align*}

\paragraph{BKL-Riesz regression.}
With $h_\alpha(x)\coloneqq \sign\p{\alpha(x)}\log\p{\frac{|\alpha(x)|-C}{|\alpha(x)|+C}}$,
\begin{align*}
\widehat{\text{BD}}_{g^{\text{BKL}}}\p{\alpha}
={}&\frac{C}{n}\sum^n_{i=1}\log\p{|\alpha(D_i,Z_i)|^2-C^2}\\
&-\frac{1}{n}\sum^n_{i=1}\frac{1-D_i}{\widehat{\pi}_0}
\p{h_\alpha(1,Z_i)-h_\alpha(0,Z_i)}.
\end{align*}

\subsection{DID Estimation}
\paragraph{Setup.}
We consider a $2\times 2$ DID design with covariates.
For each unit, we observe $W\coloneqq (Y_0,Y_1,D,Z)$, where $Y_t$ is the outcome at period $t\in\cb{0,1}$ and $D$ indicates membership in the treated group, with treatment occurring between $t=0$ and $t=1$.
Let $\Delta Y\coloneqq Y_1-Y_0$.
We target the DID estimand for the treated group
\[
\theta^{\text{DID}}_0 \coloneqq \bbE\sqb{Y_1(1)-Y_1(0)\mid D=1}.
\]
Assume conditional parallel trends and overlap, $\bbE\sqb{Y_1(0)-Y_0(0)\mid D=1,Z}=\bbE\sqb{Y_1(0)-Y_0(0)\mid D=0,Z}$ and $0<e_0(Z)<1$ almost surely.
This setup follows the semiparametric DID literature, including \citet{Abadie2005semiparametricdifference} and \citet{SantAnna2020doublyrobust}.

\paragraph{Neyman orthogonal score and Riesz representer.}
Let $X\coloneqq (D,Z)$ and define the differenced regression function $\gamma^\Delta_0(d,z)\coloneqq \bbE\sqb{\Delta Y\mid D=d,Z=z}$.
Under conditional parallel trends,
\[
\theta^{\text{DID}}_0
=
\bbE\sqb{\frac{D}{\pi_1}\p{\gamma^\Delta_0(1,Z)-\gamma^\Delta_0(0,Z)}}.
\]
Define
\[
m^{\text{DID}}\p{W,\gamma}
\coloneqq
\frac{D}{\pi_1}\p{\gamma(1,Z)-\gamma(0,Z)},
\]
where $\gamma$ is interpreted as a candidate for $\gamma^\Delta_0$.
The Riesz representer is identical to ATT,
\[
\alpha^{\text{DID}}_0(D,Z)
\coloneqq
\frac{D}{\pi_1}
-
\frac{1-D}{\pi_1}\frac{e_0(Z)}{1-e_0(Z)}.
\]
The orthogonal score is
\[
\psi^{\text{DID}}\p{W;\eta,\theta}
\coloneqq
m^{\text{DID}}\p{W,\gamma}
+
\alpha^{\text{DID}}(X)\p{\Delta Y-\gamma(X)}
-
\theta.
\]

\paragraph{SQ-Riesz regression.}
Replacing $\pi_1$ by $\widehat{\pi}_1$, the SQ-Riesz objective is
\[
\widehat{\text{BD}}_{g^{\text{SQ}}}\p{\alpha}
=
\frac{1}{n}\sum^n_{i=1}
\p{
\alpha(D_i,Z_i)^2
-
\frac{2D_i}{\widehat{\pi}_1}\p{\alpha(1,Z_i)-\alpha(0,Z_i)}
}.
\]

\paragraph{UKL-Riesz regression.}
With $f_\alpha(x)\coloneqq \sign\p{\alpha(x)}\log\p{|\alpha(x)|-C}$,
\begin{align*}
\widehat{\text{BD}}_{g^{\text{UKL}}}\p{\alpha}
={}&\frac{1}{n}\sum^n_{i=1}\Bigp{C\log\p{|\alpha(D_i,Z_i)|-C}+|\alpha(D_i,Z_i)|}\\
&-\frac{1}{n}\sum^n_{i=1}\frac{D_i}{\widehat{\pi}_1}
\p{f_\alpha(1,Z_i)-f_\alpha(0,Z_i)}.
\end{align*}

\paragraph{BP-Riesz regression.}
For $\omega\in\p{0,\infty}$, use the signed BP terms in Equation~\eqref{eq:signed_bp_terms}. Then, we have
\[
\widehat{\mathrm{BD}}_{g^{\text{BP}}}\p{\alpha}
=
\frac{1}{n}\sum^n_{i=1}
\Bigp{
A_{\omega,C,\alpha}(D_i,Z_i)
-
\frac{D_i}{\widehat{\pi}_1}\p{u_{\omega,C,\alpha}(1,Z_i)-u_{\omega,C,\alpha}(0,Z_i)}
}.
\]

\subsection{Missing Variable Mean Estimation}
\paragraph{Setup.}
We observe $W\coloneqq (R,RY,Z)$, where $R\in\cb{0,1}$ indicates whether the outcome $Y$ is observed.
The target is the population mean
\[
\theta^{\text{MISS}}_0 \coloneqq \bbE\sqb{Y}.
\]
Assume missing at random and positivity, $Y$ is independent of $R$ given $Z$, and $0<\rho_0(Z)<1$ almost surely, where $\rho_0(z)\coloneqq \bbP\p{R=1\mid Z=z}$.
This setup is standard in semiparametric missing-data theory, see \citet{Robins1994estimationregression}.

\paragraph{Neyman orthogonal score and Riesz representer.}
Let $X\coloneqq (R,Z)$ and define the regression function for the always-observed outcome $RY$,
\[
\gamma_0(r,z)\coloneqq \bbE\sqb{RY\mid R=r,Z=z}.
\]
Under missing at random, $\gamma_0(1,z)=\bbE\sqb{Y\mid Z=z}$ and $\gamma_0(0,z)=0$, so
\[
\theta^{\text{MISS}}_0
=
\bbE\sqb{\gamma_0(1,Z)}.
\]
Define
\[
m^{\text{MISS}}\p{W,\gamma}
\coloneqq
\gamma(1,Z).
\]
The Riesz representer is
\[
\alpha^{\text{MISS}}_0(R,Z)
\coloneqq
\frac{R}{\rho_0(Z)}.
\]
The orthogonal score is
\[
\psi^{\text{MISS}}\p{W;\eta,\theta}
\coloneqq
m^{\text{MISS}}\p{W,\gamma}
+
\alpha^{\text{MISS}}(X)\p{RY-\gamma(X)}
-
\theta.
\]

\paragraph{SQ-Riesz regression.}
With $g^{\text{SQ}}\p{a}\coloneqq a^2$,
\[
\widehat{\text{BD}}_{g^{\text{SQ}}}\p{\alpha}
=
\frac{1}{n}\sum^n_{i=1}
\p{
\alpha(R_i,Z_i)^2
-
2\alpha(1,Z_i)
}.
\]

\paragraph{UKL-Riesz regression.}
The representer has the structural zero $\alpha^{\mathrm{MISS}}_0(0,Z)=0$, so the signed UKL generator cannot be applied directly to $\alpha(R,Z)$. Instead, let $r_0(Z)\coloneqq 1/\rho_0(Z)$, write
\[
\alpha(R,Z)=Rr(Z),
\qquad
r(Z)>C,
\]
and fit the positive inverse-observation component $r$. With
\[
u_r(Z)\coloneqq \log\p{r(Z)-C},
\]
the reduced empirical objective is
\[
\widehat{\mathrm{BD}}_{g^{\text{UKL}}}\p{r}
=
\frac{1}{n}\sum^n_{i=1}
\Bigp{
R_i\p{C\log\p{r(Z_i)-C}+r(Z_i)}
-u_r(Z_i)
}.
\]
This expression follows from the identity $\bbE[Rr_0(Z)h(Z)]=\bbE[h(Z)]$ and never evaluates the generator at the structural zero.

\paragraph{BP-Riesz regression.}
Using the same parameterization $\alpha(R,Z)=Rr(Z)$, define
\[
u_{\omega,C,r}(Z)
\coloneqq
\p{1+1/\omega}\Bigp{\p{r(Z)-C}^\omega-1},
\]
\[
A_{\omega,C,r}(Z)
\coloneqq
r(Z)\p{r(Z)-C}^\omega
+\frac{C}{\omega}\Bigp{\p{r(Z)-C}^\omega-1}.
\]
Then, we have
\[
\widehat{\mathrm{BD}}_{g^{\text{BP}}}\p{r}
=
\frac{1}{n}\sum^n_{i=1}
\Bigp{
R_i A_{\omega,C,r}(Z_i)
-u_{\omega,C,r}(Z_i)
}.
\]

\subsection{LATE Estimation}
\paragraph{Setup.}
We observe $W\coloneqq (Y,D,V,Z)$, where $V\in\cb{0,1}$ is a binary instrument, $D\in\cb{0,1}$ is a binary treatment, and $Z$ is a vector of covariates.
Under standard instrumental-variable assumptions, the local average treatment effect is identified as a Wald ratio, see \citet{Imbens2015causalinference},
\begin{align*}
\theta^{\mathrm{LATE}}_0&\coloneqq\frac{\theta^Y_0}{\theta^D_0},\\
\theta^Y_0&\coloneqq\bbE\Bigsqb{\bbE[Y\mid V=1,Z]-\bbE[Y\mid V=0,Z]},\\
\theta^D_0&\coloneqq\bbE\Bigsqb{\bbE[D\mid V=1,Z]-\bbE[D\mid V=0,Z]}.
\end{align*}
Let $\pi_V(z)\coloneqq \bbP\p{V=1\mid Z=z}$.

\paragraph{Neyman orthogonal score and Riesz representer.}
Set $X\coloneqq (V,Z)$ and define the regression functions
\[
\gamma^Y_0(v,z)\coloneqq \bbE\sqb{Y\mid V=v,Z=z},
\qquad
\gamma^D_0(v,z)\coloneqq \bbE\sqb{D\mid V=v,Z=z}.
\]
For $Q\in\cb{Y,D}$, define the linear functional
\[
m^{\text{IV}}\p{W,\gamma}
\coloneqq
\gamma(1,Z)-\gamma(0,Z),
\]
and note that $\theta^Q_0=\bbE\sqb{m^{\text{IV}}\p{W,\gamma^Q_0}}$.
The Riesz representer for $m^{\text{IV}}$ is
\[
\alpha^{\text{IV}}_0(V,Z)
\coloneqq
\frac{V}{\pi_V(Z)}-\frac{1-V}{1-\pi_V(Z)}.
\]
The orthogonal scores for $\theta^Y_0$ and $\theta^D_0$ are
\[
\psi^Y\p{W;\eta_Y,\theta^Y}
\coloneqq
m^{\text{IV}}\p{W,\gamma^Y}
+
\alpha^{\text{IV}}(X)\p{Y-\gamma^Y(X)}
-
\theta^Y,
\]
\[
\psi^D\p{W;\eta_D,\theta^D}
\coloneqq
m^{\text{IV}}\p{W,\gamma^D}
+
\alpha^{\text{IV}}(X)\p{D-\gamma^D(X)}
-
\theta^D.
\]
Writing $\theta^{\text{LATE}}=\theta^Y/\theta^D$, a delta-method orthogonal score for $\theta^{\text{LATE}}$ is
\[
\psi^{\text{LATE}}\p{W;\eta,\theta}
\coloneqq
\frac{1}{\theta^D}\p{\psi^Y\p{W;\eta_Y,\theta^Y}-\theta\psi^D\p{W;\eta_D,\theta^D}},
\]
where $\eta$ collects $\eta_Y$, $\eta_D$, and $\alpha^{\text{IV}}$.

\paragraph{SQ-Riesz regression.}
The SQ-Riesz objective for fitting $\alpha^{\text{IV}}_0$ is
\[
\widehat{\text{BD}}_{g^{\text{SQ}}}\p{\alpha}
=
\frac{1}{n}\sum^n_{i=1}
\p{
\alpha(V_i,Z_i)^2
-
2\p{\alpha(1,Z_i)-\alpha(0,Z_i)}
}.
\]

\paragraph{UKL-Riesz regression.}
With $f_\alpha(x)\coloneqq \sign\p{\alpha(x)}\log\p{|\alpha(x)|-C}$,
\[
\widehat{\text{BD}}_{g^{\text{UKL}}}\p{\alpha}
=
\frac{1}{n}\sum^n_{i=1}
\p{
C\log\p{|\alpha(V_i,Z_i)|-C}
+
|\alpha(V_i,Z_i)|
-
\p{f_\alpha(1,Z_i)-f_\alpha(0,Z_i)}
}.
\]

\paragraph{BP-Riesz regression.}
Using the signed BP terms in Equation~\eqref{eq:signed_bp_terms},
\[
\widehat{\mathrm{BD}}_{g^{\text{BP}}}\p{\alpha}
=
\frac{1}{n}\sum^n_{i=1}
\Bigp{
A_{\omega,C,\alpha}(V_i,Z_i)
-
\p{u_{\omega,C,\alpha}(1,Z_i)-u_{\omega,C,\alpha}(0,Z_i)}
}.
\]

\subsection{Policy Path Estimation}
\paragraph{Setup.}
Let $X\coloneqq (D,Z)$ where $D\in\bbR$ is a continuous policy variable and $Z$ is a vector of covariates.
We observe i.i.d.\ samples $W\coloneqq (Y,D,Z)$ with regression function $\gamma_0(d,z)\coloneqq \bbE\sqb{Y\mid D=d,Z=z}$.
Following \citet{Kato2025scorematchingriesz}, consider translation shifts $\tau_\delta(d,z)\coloneqq (d+\delta,z)$ and define the shifted distribution $P_\delta$ as the law of $\tau_\delta(X)$ when $X\sim P_0$.
The symmetric policy path is the function $\delta\mapsto \theta_0(\delta)$ defined by
\[
\theta_0(\delta)
\coloneqq
\bbE_{X\sim P_\delta}\sqb{\gamma_0(X)}
-
\bbE_{X\sim P_{-\delta}}\sqb{\gamma_0(X)}.
\]
By a change of variables,
\[
\theta_0(\delta)
=
\bbE\sqb{\gamma_0(D+\delta,Z)-\gamma_0(D-\delta,Z)}.
\]

\paragraph{Neyman orthogonal score and Riesz representer.}
For each $\delta$, define the linear functional
\[
m^{\text{PATH}}_\delta\p{W,\gamma}
\coloneqq
\gamma(D+\delta,Z)-\gamma(D-\delta,Z).
\]
Let $\alpha^{\text{PATH}}_{0,\delta}$ be the corresponding Riesz representer, satisfying
\[
\bbE\sqb{m^{\text{PATH}}_\delta\p{W,\gamma}}
=
\bbE\sqb{\alpha^{\text{PATH}}_{0,\delta}(X)\gamma(X)}
\]
for all square-integrable $\gamma$.
The orthogonal score is
\[
\psi^{\text{PATH}}_\delta\p{W;\eta,\theta}
\coloneqq
m^{\text{PATH}}_\delta\p{W,\gamma}
+
\alpha^{\text{PATH}}_\delta(X)\p{Y-\gamma(X)}
-
\theta,
\]
with $\eta\coloneqq \p{\gamma,\alpha^{\text{PATH}}_\delta}$.

\paragraph{SQ-Riesz regression.}
For each fixed $\delta$, the SQ-Riesz objective is
\[
\widehat{\text{BD}}_{g^{\text{SQ}}}\p{\alpha}
=
\frac{1}{n}\sum^n_{i=1}
\p{
\alpha(D_i,Z_i)^2
-
2\p{\alpha(D_i+\delta,Z_i)-\alpha(D_i-\delta,Z_i)}
}.
\]

\paragraph{UKL-Riesz regression.}
With $f_\alpha(x)\coloneqq \sign\p{\alpha(x)}\log\p{|\alpha(x)|-C}$,
\begin{align*}
\widehat{\text{BD}}_{g^{\text{UKL}}}\p{\alpha}
={}&\frac{1}{n}\sum^n_{i=1}
\Bigp{C\log\p{|\alpha(D_i,Z_i)|-C}+|\alpha(D_i,Z_i)|}\\
&-\frac{1}{n}\sum^n_{i=1}
\p{f_\alpha(D_i+\delta,Z_i)-f_\alpha(D_i-\delta,Z_i)}.
\end{align*}

\paragraph{Score matching.}
Under the translation shift, the representer admits a density-ratio form.
Let $p_0$ denote the joint density of $X=(D,Z)$ and define
\[
r_\delta(d,z)\coloneqq \frac{p_0(d-\delta,z)}{p_0(d,z)}.
\]
Then, it holds that $\alpha^{\text{PATH}}_{0,\delta}(d,z)=r_\delta(d,z)-r_{-\delta}(d,z)$.
We also have
\[
\log r_\delta(d,z)
=
-\int_0^\delta \partial_u \log p_0(d-u,z)\rmd u.
\]
\citet{Kato2025scorematchingriesz} proposes to estimate the treatment-direction score $\partial_d\log p_0(d,z)$ by denoising score matching, then recover $\log r_\delta$ by numerical integration, and finally form $\widehat{\alpha}^{\text{PATH}}_\delta=\widehat{r}_\delta-\widehat{r}_{-\delta}$.
This yields the entire curve $\delta\mapsto \widehat{\theta}(\delta)$ once a single score model is learned.

\subsection{Off-Policy Evaluation Under Covariate Shift}
\paragraph{Setup.}
We consider the contextual bandit setup studied by \citet{Uehara2020offpolicy}.
Let $A\in\calA$ be an action, $X\in\calX$ be a context, and $Y\in\bbR$ be a reward.
We observe two independent datasets:
\[
\calD_{\text{hst}}\coloneqq \bigcb{(X_i,A_i,Y_i)}^{n_{\text{hst}}}_{i=1},
\qquad
\calD_{\text{evl}}\coloneqq \bigcb{Z_j}^{n_{\text{evl}}}_{j=1}.
\]
The historical data satisfy
\[
X\sim p_0,
\qquad
A\mid X \sim \pi_b(\cdot\mid X),
\qquad
Y\mid (A,X)\sim P_0(\cdot\mid A,X),
\]
where $p_0$ is the historical context density, $\pi_b$ is an unknown behavior policy, and $P_0(\cdot\mid a,x)$ is the reward distribution.
The evaluation covariates satisfy $Z\sim q_0$, where $q_0$ is the evaluation context density.
We assume that the conditional reward distribution is invariant across the two domains, namely $P_0(\cdot\mid a,x)$ is shared.

Let $\pi_e(a\mid x)$ be a known evaluation policy.
Define the regression function
\[
\gamma_0(a,x)\coloneqq \bbE\sqb{Y\mid A=a,X=x}.
\]
The off-policy evaluation (OPE) target is the evaluation-policy reward under the evaluation covariate distribution,
\[
\theta^{\text{OPE}}_0
\coloneqq
\bbE\sqb{m^{\text{OPE}}\p{Z,\gamma_0}},
\qquad
m^{\text{OPE}}\p{z,\gamma}\coloneqq \sum_{a\in\calA}\pi_e(a\mid z)\gamma(a,z).
\]
We impose overlap in the usual sense, $\pi_e(a\mid x)>0$ implies $\pi_b(a\mid x)>0$, and we assume $\theta^{\text{OPE}}_0$ is finite.

\paragraph{Neyman orthogonal score and Riesz representer.}
Although $\theta^{\text{OPE}}_0$ is an expectation over $q_0$, the orthogonal construction uses the historical sample for the residual term and the evaluation covariate sample for the plug-in term.
Consider any test function $h\colon \calA\times\calX\to\bbR$.
Define the linear functional
\[
\Gamma(h)\coloneqq \bbE\sqb{\sum_{a\in\calA}\pi_e(a\mid Z)h(a,Z)}.
\]
The corresponding Riesz representer $\alpha^{\text{OPE}}_0\in L_2(P_{(A,X)})$ is the unique function satisfying
\[
\bbE\sqb{\alpha^{\text{OPE}}_0(A,X)h(A,X)}=\Gamma(h)
\]
for all square-integrable $h$ under the historical law of $(A,X)$.
A direct calculation yields the joint density ratio form
\begin{align*}
\alpha^{\text{OPE}}_0(a,x)
&=\frac{q_0(x)\pi_e(a\mid x)}{p_0(x)\pi_b(a\mid x)}
=r_0(x)w_0(a,x),\\
r_0(x)&\coloneqq \frac{q_0(x)}{p_0(x)},
\qquad
w_0(a,x)\coloneqq \frac{\pi_e(a\mid x)}{\pi_b(a\mid x)}.
\end{align*}
This representation clarifies that one can estimate $\alpha^{\text{OPE}}_0$ directly as a single joint density ratio on $(A,X)$, rather than estimating $r_0$ and $\pi_b$ separately and multiplying.

Given generic nuisances $\eta\coloneqq (\gamma,\alpha)$, an orthogonal estimating equation for $\theta^{\text{OPE}}_0$ is
\[
0
=
\bbE\sqb{m^{\text{OPE}}\p{Z,\gamma}}
+
\bbE\sqb{\alpha(A,X)\p{Y-\gamma(A,X)}}
-
\theta.
\]
The corresponding sample estimator is
\[
\widehat{\theta}^{\text{OPE}}
\coloneqq
\frac{1}{n_{\text{evl}}}\sum^{n_{\text{evl}}}_{j=1} m^{\text{OPE}}\p{Z_j,\widehat{\gamma}}
+
\frac{1}{n_{\text{hst}}}\sum^{n_{\text{hst}}}_{i=1}\widehat{\alpha}(A_i,X_i)\p{Y_i-\widehat{\gamma}(A_i,X_i)}.
\]
If one sets $\widehat{\alpha}=\widehat r\,\widehat w$ with $\widehat r$ a density ratio estimator and $\widehat w$ a policy ratio estimator, then $\widehat{\theta}^{\text{OPE}}$ reduces to the doubly robust construction in \citet{Uehara2020offpolicy}.
The point here is that generalized Riesz regression permits estimating $\alpha^{\text{OPE}}_0$ directly.

\paragraph{SQ-Riesz regression.}
View $\alpha^{\text{OPE}}_0$ as the density ratio between the historical joint law of $(A,X)$ and the evaluation joint law of $(A,Z)$ induced by $\pi_e$.
The moment identity is
\[
\bbE\sqb{\alpha^{\text{OPE}}_0(A,X)h(A,X)}
=
\bbE\sqb{\sum_{a\in\calA}\pi_e(a\mid Z)h(a,Z)}.
\]
With the squared loss generator $g^{\text{SQ}}(\alpha)=(\alpha-1)^2$ as in the covariate shift application, the empirical objective becomes
\[
\widehat{\text{BD}}^{\text{OPE}}_{g^{\text{SQ}}}\p{\alpha}
=
\frac{1}{n_{\text{hst}}}\sum^{n_{\text{hst}}}_{i=1}\alpha(A_i,X_i)^2
-
\frac{2}{n_{\text{evl}}}\sum^{n_{\text{evl}}}_{j=1}\sum_{a\in\calA}\pi_e(a\mid Z_j)\alpha(a,Z_j).
\]
We estimate $\alpha^{\text{OPE}}_0$ by
\[
\widehat{\alpha}^{\text{OPE}}
\coloneqq
\argmin_{\alpha\in\calH}\cb{\widehat{\text{BD}}^{\text{OPE}}_{g^{\text{SQ}}}\p{\alpha}+\lambda J(\alpha)}.
\]
This is an LSIF-type objective on the augmented variable $(A,X)$, with the evaluation expectation taken under $\pi_e$ and $q_0$, see \citet{Kanamori2009aleastsquares} and \citet{Sugiyama2012densityratio}.

\paragraph{UKL-Riesz regression.}
Since $\alpha^{\text{OPE}}_0(a,x)\geq0$, we can use the UKL generator on the positive component of its domain
\[
g^{\text{UKL}}(\alpha)=\alpha\log\alpha-\alpha,
\qquad
\partial g^{\text{UKL}}(\alpha)=\log\alpha.
\]
The empirical objective is, up to constants that do not depend on $\alpha$,
\[
\widehat{\text{BD}}^{\text{OPE}}_{g^{\text{UKL}}}\p{\alpha}
=
\frac{1}{n_{\text{hst}}}\sum^{n_{\text{hst}}}_{i=1}\alpha(A_i,X_i)
-
\frac{1}{n_{\text{evl}}}\sum^{n_{\text{evl}}}_{j=1}\sum_{a\in\calA}\pi_e(a\mid Z_j)\log\alpha(a,Z_j).
\]
Imposing the normalization constraint $\frac{1}{n_{\text{hst}}}\sum^{n_{\text{hst}}}_{i=1}\alpha(A_i,X_i)=1$ yields a KLIEP-style formulation, see \citet{Sugiyama2008directimportance}.

\paragraph{BP-Riesz regression.}
BP-Riesz regression provides a continuum between SQ-type and UKL-type criteria.
For $\gamma\in(0,\infty)$, consider
\[
g^{\text{BP}}(\alpha)
\coloneqq
\frac{\alpha^{1+\gamma}-\alpha}{\gamma}-\alpha,
\qquad
\partial g^{\text{BP}}(\alpha)=\p{1+\frac{1}{\gamma}}\Bigp{\alpha^\gamma-1}.
\]
As in the covariate shift case, $\partial g^{\text{BP}}(\alpha)\alpha-g^{\text{BP}}(\alpha)=\alpha^{1+\gamma}$, so the empirical objective simplifies to
\[
\widehat{\text{BD}}^{\text{OPE}}_{g^{\text{BP}}}\p{\alpha}
=
\frac{1}{n_{\text{hst}}}\sum^{n_{\text{hst}}}_{i=1}\alpha(A_i,X_i)^{1+\gamma}
-
\p{1+\frac{1}{\gamma}}
\frac{1}{n_{\text{evl}}}\sum^{n_{\text{evl}}}_{j=1}\sum_{a\in\calA}\pi_e(a\mid Z_j)\Bigp{\alpha(a,Z_j)^\gamma-1}.
\]
This loss can be paired with a link $\alpha(a,x)=\exp\p{f(a,x)}$ to enforce nonnegativity.

\subsection{Semi-Supervised ATE Estimation With Unlabeled Covariates}
\paragraph{Setup.}
We follow the two-sample semi-supervised ATE formulation in \citet{Kato2025semisupervised}.
We observe a labeled dataset and an auxiliary unlabeled covariate dataset:
\[
\calD_{\text{L}}\coloneqq \bigcb{(X_i,D_i,Y_i)}^{n_{\text{L}}}_{i=1},
\qquad
\calD_{\text{U}}\coloneqq \bigcb{\widetilde{X}_j}^{n_{\text{U}}}_{j=1},
\]
where $D\in\cb{0,1}$ is a treatment, $X\in\calX$ is a covariate, and $Y\in\bbR$ is an outcome.
The labeled sample follows $P_0(x,d,y)$, while the unlabeled covariates follow $Q_0(x)$.
Let
\[
\gamma_0(d,x)\coloneqq \bbE\sqb{Y\mid D=d,X=x},
\qquad
e_0(x)\coloneqq \bbP\p{D=1\mid X=x}.
\]
We assume unconfoundedness and overlap as in the ATE application.

The semi-supervised ATE (SS-ATE) is defined under an evaluation covariate density $\kappa_{0,\beta}$ that is a known mixture of the labeled and unlabeled covariate densities:
\[
\kappa_{0,\beta}(x)\coloneqq \beta p_0(x)+(1-\beta)q_0(x),
\qquad
\beta\in\sqb{0,1}.
\]
Define
\[
\theta^{\text{SS-ATE}}_0
\coloneqq
\bbE_{\kappa_{0,\beta}}\sqb{\gamma_0(1,X)-\gamma_0(0,X)}.
\]
When $\beta<1$, the unlabeled covariates can reduce the efficiency bound, and \citet{Kato2025semisupervised} derives the corresponding efficient influence function under stratified sampling.

\paragraph{Neyman orthogonal score and Riesz representer.}
Define the usual ATE functional
\[
m^{\text{ATE}}\p{x,\gamma}\coloneqq \gamma(1,x)-\gamma(0,x).
\]
In the present setting,
\[
\theta^{\text{SS-ATE}}_0
=
\beta\bbE\sqb{m^{\text{ATE}}\p{X,\gamma_0}}
+
(1-\beta)\bbE\sqb{m^{\text{ATE}}\p{\widetilde{X},\gamma_0}},
\]
where the first expectation is over $X$ from the labeled covariate distribution and the second is over $\widetilde{X}$ from the unlabeled covariate distribution.

The Riesz representer $\alpha^{\text{SS-ATE}}_0(d,x)$ is defined relative to the labeled regressor distribution of $(D,X)$ by the identity
\[
\bbE\sqb{\alpha^{\text{SS-ATE}}_0(D,X)h(D,X)}
=
\beta\bbE\sqb{h(1,X)-h(0,X)}
+
(1-\beta)\bbE\sqb{h(1,\widetilde{X})-h(0,\widetilde{X})}
\]
for all square-integrable $h$.
This yields the explicit form
\[
\alpha^{\text{SS-ATE}}_0(1,x)=\frac{\kappa_{0,\beta}(x)}{p_0(x)e_0(x)},
\qquad
\alpha^{\text{SS-ATE}}_0(0,x)=-\frac{\kappa_{0,\beta}(x)}{p_0(x)\p{1-e_0(x)}}.
\]
Equivalently, if $r_{0,\beta}(x)\coloneqq \kappa_{0,\beta}(x)/p_0(x)$, then it holds that
\[
\alpha^{\text{SS-ATE}}_0(D,X)=r_{0,\beta}(X)\Bigp{\frac{D}{e_0(X)}-\frac{1-D}{1-e_0(X)}}.
\]
This is the signed version of the ratio $1/v_{0,\beta}(d,x)$ used by \citet{Kato2025semisupervised}, where $v_{0,\beta}(d,x)=p_0(d,x)/\kappa_{0,\beta}(x)$.

Given nuisances $\eta\coloneqq (\gamma,\alpha)$, the orthogonal estimating equation can be written as
\[
0
=
\beta\bbE\sqb{m^{\text{ATE}}\p{X,\gamma}}
+
(1-\beta)\bbE\sqb{m^{\text{ATE}}\p{\widetilde{X},\gamma}}
+
\bbE\sqb{\alpha(D,X)\p{Y-\gamma(D,X)}}
-
\theta,
\]
which motivates the estimator
\begin{align*}
\widehat\theta^{\mathrm{SS-ATE}}
={}&\frac{\beta}{n_{\mathrm L}}\sum_{i=1}^{n_{\mathrm L}}m^{\mathrm{ATE}}(X_i,\widehat\gamma)
+\frac{1-\beta}{n_{\mathrm U}}\sum_{j=1}^{n_{\mathrm U}}m^{\mathrm{ATE}}(\widetilde X_j,\widehat\gamma)\\
&+\frac1{n_{\mathrm L}}\sum_{i=1}^{n_{\mathrm L}}\widehat\alpha(D_i,X_i)
\{Y_i-\widehat\gamma(D_i,X_i)\}.
\end{align*}
This is the efficient estimator form in \citet{Kato2025semisupervised}, written in the Riesz representer language.

\paragraph{SQ-Riesz regression.}
A key point is that $\alpha^{\text{SS-ATE}}_0$ can be estimated directly, without separately estimating $e_0$ and the covariate ratio $r_{0,\beta}$.
Define, for any function $h(d,x)$,
\[
\widehat{\Gamma}_{\beta}(h)
\coloneqq
\beta\frac{1}{n_{\text{L}}}\sum^{n_{\text{L}}}_{i=1}\p{h(1,X_i)-h(0,X_i)}
+
(1-\beta)\frac{1}{n_{\text{U}}}\sum^{n_{\text{U}}}_{j=1}\p{h(1,\widetilde{X}_j)-h(0,\widetilde{X}_j)}.
\]
With $g^{\text{SQ}}(\alpha)=(\alpha-C)^2$, the empirical objective reduces, up to constants, to
\[
\widehat{\text{BD}}^{\text{SS-ATE}}_{g^{\text{SQ}}}\p{\alpha}
=
\frac{1}{n_{\text{L}}}\sum^{n_{\text{L}}}_{i=1}\alpha(D_i,X_i)^2
-
2\widehat{\Gamma}_{\beta}\p{\alpha}.
\]
We estimate $\alpha^{\text{SS-ATE}}_0$ by
\[
\widehat{\alpha}^{\text{SS-ATE}}
\coloneqq
\argmin_{\alpha\in\calH}\cb{\widehat{\text{BD}}^{\text{SS-ATE}}_{g^{\text{SQ}}}\p{\alpha}+\lambda J(\alpha)}.
\]

\paragraph{UKL-Riesz regression.}
For signed representers, we use the UKL generator in the main text,
\[
g^{\text{UKL}}(\alpha)=(|\alpha|-C)\log\p{|\alpha|-C}-|\alpha|,
\qquad
\partial g^{\text{UKL}}(\alpha)=\sign(\alpha)\log\p{|\alpha|-C}.
\]
Let
\[
f_\alpha(d,x)\coloneqq \sign\p{\alpha(d,x)}\log\p{|\alpha(d,x)|-C}.
\]
Then, the empirical UKL-Riesz objective becomes
\[
\widehat{\text{BD}}^{\text{SS-ATE}}_{g^{\text{UKL}}}\p{\alpha}
=
\frac{1}{n_{\text{L}}}\sum^{n_{\text{L}}}_{i=1}\Bigp{C\log\p{|\alpha(D_i,X_i)|-C}+|\alpha(D_i,X_i)|}
-
\widehat{\Gamma}_{\beta}\p{f_\alpha}.
\]
This directly targets the Riesz representer associated with the evaluation density $\kappa_{0,\beta}$.

\paragraph{BKL-Riesz regression.}
Using the BKL generator in the main text, define
\[
h_\alpha(d,x)\coloneqq \sign\p{\alpha(d,x)}\log\p{\frac{|\alpha(d,x)|-C}{|\alpha(d,x)|+C}}.
\]
Then, the empirical BKL-Riesz objective is
\begin{align*}
\widehat{\text{BD}}^{\text{SS-ATE}}_{g^{\text{BKL}}}\p{\alpha}
={}&\frac{C}{n_{\text{L}}}\sum^{n_{\text{L}}}_{i=1}
\log\p{|\alpha(D_i,X_i)|^2-C^2}\\
&-\widehat{\Gamma}_{\beta}\p{h_\alpha}.
\end{align*}
As in the main text, this loss is paired with a logistic-style link for the magnitude of $\alpha$, while $h_\alpha$ uses the sign determined by treatment status.

\section{Bayesian Interpretation of Generalized Riesz Regression}
\label{appdx:bayesianinterpretation}

This appendix discusses Bayesian perspectives on generalized Riesz regression \eqref{eq:empbregman}. Its purpose is to explain how the empirical Bregman objective and its regularization can be read as a negative log posterior under a suitably chosen pseudo-likelihood and to relate that interpretation to the Bayesian discussions in \citet{Zhao2019covariatebalancing} and \citet{BrunsSmith2025augmentedbalancing}.

\subsection{Generalized Posterior and Maximum a Posteriori Estimation}
Generalized Riesz regression solves the regularized empirical problem
\[
\widehat{\alpha}
\in
\argmin_{\alpha\in\calH}
\Biggcb{
\frac{1}{n}\sum^n_{i=1}
\Bigp{
-g\p{\alpha\p{X_i}}
+
\partial g\p{\alpha\p{X_i}}\alpha\p{X_i}
-
m\bigp{W_i,\p{\partial g}\circ \alpha}
}
+
\lambda J\p{\alpha}
}.
\]
For each $x$, let $\calA(x)$ be the interval allowed by the representer model and define $g_x^*(u)=\sup_{\alpha\in\calA(x)}\{\alpha u-g(\alpha)\}$. A convenient reparameterization uses the dual coordinate $u\coloneqq\partial g\circ\alpha$. On the range of $\partial g$ over $\calA(x)$, the Fenchel--Young identity gives $g_x^*(u)=\alpha u-g(\alpha)$ at $u=\partial g(\alpha)$. Up to additive constants that do not depend on $\alpha$, the empirical objective can therefore be written as
\[
\widehat{\alpha}
\in
\argmin_{\alpha\in\calH}
\Biggcb{
\frac{1}{n}\sum^n_{i=1}
\Bigp{
g_{X_i}^*\p{u\p{X_i}}
-
m\p{W_i,u}
}
+
\lambda J\p{\alpha}
},
\qquad
u=\partial g\circ \alpha.
\]
This form suggests the following generalized posterior on $\alpha$:
\begin{align}
\Pi\p{d\alpha\mid \cb{W_i}_{i=1}^n}
\propto
\exp\Bigp{
-\sum^n_{i=1}\Bigp{g_{X_i}^*\p{u\p{X_i}}-m\p{W_i,u}}
-n\lambda J\p{\alpha}
}
\Pi_{\mathrm{ref}}\p{d\alpha},
\label{eq:appdx_generalizedposterior}
\end{align}
where $\Pi_{\mathrm{ref}}$ is an arbitrary reference measure on $\calH$ and $u=\partial g\circ \alpha$.
Under \eqref{eq:appdx_generalizedposterior}, the generalized Riesz regression solution is the maximum a posteriori estimator of $\alpha$.

\paragraph{Regularization as a prior.}
The factor $\exp\p{-n\lambda J\p{\alpha}}$ plays the role of a prior density, so $\lambda$ acts as a prior scale parameter.
For example, if $J(\alpha)=\|\alpha\|_{\calH}^2$ for a Hilbert space $\calH$, then $\lambda$ controls the concentration of the implied Gaussian prior on $\alpha$.
If we specify the model in dual coordinates as $u=u_\bmbeta$ and take $J\p{\alpha_\bmbeta}=\tfrac{1}{a}\|\bmbeta\|_a^a$, then $a=2$ corresponds to a Gaussian prior on $\bmbeta$ and $a=1$ corresponds to a Laplace prior, up to scale.

\paragraph{Loss as a pseudo-likelihood.}
The remaining factor in Equation~\eqref{eq:appdx_generalizedposterior} is
\[
\exp\left[-\sum_{i=1}^n\{g_{X_i}^*(u(X_i))-m(W_i,u)\}\right].
\]
This factor acts as a pseudo-likelihood constructed to target the Riesz equations through $m(W,u)$ rather than to model the conditional distribution of the treatment. This interpretation also agrees with the decision-theoretic motivation in \citet{Zhao2019covariatebalancing}, where the loss is tailored to covariate balance.

\subsection{SQ-Riesz and Gaussian Priors}
For the squared loss choice $g^{\mathrm{SQ}}(\alpha)=\p{\alpha-C}^2$, SQ-Riesz reduces to a quadratic objective, up to constants,
\[
\widehat{\alpha}
\in
\argmin_{\alpha\in\calH}
\Biggcb{
\frac{1}{n}\sum^n_{i=1}
\Bigp{\alpha\p{X_i}^2-2m\p{W_i,\alpha}}
+
\lambda J\p{\alpha}
}.
\]
If $\calH$ is parameterized as $\alpha_\bmbeta(x)=\bmphi(x)^\top\bmbeta$ and $J\p{\alpha_\bmbeta}=\|\bmbeta\|_2^2$, the objective is quadratic in $\bmbeta$.
Consequently, the generalized posterior \eqref{eq:appdx_generalizedposterior} is Gaussian under a Gaussian pseudo-likelihood, and the SQ-Riesz solution is both the posterior mode and posterior mean.
Equivalently, the penalty term corresponds to a Gaussian prior on $\bmbeta$ whose precision scales with $n\lambda$.

If $\calH$ is an RKHS and $J\p{\alpha}=\|\alpha\|_{\calH}^2$, the same quadratic structure yields the familiar Gaussian random field interpretation of RKHS penalties, where the RKHS norm defines the Cameron--Martin space of the prior.
In this sense, SQ-Riesz with an RKHS penalty admits the same conjugate Bayesian interpretation as kernel ridge regression, although the sample enters through the Riesz functional $m(W,\alpha)$ rather than through direct observations of $\alpha_0(X)$.

\subsection{Empirical Bayes Connection Through Augmented Balancing Weights}
When the dual coordinate is modeled linearly, SQ-Riesz also connects to regressor balancing and to the linear regression equivalence results in \citet{BrunsSmith2025augmentedbalancing}.
Appendix~J of \citet{BrunsSmith2025augmentedbalancing} studies augmented $\ell_2$ balancing weights and provides an empirical Bayes interpretation in a Gaussian linear model.
In the simple diagonal design case they consider, ridge regression coefficients arise as posterior means under a Normal prior, and centering that prior at an externally supplied estimate of the outcome regression coefficients yields the augmented balancing form as a posterior mean shrinkage rule.
Concretely, for each coefficient index $j$, their posterior mean has the convex combination form
\[
\widetilde{\beta}_j
=
\kappa_j\widehat{\beta}_{\mathrm{ols},j}
+
\p{1-\kappa_j}\widehat{\beta}_{\mathrm{reg},j},
\qquad
\kappa_j\in \p{0,1},
\]
where $\widehat{\beta}_{\mathrm{ols},j}$ is the OLS coefficient and $\widehat{\beta}_{\mathrm{reg},j}$ is the outcome model coefficient that centers the prior.
Through the correspondence between SQ-Riesz and stable $\ell_2$ balancing under an appropriate loss and link, this argument can be read as a Bayesian motivation for augmenting the representer based estimator with an outcome regression fit, where the augmentation acts as prior centering.

\subsection{Outcome Priors, KL-Type Regularization, and Dual Weights}
For KL-type losses, generalized Riesz regression is naturally expressed in dual coordinates.
Under loss and link choices that make $u=\partial g\circ \alpha$ linear in parameters, the empirical objective takes the form
\[
\frac{1}{n}\sum^n_{i=1} g_{X_i}^*\p{u_\bmbeta\p{X_i}}
-
\frac{1}{n}\sum^n_{i=1} m\p{W_i,u_\bmbeta}
+
\lambda J\p{\alpha_\bmbeta}.
\]
In treatment effect applications, the corresponding dual problem is a balancing weight program with an entropy type regularizer, which is the same dual structure emphasized by \citet{Zhao2019covariatebalancing}.
Section~5.4 of \citet{Zhao2019covariatebalancing} makes the Bayesian content of this duality explicit.
In their RKHS setting, if the outcome regression $g_0$ is assigned a Gaussian random field prior $g_0(\cdot)\sim G\p{0,K}$, then the design mean squared error of a weighting estimator decomposes as
\begin{align*}
&\bbE_{g_0}\left[\left\{\sum_{i:T_i=1}w_i g_0(X_i)-\sum_{i:T_i=0}w_i g_0(X_i)\right\}^2\right]
+\sum_{i=1}^n w_i^2\mathrm{Var}(Y_i\mid X_i,T_i)\\
&\hspace{3cm}=\widetilde w^\top K\widetilde w
+\sum_{i=1}^n w_i^2\mathrm{Var}(Y_i\mid X_i,T_i).
\end{align*}
Here, $\widetilde w_i=\p{2T_i-1}w_i$ and $K$ is the sample covariance matrix induced by the kernel.
They also show that the dual optimization, for example for ATT with a logistic link, takes the schematic form
\[
\min_{w}
\Biggcb{
\lambda\,\widetilde w^\top K\widetilde w
+
\sum^n_{i=1} w_i\log\p{w_i}
},
\]
so that changing $\lambda$ corresponds to changing prior beliefs about the conditional variance $\mathrm{Var}\p{Y\mid X,T}$.
The same calculation applies when KL-type generalized Riesz regression uses an RKHS model for the dual coordinate: the RKHS penalty corresponds to a Gaussian random field prior, while the dual retains the entropy term.

\paragraph{Interpretation of $\lambda$.}
From this viewpoint, $\lambda$ controls both the tolerance in the empirical Riesz equations and the scale of the implied prior, which determines how much imbalance in a rich function class is regarded as plausible relative to sampling variability. This interpretation is consistent with \citet{Zhao2019covariatebalancing}, where covariate-balance diagnostics correspond to assumptions about the unknown outcome regression function.

\section{Bayesian Perspectives on Density Ratio Estimation}
\label{app:bayes_dre}
Let $P_0$ and $P_1$ be distributions on $\calZ$ with densities $p_0$ and $p_1$ with respect to a common dominating measure, and define the density ratio
\[
r_0(z)\coloneqq \frac{p_1(z)}{p_0(z)}.
\]
As discussed in Appendix~\ref{appdx:rieszdens}, density ratio estimation is a special case of Riesz representer fitting, and Bregman--Riesz objectives induce a family of density ratio estimators, including LSIF and KLIEP. This section organizes Bayesian connections into four types: Bayesian estimation of the density ratio itself, constrained Bayesian nonparametric priors on the density ratio, density ratio estimation as a general Bayesian update, and reducing Bayesian inference to density ratio estimation.

\subsection{Bayesian Estimation of the Density Ratio Itself}
A standard model-based route is the semiparametric density ratio model \citep{Qin1998inferencesfor,Cheng2004semiparametricdensity},
\[
p_1(z) = r_{\bmbeta}(z)p_0(z),
\qquad
r_{\bmbeta}(z) \ge 0,
\qquad
\bbE_{P_0}\sqb{r_{\bmbeta}(Z)}=1,
\]
where $p_0$ is an unknown baseline density and $r_{\bmbeta}$ is a parametric ratio model. A common choice is exponential tilting,
\[
r_{\bmbeta}(z)
=
\frac{\exp\p{\bmphi(z)^\top \bmbeta}}{\bbE_{P_0}\sqb{\exp\p{\bmphi(Z)^\top \bmbeta}}}.
\]
Replacing the unknown normalizer by the empirical average over a sample $\cb{Z_i^0}^{n_0}_{i=1}\sim P_0$ yields the plug-in normalizer
\[
\widehat{c}\p{\bmbeta}
\coloneqq
\frac{1}{n_0}\sum^{n_0}_{i=1}\exp\p{\bmphi\p{Z_i^0}^\top \bmbeta},
\qquad
\widehat{r}_{\bmbeta}(z)
=
\frac{\exp\p{\bmphi(z)^\top \bmbeta}}{\widehat{c}\p{\bmbeta}}.
\]
Then, the likelihood contribution of a $P_1$ sample $\cb{Z_j^1}^{n_1}_{j=1}\sim P_1$ is
\[
\prod^{n_1}_{j=1}\widehat{r}_{\bmbeta}\p{Z_j^1}
=
\exp\Bigp{
\sum^{n_1}_{j=1}\bmphi\p{Z_j^1}^\top\bmbeta
-
n_1\log\p{\widehat{c}\p{\bmbeta}}
},
\]
whose negative log is the KLIEP objective \citep{Sugiyama2008directimportance, Sugiyama2012densityratio}, up to constants. Therefore, a Bayesian KLIEP is obtained by placing a prior $\pi_0(\bmbeta)$ and defining the posterior
\[
\pi\p{\bmbeta \mid \cb{Z_i^0}_{i=1}^{n_0}, \cb{Z_j^1}_{j=1}^{n_1}}
\propto
\pi_0(\bmbeta)
\prod^{n_1}_{j=1}\widehat{r}_{\bmbeta}\p{Z_j^1}.
\]
\citet{Sadeghkhani2019aparametric} study parametric Bayesian density ratio estimation, including robust Bayes risk motivations, providing an explicit example of the model-based Bayesian route (A).

\paragraph{Constrained Bayesian nonparametric priors on the density ratio.}
A Bayesian nonparametric route is to place priors on $p_0$ and $p_1$ separately and induce a prior on $r_0=p_1/p_0$. For instance, Dirichlet process mixture priors on densities yield posterior draws of $p_0$ and $p_1$, hence posterior draws of $r_0$ \citep{Tchetgen2011semiparametrictheory,Ferguson1973bayesiananalysis, Ghosal2017fundamentalsnonparametric}.

A more direct construction places a prior on the log density ratio $f\colon\calZ\to\bbR$ and enforces the normalization constraint by defining
\[
r_f(z)
\coloneqq
\frac{\exp\p{f(z)}}{\bbE_{P_0}\sqb{\exp\p{f(Z)}}},
\]
so that $\bbE_{P_0}\sqb{r_f(Z)}=1$ holds by construction. In practice, $\bbE_{P_0}\sqb{\exp\p{f(Z)}}$ is replaced by $\frac{1}{n_0}\sum^{n_0}_{i=1}\exp\p{f\p{Z_i^0}}$, yielding a nonparametric posterior over $r_f$ that satisfies the normalization constraint. This construction is particularly compatible with RKHS or Gaussian-process priors that are already natural in Riesz representer estimation.

\subsection{Generalized Bayesian Update for Density Ratio Estimation}
Even when an objective is not taken literally as a likelihood, it can be used as a loss for generalized Bayesian updating \citep{Bissiri2016generalframework}. In our setting, the same Bregman objective that defines density ratio estimation also defines a generalized posterior.

Let $r$ be a candidate density ratio in a model class, and let $\widehat{\text{BD}}_g\bigp{r}$ be the empirical Bregman objective corresponding to Appendix~\ref{appdx:rieszdens}. Then, a loss-based posterior is
\[
\Pi_n\bigp{dr \mid \cb{Z_i^0}_{i=1}^{n_0}, \cb{Z_j^1}_{j=1}^{n_1}}
\propto
\exp\Bigp{-\eta\p{n_0+n_1}\widehat{\text{BD}}_g\bigp{r}}
\Pi_0\bigp{dr}.
\]
Under $g^{\text{UKL}}$, this generalized posterior agrees with the exponential tilting likelihood described above. Under $g^{\text{SQ}}$, it yields a Gaussian-type pseudo-likelihood when $r$ is linear in parameters, so conjugacy arises. This gives a Bayesian interpretation of the correspondence in the main text: UKL-Riesz is related to calibrated estimation and tailored loss minimization, whereas SQ-Riesz is related to Riesz regression and LSIF.

This also connects directly to generalized Bayesian treatments of covariate balancing. \citet{Orihara2025generalbayesian} propose a general Bayesian framework for causal effects using covariate balancing losses, which matches the UKL-style losses that appear in our Table~\ref{tbl:dre_rre} and Appendix~\ref{appdx:rieszdens}. In particular, the UKL equivalence between KLIEP and the covariate balancing objectives \citep{Zhao2019covariatebalancing, Tan2019regularizedcalbrated} implies that the general Bayesian posterior of \citet{Orihara2025generalbayesian} can be read as a Bayesian analogue of UKL-based density ratio estimation.

\subsection{Bayesian Inference Through Density Ratio Estimation}
A separate and influential line of work reduces Bayesian inference itself to density ratio estimation. Let $\theta\in\Theta$ denote a parameter, $x\in\calX$ denote data, and consider Bayes' rule
\[
p\p{\theta \mid x}
=
\frac{p(\theta)p\p{x\mid\theta}}{p(x)}.
\]
Define the likelihood-to-evidence ratio
\[
r(x,\theta)\coloneqq \frac{p\p{x\mid\theta}}{p(x)}.
\]
Bayes' rule gives $p\p{\theta \mid x}\propto p(\theta)r(x,\theta)$, so estimating $r$ is sufficient to construct the posterior. \citet{Thomas2016likelihoodfree} formalize this idea and estimate $r$ via logistic regression by reframing ratio estimation as probabilistic classification. This is exactly the BKL viewpoint of density ratio estimation \citep{Qin1998inferencesfor, Cheng2004semiparametricdensity}, and it sits naturally in our Bregman--Riesz framework via the BKL loss.

Ratio-based simulation-based inference methods use the same identity. \citet{Hermans2020likelihoodfreeMCMC} and \citet{Greenberg2019automaticposterior} build likelihood-free Bayesian procedures around learned likelihood-to-evidence ratios, connecting modern amortized inference directly to density ratio estimation.

Finally, density ratio estimation also appears inside kernelized Bayesian updating. Kernel Bayes' rule updates prior embeddings into posterior embeddings, and importance weighting formulations explicitly rely on estimating ratios in RKHS \citep{Fukumizu2013kernelbayesrule, Xu2022importanceweightingkbr}. This provides another instance of (C), Bayesian updating expressed as a density ratio estimation subproblem.

\subsection{Correspondence Table}
\label{app:bayes_table}
Table~\ref{tbl:bayes_correspondence} summarizes the main correspondences among (i) the loss choices in generalized Riesz regression, (ii) the associated density ratio estimators in Table~\ref{tbl:dre_rre}, and (iii) the four Bayesian perspectives.

\begin{table*}[t]
\centering
\resizebox{1.0\linewidth}{!}{
\begin{tabular}{lll}
\toprule
Loss choice in generalized Riesz regression
& Bayesian interpretation of ratio or representer
& Representative references
\\
\midrule
SQ-Riesz
& Gibbs posterior with quadratic loss, conjugate Gaussian update under linear models
& \citep{Kanamori2009aleastsquares, Sugiyama2012densityratio, Bissiri2016generalframework}
\\
\midrule
UKL-Riesz
& Exponential tilting likelihood for the density ratio, Bayesian KLIEP
& \citep{Sugiyama2008directimportance, Tan2019regularizedcalbrated, Zhao2019covariatebalancing, Sadeghkhani2019aparametric, Orihara2025generalbayesian}
\\
\midrule
BKL-Riesz
& Posterior computation via likelihood-to-evidence ratio estimation
& \citep{Qin1998inferencesfor, Cheng2004semiparametricdensity, Thomas2016likelihoodfree, Hermans2020likelihoodfreeMCMC, Greenberg2019automaticposterior}
\\
\midrule
General Bregman--Riesz
& Constrained Bayesian nonparametric priors for densities or log ratios
& \citep{Ferguson1973bayesiananalysis, Ghosal2017fundamentalsnonparametric}
\\
\bottomrule
\end{tabular}
}
\caption{Correspondence between generalized Riesz regression losses, density ratio estimation, and Bayesian interpretations. SQ, UKL, and BKL refer to the losses defined in Section~\ref{sec:generalizedrieszregression}. The Bayesian perspectives are Bayesian estimation of the density ratio, constrained Bayesian nonparametric density ratio estimation, generalized Bayesian updating based on the empirical Bregman objective, reducing Bayesian inference to density ratio estimation.}
\label{tbl:bayes_correspondence}
\end{table*}

\section{Nonlinear Estimands in Regression Functions}
\label{appdx:nonlinear_estimands_gamma}

This appendix extends our baseline setup to estimands that are nonlinear in the regression nuisance $\gamma_0$. The main text focuses on linear and continuous maps $\gamma \mapsto \bbE\sqb{m(W,\gamma)}$, which yield a Riesz representer $\alpha_0$ that does not depend on $\gamma_0$ and enable generalized Riesz regression via Bregman divergence minimization. Many causal and structural targets are instead smooth nonlinear functionals of one or more regressions. \citet{Chernozhukov2022automaticdebiased} propose an automatic debiased machine learning construction for such nonlinear effects based on orthogonal scores whose bias corrections involve derivatives of the target map. We show that the same construction fits naturally into our framework by replacing the original linear map with its Gateaux derivative at $\gamma_0$, which is itself a linear functional and therefore admits a (local) Riesz representer \citep{Chernozhukov2022locallyrobust,Ichimura2022influencefunction}.

\subsection{General Construction Via Gateaux Derivatives}
\label{appdx:nonlinear_general}

\paragraph{Setup.}
We keep the notation of Section~\ref{sec:setup}. Let $\gamma_0(x)=\bbE\sqb{Y\mid X=x}$ and consider a parameter
\begin{align}
\label{eq:nonlinear_theta_def}
\theta_0 \coloneqq \bbE\sqb{m(W,\gamma_0)},
\end{align}
where $m(W,\gamma)$ may be nonlinear in $\gamma$. We assume $m(W,\gamma)$ is well-defined for $\gamma \in \calH$, where $\calH=\cb{\gamma\colon \calX\to\bbR\mid \bbE\sqb{\gamma(X)^2}<\infty}$.

\paragraph{Gateaux derivative and local Riesz representer.}
For $\gamma \in \calH$ and a direction $h \in \calH$, define the Gateaux derivative
\begin{align}
\label{eq:gateaux_derivative_m}
\dot m_{\gamma}\p{W,h} \coloneqq \partial_{\tau} m\p{W,\gamma+\tau h}\Big|_{\tau=0}.
\end{align}
We impose a mean-square continuity condition for the derivative map at $\gamma_0$, analogous to the linear continuity condition used for the standard Riesz representer.

\paragraph{Derivative continuity.}
Assume that $h \mapsto \bbE\sqb{\dot m_{\gamma_0}\p{W,h}}$ is linear and continuous on $\calH$, equivalently, there exists $C>0$ such that
\begin{align}
\label{eq:derivative_continuity}
\bbE\sqb{\dot m_{\gamma_0}\p{W,h}}^2 \leq C\,\bbE\sqb{h(X)^2}
\end{align}
for all $h \in \calH$.

Under \eqref{eq:derivative_continuity}, by the Riesz representation theorem, there exists a function $\alpha_0^{\text{NL}}\in\calH$ such that
\begin{align}
\label{eq:local_riesz}
\bbE\sqb{\dot m_{\gamma_0}\p{W,h}}=\bbE\sqb{\alpha_0^{\text{NL}}(X)h(X)}
\end{align}
for all $h\in\calH$. We refer to $\alpha_0^{\text{NL}}$ as the local Riesz representer for the nonlinear target \eqref{eq:nonlinear_theta_def}. When $m(W,\gamma)$ is linear in $\gamma$, $\dot m_{\gamma_0}\p{W,h}=m(W,h)$ and \eqref{eq:local_riesz} reduces to the standard Riesz representer identity in Section~\ref{sec:setup}.

\paragraph{Orthogonal score and Riesz representer.}
Let $\eta \coloneqq (\gamma,\alpha)$ with $\alpha$ intended to estimate $\alpha_0^{\text{NL}}$. Define the score
\begin{align}
\label{eq:nonlinear_orthogonal_score}
\psi^{\text{NL}}\p{W;\eta,\theta}
\coloneqq
m(W,\gamma) + \alpha(X)\p{Y-\gamma(X)} - \theta.
\end{align}
Let $\eta_0^{\text{NL}} \coloneqq (\gamma_0,\alpha_0^{\text{NL}})$. Then, $\bbE\sqb{\psi^{\text{NL}}\p{W;\eta_0^{\text{NL}},\theta_0}}=0$ holds by \eqref{eq:nonlinear_theta_def} and $\bbE\sqb{Y-\gamma_0(X)\mid X}=0$.

Equation~\eqref{eq:nonlinear_orthogonal_score} is also Neyman orthogonal at $\eta_0^{\text{NL}}$ under \eqref{eq:local_riesz}. Indeed, for $h\in\calH$ and the path $\gamma_t=\gamma_0+t h$, differentiation gives
\[
\partial_t \bbE\sqb{\psi^{\text{NL}}\p{W;\p{\gamma_t,\alpha_0^{\text{NL}}},\theta_0}}\Big|_{t=0}
=
\bbE\sqb{\dot m_{\gamma_0}\p{W,h}}-\bbE\sqb{\alpha_0^{\text{NL}}(X)h(X)}=0,
\]
where the last equality follows from \eqref{eq:local_riesz}. The derivative with respect to $\alpha$ vanishes because $\bbE\sqb{Y-\gamma_0(X)}=0$.

\paragraph{Second-order remainder and rate conditions.}
A key difference from the linear case is the presence of a second-order remainder term. For any candidate $\gamma\in\calH$, define
\begin{align}
\label{eq:nonlinear_remainder}
R_m\p{\gamma,\gamma_0}
\coloneqq
\bbE\sqb{
m(W,\gamma)-m(W,\gamma_0)-\dot m_{\gamma_0}\p{W,\gamma-\gamma_0}
}.
\end{align}
Using \eqref{eq:local_riesz}, we can write the score drift as
\begin{align}
\label{eq:nonlinear_score_drift}
\bbE\sqb{\psi^{\text{NL}}\p{W;\eta,\theta_0}}
=
\bbE\sqb{\p{\alpha_0^{\text{NL}}(X)-\alpha(X)}\p{\gamma(X)-\gamma_0(X)}} + R_m\p{\gamma,\gamma_0}.
\end{align}
For linear $m$, $R_m\p{\gamma,\gamma_0}=0$, recovering the purely product-form drift. For nonlinear $m$, a sufficient condition for $R_m\p{\gamma,\gamma_0}$ to be second order is a Lipschitz property of the derivative map, for example, the existence of $L>0$ such that
\[
\bbE\sqb{
\dot m_{\gamma}\p{W,h} - \dot m_{\gamma_0}\p{W,h}
}
\leq
L\Bigp{\bbE\sqb{\p{\gamma(X)-\gamma_0(X)}^2}}^{1/2}
\Bigp{\bbE\sqb{h(X)^2}}^{1/2}.
\]
Then, $|R_m\p{\gamma,\gamma_0}|$ is controlled by $\bbE\sqb{\p{\gamma(X)-\gamma_0(X)}^2}$ via a standard Taylor remainder argument, which yields the familiar requirement $\bbE\sqb{\p{\widehat{\gamma}(X)-\gamma_0(X)}^2}^{1/2}=o_p\p{n^{-1/4}}$ for the remainder to be $o_p\p{n^{-1/2}}$.

Under an appropriate Donsker condition, or when the nuisance parameters are estimated via cross-fitting, the ARW estimator built from \eqref{eq:nonlinear_orthogonal_score} is root-$n$ consistent and asymptotically normal if the following conditions hold:
\[
\bbE\sqb{\p{\widehat{\alpha}(X)-\alpha_0^{\text{NL}}(X)}^2}^{1/2}
\bbE\sqb{\p{\widehat{\gamma}(X)-\gamma_0(X)}^2}^{1/2}
=
o_p\p{n^{-1/2}},
\]
together with $R_m\p{\widehat{\gamma},\gamma_0}=o_p\p{n^{-1/2}}$. This parallels the nonlinear-rate discussion in \citet{Chernozhukov2022automaticdebiased}.

\subsection{Generalized Riesz Regression for Local Riesz Representers}
Our generalized Riesz regression estimator in Section~\ref{sec:generalizedrieszregression} relies on the identity
\[
\bbE\sqb{u(X)\alpha_0(X)}=\bbE\sqb{m\p{W,u}}
\]
for linear $u$, which follows from linearity of $\gamma \mapsto \bbE\sqb{m(W,\gamma)}$. For nonlinear $m$, we instead use the derivative identity \eqref{eq:local_riesz}. Define the linear-in-$h$ map
\begin{align}
\label{eq:linearized_m_map}
m^{\text{lin}}_0\p{W,h}\coloneqq \dot m_{\gamma_0}\p{W,h}.
\end{align}
Then, for any $h\in\calH$,
\[
\bbE\sqb{m^{\text{lin}}_0\p{W,h}}=\bbE\sqb{\alpha_0^{\text{NL}}(X)h(X)}.
\]
Therefore, generalized Riesz regression applies verbatim with $m$ replaced by $m^{\text{lin}}_0$.

Specifically, let $g\colon \calA\to\bbR$ be a differentiable strictly convex generator and consider the same Bregman objective as in Section~\ref{sec:generalizedrieszregression}, but with $m^{\text{lin}}_0$:
\begin{align}
\label{eq:nonlinear_bd_objective}
\text{BD}^{\text{NL}}_g\bigp{\alpha}
\coloneqq
\BigExp{
-g\p{\alpha(X)} + \partial g\p{\alpha(X)}\alpha(X) - m^{\text{lin}}_0\Bigp{W,\p{\partial g}\circ\alpha}
}.
\end{align}
In practice, $m^{\text{lin}}_0$ is unknown because it depends on $\gamma_0$. Following \citet{Chernozhukov2022automaticdebiased}, we plug a preliminary estimator $\widehat{\gamma}$ into the derivative; when a Donsker condition is not imposed, $\widehat{\gamma}$ may be estimated via cross-fitting, and we define
\[
\widehat m^{\text{lin}}\p{W,h}
\coloneqq
\partial_{\tau} m\p{W,\widehat{\gamma}+\tau h}\Big|_{\tau=0}.
\]
We then solve the empirical analogue of \eqref{eq:nonlinear_bd_objective} with regularization,
\begin{align}
\label{eq:nonlinear_grr}
\widehat\alpha^{\mathrm{NL}}
\in\argmin_{\alpha\in\calH}\quad
&\frac1n\sum_{i=1}^n\Bigp{-g(\alpha(X_i))+\partial g(\alpha(X_i))\alpha(X_i)}\nonumber\\
&-\frac1n\sum_{i=1}^n\widehat m^{\mathrm{lin}}\Bigp{W_i,(\partial g)\circ\alpha}
+\lambda J(\alpha).
\end{align}
When $g=g^{\text{SQ}}$ and $\calH$ is linear-in-parameters, \eqref{eq:nonlinear_grr} recovers the squared-loss Riesz regression structure used in automatic debiased machine learning for nonlinear effects, where the key difference from the linear case is that the moment vector involves derivatives of $m$ evaluated at $\widehat{\gamma}$, see Section 5 of \citet{Chernozhukov2022automaticdebiased}.

Because $\widehat m^{\text{lin}}$ itself uses $\widehat{\gamma}$, an additional layer of sample splitting can be helpful so that the derivative evaluation used to construct $\widehat{\alpha}^{\text{NL}}$ is conditionally independent of the sample used to fit $\widehat{\alpha}^{\text{NL}}$, following the multiple cross-fitting construction in \citet{Chernozhukov2022automaticdebiased}. The same construction can be used here: one may estimate $\widehat{\gamma}$ on one fold, construct $\widehat m^{\text{lin}}$ on a second fold, and fit $\widehat{\alpha}^{\text{NL}}$ on a third fold, then rotate folds and aggregate.

\subsection{Applications}
\label{appdx:nonlinear_applications}

\subsubsection{Smooth Pointwise Transforms of Regression Functions}
Let $\phi\colon \bbR\to\bbR$ be differentiable and define
\begin{align}
\label{eq:phi_gamma}
\theta_0 = \bbE\sqb{\phi\p{\gamma_0(X)}},
\qquad
m\p{W,\gamma} \coloneqq \phi\p{\gamma(X)}.
\end{align}
Then, the Gateaux derivative is
\[
\dot m_{\gamma_0}\p{W,h} = \phi'\p{\gamma_0(X)}h(X),
\]
so the local Riesz representer is
\[
\alpha_0^{\text{NL}}(X)=\phi'\p{\gamma_0(X)}.
\]
The orthogonal score \eqref{eq:nonlinear_orthogonal_score} becomes
\[
\psi^{\text{NL}}\p{W;\eta,\theta}
=
\phi\p{\gamma(X)} + \alpha(X)\p{Y-\gamma(X)} - \theta,
\]
with $\alpha_0^{\text{NL}}(X)=\phi'\p{\gamma_0(X)}$. If $\phi'$ is Lipschitz, the remainder $R_m\p{\gamma,\gamma_0}$ in \eqref{eq:nonlinear_remainder} is quadratic in $\gamma-\gamma_0$, which leads to the familiar $n^{-1/4}$ requirement on the regression rate for root-$n$ inference.

This class contains many targets as special cases, including quadratic, logarithmic, and exponential functionals. Below we spell out representative examples.

\paragraph{Second moment of the regression.}
Take $\phi(u)=u^2$, so $\theta_0=\bbE\sqb{\gamma_0(X)^2}$. Then, we have $\alpha_0^{\text{NL}}(X)=2\gamma_0(X)$, and one convenient orthogonal score is obtained by setting $\alpha(X)=2\gamma(X)$,
\[
\psi^{\text{NL}}\p{W;\gamma,\theta}
=
\gamma(X)^2 + 2\gamma(X)\p{Y-\gamma(X)} - \theta.
\]
This estimator requires only an estimator of $\gamma_0$ before the orthogonal score is evaluated.

\paragraph{Log-welfare functional.}
Assume $\gamma_0(X)>0$ almost surely and take $\phi(u)=\log(u)$, so that $\theta_0=\bbE\sqb{\log\p{\gamma_0(X)}}$ and $\alpha_0^{\text{NL}}(X)=1/\gamma_0(X)$. The score is
\[
\psi^{\text{NL}}\p{W;\eta,\theta}
=
\log\p{\gamma(X)} + \alpha(X)\p{Y-\gamma(X)} - \theta,
\]
with $\alpha_0^{\text{NL}}(X)=1/\gamma_0(X)$. This is an example where the representer depends on an inverse regression, and estimating $\alpha_0^{\text{NL}}$ directly via \eqref{eq:nonlinear_grr} can be preferable to naive inversion when $\widehat{\gamma}$ is close to zero in some regions.

\subsubsection{Nonlinear Maps of Finite-Dimensional Linear Functionals}
Many parameters are smooth maps of a finite number of linear functionals. Let $\bmtheta_0\in\bbR^J$ collect $J$ linear functionals,
\[
\bmtheta_{0,j}\coloneqq \bbE\sqb{m_j\p{W,\gamma_0}},
\qquad
j\in\cb{1,\dots,J},
\]
where each map $\gamma \mapsto \bbE\sqb{m_j(W,\gamma)}$ is linear and continuous, with Riesz representer $\alpha_{0,j}(X)$ in the sense of Section~\ref{sec:setup}. Let $\varphi\colon \bbR^J\to\bbR$ be differentiable and define the nonlinear parameter
\[
\theta_0\coloneqq \varphi\p{\bmtheta_0}.
\]
An asymptotically linear estimator is obtained by estimating each $\bmtheta_{0,j}$ by an ARW estimator and then applying $\varphi$. The influence function for $\theta_0$ is given by the delta method,
\[
\sum^J_{j=1}\partial_j \varphi\p{\bmtheta_0}
\Bigp{
m_j\p{W,\gamma_0} + \alpha_{0,j}(X)\p{Y-\gamma_0(X)} - \bmtheta_{0,j}
}.
\]
This construction is useful for ratio and variance-type effects built from ATE-like building blocks.

\paragraph{Ratio-type effects.}
As a concrete example, suppose $X=(D,Z)$ and define $\mu_{d,0}\coloneqq \bbE\sqb{\gamma_{d,0}(Z)}$, where $\gamma_{d,0}(z)=\bbE\sqb{Y\mid D=d,Z=z}$. Consider the risk ratio $\theta_0=\mu_{1,0}/\mu_{0,0}$, for which $J=2$ and $\varphi\p{u_1,u_0}=u_1/u_0$. The influence function is
\[
\frac{1}{\mu_{0,0}}\text{EIF}_{\mu_{1,0}}(W) - \frac{\mu_{1,0}}{\mu_{0,0}^2}\text{EIF}_{\mu_{0,0}}(W),
\]
where each $\text{EIF}_{\mu_{d,0}}(W)$ is the efficient influence function for $\mu_{d,0}$ in the corresponding model. Estimation proceeds by estimating the two Riesz representers (or using their closed forms when available) and then applying the delta method.

\subsubsection{Causal Mediation as a Nonlinear Effect of Multiple Regressions}
Nonlinear causal targets often depend on multiple regressions. A representative example is causal mediation \citep{Farbmacher2022causalmediation}, which \citet{Chernozhukov2022automaticdebiased} study as a nonlinear effect of several regressions.

Let $W=(Y,D,Q,Z)$, where $D\in\cb{0,1}$ is treatment, $Q$ is a mediator, and $Z$ are covariates. Define
\[
\rho_0(k,z)\coloneqq
\gamma_{Y,0}(d,k,z)\gamma_{Q,0}(k,d',z).
\]
Consider the mediation-type parameter
\[
\theta_0=\bbE\left[\sum_k\rho_0(k,Z)\right].
\]
Here,
\[
\gamma_{Y,0}(d,q,z)=\bbE\sqb{Y\mid D=d,Q=q,Z=z},
\]
and
\[
\gamma_{Q,0}(k,d',z)=\bbE\sqb{\bbI\p{Q=k}\mid D=d',Z=z}.
\]
The parameter is nonlinear in $\gamma_0=(\gamma_{Y,0},\gamma_{Q,0})$.

The orthogonal score has the general multi-regression form in \citet{Chernozhukov2022automaticdebiased},
\[
m\p{W,\gamma}
+ \alpha_Y\p{X_Y}\p{Y-\gamma_Y\p{X_Y}}
+ \sum_k \alpha_{Q,k}\p{X_Q}\p{\bbI\p{Q=k}-\gamma_{Q,k}\p{X_Q}}
- \theta,
\]
where $X_Y=(D,Q,Z)$ and $X_Q=(D,Z)$, and the representers are the local Riesz representers for the Gateaux derivatives with respect to each regression component. In this example, the derivative of $m$ with respect to $\gamma_Y$ produces weights proportional to $\gamma_Q$, and the derivative with respect to $\gamma_Q$ produces weights proportional to $\gamma_Y$. This yields a concrete instance where local Riesz representers depend on the other regression nuisances, and it motivates estimating the representers directly, using derivative-based moments, rather than relying on plug-in inversions.

Generalized Riesz regression extends this logic beyond squared loss by applying Equation~\eqref{eq:nonlinear_grr} separately to each component representer with the appropriate linearized map $m^{\text{lin}}_0$ defined by the partial Gateaux derivatives, thereby preserving the automatic representer-fitting viewpoint while accommodating the nonlinear structure of the target.

\section{Function Spaces and Generator Curvature}
\label{appdx:function_spaces_curvature}

\subsection{Projection Onto Functions of a Summary}
To compare the fixed regressor span with functions of a measurable summary, let $S=s(X)$ be fixed and define
\[
\calF_S\coloneqq L_2(\sigma(S)),
\qquad
\calF_{\bmphi}\coloneqq\operatorname{span}\{\widetilde\phi_1,\ldots,\widetilde\phi_p\}.
\]

\begin{proposition}[Projection onto functions of a summary]
\label{prop:summary_space_projection}
For every $f\in L_2(P_X)$, the orthogonal projection onto $\calF_S$ is
\[
\Pi_S f=\bbE[f(X)\mid S],
\]
and
\[
\|(I-\Pi_S)f\|_{L_2(P_X)}^2
=\bbE\sqb{\Var(f(X)\mid S)}.
\]
If $\bbE[\Var(\widetilde\phi_j(X)\mid S)]>0$ for some $j$, then $\calF_{\bmphi}\not\subseteq\calF_S$. If $\dim(\calF_S)>p$, then $\calF_S\not\subseteq\calF_{\bmphi}$. Finally, for the closed sum $\calF_{\bmphi,S}\coloneqq\overline{\calF_{\bmphi}+\calF_S}$,
\[
\|(I-\Pi_{\bmphi,S})f\|_{L_2(P_X)}
\leq
\min\left(
\|(I-\Pi_{\bmphi})f\|_{L_2(P_X)},
\|(I-\Pi_S)f\|_{L_2(P_X)}
\right).
\]
\end{proposition}

\begin{proof}
Conditional expectation is the orthogonal projection from $L_2(P_X)$ onto $L_2(\sigma(S))$, which gives the first two displays. The two non-inclusion statements follow from the positive projection residual and the dimension condition, respectively. The final inequality follows because both $\calF_{\bmphi}$ and $\calF_S$ are contained in $\calF_{\bmphi,S}$.
\end{proof}

\subsection{Curvature Condition Numbers}
To quantify inverse-link sensitivity on a common admissible range, write $u=|\alpha|$ on either sign-specific component and suppose that $u\in[m,M]$ with $0<C<m<M$. Define
\[
\kappa_g(m,M)\coloneqq
\frac{\sup_{u\in[m,M]}g''(u)}{\inf_{u\in[m,M]}g''(u)}.
\]

\begin{proposition}[Curvature condition numbers]
\label{prop:generator_condition_numbers}
For $0<\omega\leq1$, the principal generators satisfy
\[
\begin{aligned}
\kappa_{\mathrm{SQ}}&=1, &
\kappa_{\mathrm{BP}(\omega)}&=\left(\frac{M-C}{m-C}\right)^{1-\omega},\\
\kappa_{\mathrm{UKL}}&=\frac{M-C}{m-C}, &
\kappa_{\mathrm{BKL}}&=\frac{M^2-C^2}{m^2-C^2}.
\end{aligned}
\]
If $0<\omega<1$ and $M>m$, then
\[
\kappa_{\mathrm{SQ}}<\kappa_{\mathrm{BP}(\omega)}<\kappa_{\mathrm{UKL}}<\kappa_{\mathrm{BKL}}.
\]
For the ATE, if $e_0(Z)\in[\eta,1-\eta]$ with $0<\eta<1/2$, then $m=1/(1-\eta)$ and $M=1/\eta$. As $\eta\downarrow0$, fixed $C<1$ gives orders $1$, $\eta^{-(1-\omega)}$, $\eta^{-1}$, and $\eta^{-2}$, whereas $C=1$ gives $1$, $\eta^{-2(1-\omega)}$, $\eta^{-2}$, and $\eta^{-3}$.
\end{proposition}

\begin{proof}
The formulas follow from the second derivatives $2$, $(1+\omega)(u-C)^{\omega-1}$, $(u-C)^{-1}$, and $2C/(u^2-C^2)$ and their monotonicity on $[m,M]$. The ATE range follows from the inverse propensity representation. If $C=1$, then $m-C=\eta/(1-\eta)$, which yields the stated orders.
\end{proof}